\documentclass[twocolumn]{aastex631}
\usepackage{comment} 
\usepackage{xcolor}
\usepackage{graphicx}
\usepackage{float}
\usepackage{soul}
\shorttitle{Star Formation and AGNs in $CLASH$ BCGs}
\shortauthors{Levitskiy et al.}
\graphicspath{{./}{figures/}}

\begin{document}

\title{Star Formation, Nebulae, and Active Galactic Nuclei in $CLASH$ Brightest Cluster Galaxies: \\ I.  Dependence on Core Entropy of Intracluster Medium}

\author{Arsen Levitskiy}
\affiliation{Department of Physics, The University of Hong Kong, Pokfulam Road, Hong Kong}

\author{Jeremy Lim}
\affiliation{Department of Physics, The University of Hong Kong, Pokfulam Road, Hong Kong}

\author{Youichi Ohyama}
\affiliation{Academia Sinica, Institute of Astronomy and Astrophysics, 11F of Astronomy-Mathematics Building, No.1, Section\,4, Roosevelt Rd., Taipei 10617, Taiwan, R.O.C.}

\author{Juno Li}
\affiliation{Department of Physics, The University of Hong Kong, Pokfulam Road, Hong Kong}
\affiliation{International Centre for Radio Astronomy Research (ICRAR) and the International Space Centre (ISC), The University of Western Australia, M468, 35 Stirling Highway, Crawley, WA 6009, Australia}

\author{Megan Donahue}
\affiliation{Physics and Astronomy Dept., Michigan State University, East Lansing, MI, 48824 USA}



\begin{abstract}
We set the stage for reassessing how star formation, emission-line nebulae, and active galactic nuclei (AGNs) in Brightest Cluster Galaxies (BCGs) depend on the thermodynamics of the intracluster medium (ICM).  Our work is based on the 25 clusters observed in the $CLASH$ program for which the aforementioned attributes in their BCGs can be well scrutinised, as has the thermodynamics of their ICM.  Nine of these BCGs display complex UV morphologies tracing recent star formation, whereas the remaining sixteen are characterised by a relatively compact central UV enhancement.  Here, we show definitively that three of the latter BCGs also display star formation, whereas the diffuse UV of the remaining thirteen is entirely consistent with old low-mass stars. The overall results support the previously established dependence of star formation and nebulae in BCGs on an ``excess core entropy'', $K_0$, for the ICM: all eleven clusters having $K_0 \le 24 \rm \, keV \, cm^2$, but only one of fourteen clusters having $K_0 \ge 42 \rm \, keV \, cm^2$, host star-forming BCGs that almost if not always possess nebulae.   Instead of an entropy floor, we show that $K_0$ reflects the degree to which the radial entropy profile decreases inwards within $\sim$100\,kpc rather than (except perhaps at large $K_0$) actually flattening: clusters having lower ICM entropies and hence shorter cooling times at their cores preferentially host BCGs displaying star formation, nebulae, and more radio-luminous AGNs.  Nearly all BCGs possess detectable AGNs, however, indicating multiple pathways for fuelling their AGNs.

\end{abstract}


\section{Introduction} \label{sec:Intro}
The brightest, largest, and most massive galaxies in the universe reside at the centres of galaxy clusters.  The manner by which these Brightest Cluster Galaxies (BCGs) grew to their enormous masses and, especially, sizes challenges cosmological models for stellar assembly at the largest extremes.  Early ideas imagined a primarily two-phase process:
(i) an initial collapse with rapid cooling and star formation at high redshift \citep{Eggen1962}; followed by (ii) cannibalisms of relatively low-mass galaxies that are largely devoid of cool gas \citep{White1976, Ostriker1977}.  Such dry minor mergers are especially effective at promoting galaxy growth in stellar size, as the cannibalised galaxy is disrupted at the outskirts of the dominant galaxy.  More recent studies, however, suggest a more complicated history of BCG stellar growth \citep[e.g., see summary in][]{COoke2019}, with additional contributions from mergers with relatively massive galaxies \citep[e.g.,][]{Kluge2023}, in situ star formation fuelled by either mergers with gas-rich galaxies \citep[e.g.,][]{Bonaventura2017} or cooling of the intracluster medium (this manuscript and the references cited), and tidal interactions that strip stars from cluster members to grow both the BCG and its surrounding intracluster light \citep[e.g.,][]{Marx2023}.  The relative contributions of the different proposed mechanisms for BCG stellar growth during different epochs of cosmic history remain contentious.

In this paper, we focus on whether ICM cooling is responsible for fuelling star formation in BCGs, and the conditions under which such cooling and star formation occurs.  While evidence for star formation have been found in BCGs up to high redshifts \citep[e.g.,][]{Bonaventura2017}, our focus here is on BCGs at redshifts up to $z \approx 0.5$ that are hosted by relatively massive clusters ($\gtrsim 10^{14} \, \rm M_\sun$).  There are three key arguments for why ICM cooling rather than galaxy mergers is responsible for fuelling star formation in these particular BCGs:

\begin{itemize}
\item[I.]  For cluster members to merge with the BCG, they first need to be slowed dramatically by dynamical friction -- by which time ram-pressure stripping may have largely depleted these galaxies of cool gas.  By contrast, BCGs engaged in star formation have been inferred to contain as much as $\sim$$10^{10}$--$10^{11} \rm \, M_\sun$ of molecular gas \citep[e.g.,][]{Edge2001, Salome2003, Salome2011}.  The situation for BCGs is therefore very different than field elliptical galaxies, whereby cool gas acquired through close encounters or mergers with other galaxies can temporarily fuel their star formation.

\item[II.]  BCGs engaged in star formation and/or displaying luminous emission-line nebulae have been found to preferentially reside in clusters having relatively low ICM entropies at their cores (central radii of order tens of kpc) -- providing circumstantial evidence that both phenomena are fuelled by cooling of their surrounding ICM (see Section\,\ref{subsec:connecting activity entropy}).  We note here that emission-line nebulae in BCGs can have spectra very different from H\,II regions \citep[][]{Crawford1999,Ferland2009}, and therefore ought not to be automatically equated to star-forming regions (see Section\,\ref{sec:Results}).  This situation is most clearly illustrated by NGC\,1275, the BCG in the Perseus cluster, in which the majority of newly-formed star clusters are spatially displaced from its emission-line nebula.  \citet{Ferland2009} argue that energetic particles are responsible for powering the emission-line nebula in NGC\,1275; \citet{Fabian2011} associate the energetic particles with electrons from the ICM that penetrate into the nebula.  In this viewpoint, the emission-line nebulae of BCGs constitute gas deposited by ICM cooling, rather than gas photoionised by newly-formed massive stars.

\item[III.]  If ICM cooling is responsible for fuelling star formation in BCGs, then their star formation ought to be persistent over time.  Indeed, prolonged star formation that is still ongoing has been found \citep{Lim2020} or inferred \citep{Mittal2015, Fogarty2017} for a number of BCGs, the clearest example of which is (again) NGC\,1275.  This galaxy has formed over a thousand star clusters at an approximately constant rate over the past, at least, $\sim$1\,Gyr \citep{Lim2020}.  The stars clusters formed have masses and sizes comparable to globular clusters: at ages older than $\sim$1\,Gyr, these star clusters become difficult to differentiate from globular clusters, and may therefore contribute to the enormous populations of globular clusters (numbering tens of thousands) seen around BCGs.  Those that are disrupted by strong tidal fields as they orbit close to the centre of the BCG contribute to the stellar growth of the galaxy.

\end{itemize}

Over the past decade, studies of galaxy clusters at redshifts of up to $z \sim 1$ have brought about a radical revision in our understanding of their ICM thermodynamics.  Rather than displaying different ICM core entropies, the vast majority of clusters appear to display a steady decrease in ICM entropy inwards (see Section\,\ref{subsec:entropy floor}) -- thus calling into question a dependence between star formation and/or emission-line nebulae in BCGs and the ICM core entropies of their host clusters \citep[e.g., see][]{Hogan2017b}.  In the remainder of this Section, we first set the stage for any debate linking ICM cooling with star formation and/or emission-line nebulae in BCGs: briefly summarising familiar evidence against catastrophic ICM cooling at a rate commensurate with its cooling timescale (Section\,\ref{subsec:core entropy}).  We then describe in more granular detail the relationships found between star formation and/or emission-line nebulae in BCGs and the ICM core entropy of their host clusters (Section\,\ref{subsec:connecting activity entropy}) -- before calling into the question the validity of these relationships based on our revised understanding of ICM thermodynamics (Section\,\ref{subsec:entropy floor}).  In the last part of this Section, we introduce our work to re-examine the relationship between star formation and emission-line nebulae in BCGs and the ICM thermodynamics of their host clusters (Section\,\ref{subsec:goals}).

\subsection{ICM Entropy and Gas Cooling} \label{subsec:core entropy}

The entropy of the hot X-ray emitting gas that permeates galaxy groups and clusters is (most commonly) defined as $K \equiv k_{\rm B} \, T_{\rm X} \, n_{\rm e}^{-2/3}$, where $k_{\rm B}$ is Boltzman's constant, $T_{\rm X}$ the gas temperaure, and $n_{\rm e}$ the electron density.  In the gravitational potential well of a cluster, higher-entropy (less dense and warmer) gas rises whereas lower-entropy (denser and cooler) gas sinks, thus leaving only the lowest-entropy gas at the cluster core by the time the ICM reaches hydrostatic equilibrium.  As the X-ray luminosity $L_{\rm X} \propto n_{\rm e}^{2} \, T_{\rm X}^{1/2}$, X-ray gas with lower entropies radiates more prodigiously and therefore cools faster:\,\,for emission purely through thermal bremsstrahlung (i.e., no significant line emission, as is the case when $T_{\rm X} \gtrsim 3 \times 10^7 \rm \, K$), the timescale for such gas to radiate away all its thermal energy is $t_{\rm cool} \approx 10^{8} {\rm \, yr} \, [K/(10 {\rm \, keV \, cm^2})]^{3/2} \, [k \, T_{\rm X} / (5 {\rm \, keV})]^{-1}$ \citep{Donahue2005}.

In some clusters, the cooling time of the ICM at the cluster core is much shorter than a Hubble time \citep[][]{Cowie1977,Fabian1977}.  Yet, no gas radiating in X-rays at temperatures significantly below the bulk temperature of the ICM at the cluster core has been detected \citep[beginning with work by][]{Molendi2001,David2001,Peterson2003}, placing upper limits on the ICM cooling rate that are about an order of magnitude lower than the levels predicted \citep{Peterson2003}.  Radiative cooling of the gas must therefore be mitigated by reheating, for which the most widely-accepted source is jets from an AGN in BCGs \citep[e.g., review by][]{Fabian2012}.  Such jets are observed to inflate large cavities (buoyant bubbles of relativistic plasma) in the ICM around BCGs, and via a process that remains contentious convert their mechanical energy into ICM thermal energy.  
AGN reheating, however, may not completely balance ICM cooling in all clusters: in such cases, some gas presumably cools catastrophically to produce emission-line nebulae in BCGs and fuel their star formation.  The manner by which such cooling occurs remain contentious, with ideas ranging from thermal instabilities owing to gas cooling times shorter than their free-fall times \citep[see][and references therein]{Voit2017} to a dynamic interplay between AGN jets and uplifted ICM gas \citep[][]{Qiu2020}.

\subsection{Dependence of BCG activity on ICM Core Entropy}\label{subsec:connecting activity entropy}

A number of studies have found a clear relationship between star formation, emission-line nebulae, and radio AGNs in BCGs -- hereafter, collectively referred to as BCG activity -- and the ICM core entropy of their host clusters.  First, based on a study of optical colour gradients in forty-six BCGs, \citet{Rafferty2008} classified these galaxies as:\,(i) star forming if they exhibit increasing bluer colours inward; and (ii) non-star-forming if they exhibit no detectable change or increasingly redder colours inward (the latter as would be expected owing to a well known metallicity gradient in elliptical galaxies).  All twenty of the BCGs classified as star forming inhabit clusters having entropies as measured over a central radius of 12\,kpc of $K_{\rm 12 \, kpc} \lesssim 30 \rm \, keV \, cm^2$.  Among the remaining twenty-six classified as non-star-forming, fourteen also inhabit clusters having $K_{\rm 12 \, kpc} \lesssim 30 \rm \, keV \, cm^2$, whereas the remaining twelve inhabit clusters having $K_{\rm 12 \, kpc} > 30 \rm \, keV \, cm^2$.  \citet{Rafferty2008} suggest that the lack of star formation among BCGs inhabiting clusters having $K_{\rm 12 \, kpc} > 30 \rm \, keV \, cm^2$ is caused by excess AGN re-heating of the surrounding ICM, thus preventing any net cooling.

Second, based on a study of two hundred and twenty-two clusters, \citet{Cavagnolo2008} reported a relationship between the ``excess core entropy'' of the ICM, $K_0$, and the onset of luminous H$\alpha$+[N\,II] nebulae in BCGs.  The concept of an excess core entropy was introduced by \citet{Cavagnolo2009}, who found that the ICM entropy decreases radially inwards as a power law before seemingly levelling off to a constant value at small radii: $K(r) = K_0 + K_{100} (r/100 {\rm \, kpc})^\alpha$, where $K_0$ quantifies the excess in core entropy above the best-fitting power law found at larger radii.  (All values of $K_0$ quoted in this manuscript are from \cite{Cavagnolo2009}, except for MACS J2129.4-0741 and MACS J0416.1-2403 as reported by \citealt{Donahue2015}.)  \citet{Cavagnolo2008} found that nearly all the BCGs inhabiting clusters having $K_0 \lesssim 30 \rm \, keV \, cm^2$ exhibit luminous H$\alpha$+[N\,II] nebulae, whereas none of the clusters having $K_0 > 30 \rm \, keV \, cm^2$ exhibit detectable H$\alpha$+[N\,II].  In addition, \citet{Cavagnolo2008} found that BCGs in clusters having $K_0 \lesssim 30 \rm \, keV \, cm^2$ preferentially exhibit more radio-luminous AGNs than those in clusters having $K_0 > 30 \rm \, keV \, cm^2$, although AGNs spanning a broad range of moderate radio luminosities were found in BCGs irrespective of $K_0$ for their host clusters.

Subsequent studies have only buttressed the dependence of BCG activity on ICM entropy at the cluster core as found by \citet{Rafferty2008} and  \citet{Cavagnolo2008}.  First, based on observations in the near-UV (NUV) from {\it GALEX} and near-IR K-band from {\it Spitzer}, \citet{Hoffer2012} found that just over one-third of BCGs hosted by clusters having $K_0 \lesssim 30 \rm \, keV \, cm^2$ display NUV$-$K colours that are significant bluer -- attributed to recently-formed stars -- than the remainder, which display NUV$-$K colours scattered around the same mean value; by contrast, no BCGs hosted by clusters spanning $30 < K_0 < 700 \rm \, keV \, cm^2$ display bluer NUV$-$K colours.  The onset of star formation in BCGs at $K_0 \lesssim 30 \rm \, keV \, cm^2$ is similar to that found by \citet{Cavagnolo2008} for the onset of emission-line nebulae as well as more radio-luminous AGNs among BCGs.  Strongly reinforcing these results and pointing to an even stronger dependence, all eleven of the $CLASH$ BCGs reported by \citet{Donahue2015} and/or \citet{Fogarty2015} to be engaged in star formation and to possess optical emission-line nebulae are hosted by clusters having $K_0 \leq 24 \rm \, keV \, cm^2$.  By contrast, the remaining fourteen $CLASH$ BCGs are hosted by clusters having $K_0 \geq 42 \rm \, keV \, cm^2$.

\subsection{No ICM Entropy Floor?}  \label{subsec:entropy floor}

Bolstered by the reported relationship between BCG activity and $K_0$, the excess core entropy introduced by \citet{Cavagnolo2009} has since been widely interpreted as an inherent entropy floor for the ICM at the cluster core -- despite \citet{Cavagnolo2009} warning that ``$K_0$ is not intended to represent the minimum core entropy or the entropy at $r = 0$,'' where $r$ is the radius measured from the cluster center (centroid of the X-ray emission from the ICM).  Presumably when $K_0 \lesssim 30 \rm \, keV \, cm^2$, ICM cooling becomes so strong as not to be entirely rebalanced by AGN reheating, thereby depositing cool gas visible as emission-line nebulae and which fuels star formation in BCGs.  Furthermore, the cooled ICM provides additional fuel for the central supermassive black holes (SMBHs) of BCGs, resulting in their enhanced AGN radio luminosities.  Absent cool gas, AGNs can still be fuelled by accretion from the surrounding ICM within the sphere of influence of the central SMBH (Bondi, or Bondi-Hoyle-Lyttleton, accretion).  The Bondi accretion rate $\dot M_{\rm Bondi} \propto M_{\rm SMBH}^2 K^{-3/2}$ \citep[e.g.,][]{Panagoulia2014}, where $M_{\rm SMBH}$ is the mass of the SMBH, and so for a given $M_{\rm SMBH}$ is higher for gas having a lower entropy $K$.

A number of studies, however, have called into question the existence of an ICM entropy floor in galaxy clusters.  From a study of sixty-six nearby galaxy groups and clusters within 300\,Mpc ($z \leq 0.071$) as observed by the {\it Chandra} and/or X-ray Multi-Mirror Mission ({\it XMM-Newton}), \citet{Panagoulia2014} showed that their radial entropy profiles decrease monotonically inward down to the smallest measurable radii (typically between $\sim$0.2\,kpc and a few kpc).  They attribute the apparent inward flattening in radial entropy profiles as inferred by \citet{Cavagnolo2009} to a coarser sampling in temperature compared with density at small radii.  Because derivations of temperature (from X-ray spectra) require far more X-ray counts than derivations of density (from X-ray intensity), in a deprojection analysis\footnote{In such an analysis, the radial temperature profile of the ICM is first derived by fitting a single-temperature thermal model to X-ray spectra extracted from concentric annuli centred on the cluster X-ray centre.  To then derive the gas density profile, the X-ray surface brightness profile is deprojected to attain a volume emission density, before being used along with other X-ray information (including the inferred temperature) to derive the gas density profile.} such as that used by \citet{Cavagnolo2008} as was commonly adopted at the time, wider annular strips are employed in computing temperature compared with density.  Indeed, an inspection of the results reported by \citet{Cavagnolo2009} (compiled at https://web.pa.msu.edu/astro/MC2/accept/) for twenty-three of the clusters in common with those observed in the $CLASH$ program reveal only three to five temperature measurements throughout their entire radial profiles in the majority of cases; moreover, the central temperature bin spans an outer radius\footnote{ The innermost regions being masked out for those clusters having X-ray point sources or known AGNs associated with their BCGs.} of typically 20--50\,kpc if not larger.  By comparison, \citet{Cavagnolo2009} provide typically twenty to thirty density -- and therefore also corresponding entropy -- measurements over their radial profiles reaching inward to radii as small as typically $\sim$10--20\,kpc. More modern inferences of radial entropy profiles, on the other hand, solve simultaneously for both the temperature and density of the ICM \citet{Donahue2014}.

Reinforcing the findings by \citet{Panagoulia2014}, \citet{Hogan2017a} reported the lack of an entropy floor for four nearby ($0.055 < z < 0.077$) and massive galaxy clusters having deep {\it Chandra} observations, with their radial entropy profiles decreasing monotonically inward to the smallest measurable radii of $\sim$1\,kpc.  Based also on {\it Chandra} observations, \citet{Hogan2017b} reported similar results for fifty-seven clusters selected for having been previously observed (but not necessarily detected) in either H$\alpha$ or CO, and \citet{Sanders2018} for eighty-three clusters at $0.28 < z < 1.22$ selected via their Sunyaev-Zel'dovich effect through observations with the South Pole Telescope ({\it SPT}).

 \subsection{Our Goals} \label{subsec:goals}
The revision in our understanding of the ICM entropy at cluster cores calls for a re-examination of the dependence between star formation, emission-line nebulae, and radio-luminous AGNs in BCGs on the ICM thermodynamics of their host clusters.  Owing to the quality and breadth of the data available, we base our re-examination on the twenty-five clusters observed in the Cluster Lensing And Supernova survey with Hubble ($CLASH$) program.  The criteria used to select these clusters, which span the redshift range $0.19 \leq z \leq 0.89$, have been described by \citet{Postman2012}. 
Twenty of the clusters were assessed by the $CLASH$ team to be dynamically relaxed \citep{Postman2012,Donahue2016}; we shall henceforth refer to these clusters as the dynamically-relaxed subsample.  By contrast, the remaining five clusters were selected solely for their exceptional strength as gravitational lenses, many if not all of which are obviously dynamically disturbed; i.e., in the midst or aftermath of a merger with another galaxy cluster.  We refer to these five clusters as the high-lensing subsample.  All these clusters were imaged in sixteen broadband filters spanning the ultraviolet (UV) to the near-infrared (near-IR), providing continuous wavelength coverage over the range $\sim$2000--17000\,\AA.  This broad wavelength coverage, together with the exceptional depth and high angular resolution of the images, permit an especially sensitive search for star formation and candidate optical AGNs, as well as a search for optical emission-line nebulae, in BCGs.


Conveniently given the purpose of our work, the clusters in the $CLASH$ program are quite evenly split between those on either sides of $K_0 \simeq 30 \rm \, keV \, cm^2$ as determined by \citet{Cavagnolo2009} or, for the two clusters not studied by \citet{Cavagnolo2009}, as reported in \citet{Donahue2015}.  They span a higher redshift range than the clusters studied by \citet{Rafferty2008} and are weighted toward higher redshifts than those studied by \citet{Hoffer2012}, making the depth and angular resolution of the images provided by the $CLASH$ program essential for a sensitive search for recently-formed stars in their BCGs.  Necessarily for our purposes, the physical parameters of the ICM in all of the $CLASH$ clusters have been re-derived by \citet{Donahue2014} by simultaneously fitting for the temperature and density of the X-ray spectra extracted in concentric annular strips, thus providing the same radial sampling in both temperature and density.  Their results provide a crucial check on the results reported by \citet{Cavagnolo2009} for, especially, the existence of an excess core entropy in the ICM.

\citet[][see their Fig.\,1]{Donahue2015} reported that nine of the $CLASH$ BCGs display highly complex UV morphologies with projected linear extents of up to $\sim$100\,kpc, a clear signature of recent and perhaps ongoing star formation.
By contrast, \citet[][see their Fig.\,2]{Donahue2015} show that the remaining sixteen $CLASH$ BCGs appear to display simple UV morphologies characterised by a relatively compact central enhancement.  One of the central goals of this paper is to definitively identify the sources of the UV continuum in these particular BCGs, for which the possibilities are either: (i) newly-formed massive stars; or (ii) old low-mass stars that have evolved away from the main sequence.  In the latter case, the UV continuum should be centrally concentrated in the same way as old low-mass stars, thus giving rise to a relatively compact central enhancement.  Table\,\ref{BCG parameters} lists the names and redshifts of all the $CLASH$ clusters.  Those scrutinised for BCG activity in this paper -- comprising all sixteen shown in Figure\,2 of \citet[][]{Donahue2015} that exhibit apparently simple UV morphologies characterised by a relatively compact central enhancement -- are accompanied by information on the overall shapes of their BCGs in the near-IR, which best traces light from old stars with minimal contribution from any newly-formed stars (see Section\,\ref{subsec:BCG parameters} for how these shapes are derived, and the reasons why).

\begin{deluxetable}{lccccc}[htb!]
\tabletypesize{\footnotesize}
\tablewidth{0pt}
\tablecolumns{5}
\tablecaption{$CLASH$ BCG Parameters}
\tablehead{
\colhead{Host Cluster} & \colhead{Redshift} &\colhead{$R_{\rm e}$ (kpc)} & \colhead{$PA$} & \colhead{$e$}}
\startdata
Abell 383 & 0.187 & - & - & - \\
Abell 209 & 0.206 & 53.55 $\pm$ 2.5 & 133.05 & 0.19 \\
Abell 1423 & 0.213 & 30.67 $\pm$ 6.1 & 56.31 & 0.26\\
Abell 2261 & 0.224 & 18.90 $\pm$ 4.1 & 174.99 & 0.12\\
RXC\,J2129.7+0005 & 0.234 & 44.51 $\pm$ 6.7 & 67.02 & 0.39\\
Abell 611 & 0.288 & 26.66 $\pm$ 4.2 & 42.01 & 0.28\\
MS2137.3-2353 & 0.313 & 12.17 $\pm$ 2.9 & 76.01 & 0.14\\
MACS\,J1532.8+3021 & 0.345 & - & - & - \\
RXC\,J2248.7-4431 & 0.348 & 31.42 $\pm$ 4.2 & 50.01 & 0.22\\
MACS\,J1115.8+0129 & 0.352 & - & - & - \\
MACS\,J1931.8-2635 & 0.352 & - & - & - \\
MACS\,J1720.2+3536 & 0.391 & - & - & - \\
MACS\,J0416.1-2403 & 0.396 & 13.25 $\pm$ 5.2 & 56.16 & 0.19\\
MACS\,J0429.6-0253 & 0.399 & - & - & - \\
MACS\,J1206.2-0847 & 0.44 & 22.38 $\pm$ 6.1 & 131.76 & 0.46\\
MACS\,J0329.6-0211 & 0.45 & - & - & - \\
RXC\,J1347.5-1145 & 0.451 & - & - & - \\
MACS\,J1311.0-0310 & 0.494 & 14.15 $\pm$ 4.7 & 131.76 & 0.09\\
MACS\,J1149.6+2223 & 0.544 & 57.19 $\pm$ 6.9 & 131.99 & 0.23\\
MACS\,J1423.8+2404 & 0.545 & - & - & - \\
MACS\,J0717.5+3745 & 0.548 & 15.75 $\pm$ 4.9 & 118.99 & 0.04\\
MACS\,J2129.4-0741 & 0.570 & 49.62 $\pm$ 7.1 & 0.01 & 0.22\\
MACS\,J0647.8+7015 & 0.584 & 83.24 $\pm$ 6.3 & 111.42 & 0.33\\
MACS\,J0744.9+3927 & 0.686 & 14.31 $\pm$ 6.1 & 18.51 & 0.15\\
CL\,J1226.9+3332 & 0.890 & 41.21 $\pm$ 10.2 & 94.297 & 0.28\\
\enddata
\tablecomments{Derived parameters of the sixteen BCGs studied here that exhibit apparently simple UV morphologies characterised by a relatively compact central enhancement.  The remaining clusters in the $CLASH$ program are listed for completeness.}
\label{BCG parameters}
\end{deluxetable}

\vspace{0.2cm}
The remainder of this paper is organised as follows.  Section\,\ref{sec:DataProc} lays out the technical aspects of our work.  The results are presented in Section\,\ref{sec:Results}, and our interpretation of the results in Section\,\ref{sec:Interpretation}.  Section\,\ref{sec:Discussion} provides a concise summary of the results for all the $CLASH$ BCGs, compiled by combining the results obtained here with those previously reported by \citet{Donahue2015} and \citet{Fogarty2015}.  We recommend reading this section first before seeking details of particular interest in the preceding sections.  Also in Section\,\ref{sec:Discussion}, we offer an explanation for what the entropy floor of the ICM as inferred from the work by \citet{Cavagnolo2009} actually reflects, and therefore how the strong link previously established between BCG activity and the apparent entropy floors of their host clusters ought to be interpreted.  Finally, in Section\,\ref{sec:Summary}, we summarise the work conducted and the key results presented in this paper.  Throughout this manuscript, we adopt the cosmological parameters $H_0 = 70 {\rm \, km \, s^{-1} \, Mpc^{-1}}$, $\Omega_m = 0.3$, and $\Omega_\Lambda = 0.7$ to compute distances and hence physical scales.

\section{Data Processing and Analysis} \label{sec:DataProc} 
We downloaded, from the $CLASH$ archive (https://archive.stsci.edu/prepds/clash/), images in all sixteen broadband filters employed in the $CLASH$ program for the sixteen galaxy clusters listed in Table\,\ref{BCG parameters} accompanied by information on their BCGs.  As mentioned above, these particular clusters host BCGs that appear to exhibit simple UV morphologies characterised by a relatively compact central enhancement \citep[see Fig.\,2 of][]{Donahue2015}.  The remaining clusters host BCGs with complex and spatially-extended UV morphologies, a clear signature of recent and widespread star formation \citep[see Fig.\,1 of][]{Donahue2015}.  All the images retrieved have a pixel size of 65\,mas.

\subsection{PSF homogenisation}\label{PSF}
To make the SED plots, colour images, and continuum-subtracted images described in Section\,\ref{sec:Results}, it is imperative that all the images for each cluster be convolved to the same angular resolution.  For this reason, we fitted two-dimensional Gaussian functions to stars visible in the images so as to determine the full-width half-maximum (FWHM) of the point-spread-function (PSF) of the telescope in each filter.  In general, the PSF becomes broader with increasing wavelength, attaining its maximum size in the F160W filter (the longest wavelength filter employed in the $CLASH$ program) of typically 0\farcs22 at FWHM.  The images were then convolved, where necessary, to provide a uniform PSF with a FWHM of 0\farcs22 in all filters for all the clusters.  

\subsection{Additional UV Background Subtraction}
As part of the pipeline processing, backgrounds were removed from the images posted on the $CLASH$ archive.  Nonetheless, we found that the UV images can exhibit a small non-zero background: unlike images at optical and near-infrared wavelengths, images in the UV contain relatively few sources, and so their backgrounds are easy to accurately measure and hence even small offsets simple to detect.  Although similarly small offsets may well plague the images in every filter, such offsets are especially detrimental when measuring the brightnesses of faint sources -- as is the case at UV wavelengths for all the BCGs considered in our work, albeit not at optical or near-infrared wavelengths where the BCGs are typically the brightest extragalactic objects in the fields.  To check and correct for any non-zero background in the images at UV wavelengths, we performed aperture photometry on empty regions of sky near the BCG.

\subsection{Noise level}\label{error}

As explained by \citet{Li2019}, the root-mean-square (rms) noise fluctuation in nearby pixels are correlated as each images are re-sampled to a common pixel scale (i.e. 65\,mas) from different physical pixel scales (i.e. 50\,mas for ACS, 128\,mas and 40\,mas for WFC3 IR and UVIS respectively). As a consequence, pixels in these images are not statistically independent, worse among pixels within a smaller area.  Nonetheless, as demonstrated by \citet[][see their Fig.\,6]{Li2019} for the cluster MACS\,J0329.6-0211, measurements of rms noise fluctuations over regions encompassing $\gtrsim 100$\,pixels correspond closely to the true rms noise fluctuations over regions of these sizes.  For other clusters, the ranges over which the same applies may be different depending on the actual observation strategies. Because all the SEDs shown below were extracted from regions spanning areas no smaller than 70\,pixels and typically much larger, we determined the rms noise fluctuation at each wavelength directly from blank regions of the image encompassing an area of equal size; the uncertainties over the regions considered are, in any case, dominated by Poisson noise.  This sky noise was added in quadrature with Poisson noise as determined from the total count rate over the targeted region.

\subsection{BCG Parameters}\label{subsec:BCG parameters} 
So as to extract SEDs from annular zones having the same ellipticity and position angle for the major axis as for the individual BCGs (see Section\,\ref{subsec:SEDs}), we fitted a two-dimensional (2-D) S\'ersic function to each BCG at the three longest wavelength filters in the near-IR, where all the BCGs are at their brightest.  Fits were made to images at multiple wavelengths to ensure a consistent and hence reliable solution for their morphological parameters.  In this way, we determined the ellipticity, $e$, and position angle, $PA$, for the major axes of each BCG, and as a by-product also their individual effective radius, $R_{\rm e}$, as listed in Table\,\ref{BCG parameters} averaged over the individual values determined from each filter image.  The position angles thus derived are generally similar to, although the ellipticities significantly larger than, those reported by \citet{Donahue2015} based on a moments analysis also from $CLASH$ near-IR images but over larger apertures.  
We stress here that our primary goal is to determine the overall shape of each BCG, rather than to obtain a precise fit to their morphologies -- as two or more S\'ersic components are often required to provide an acceptable fit.  The values derived for the ellipticity and position angle of each BCG were used to construct central apertures and annular zones for extracting their spectral energy distributions (SEDs) as described next.

\section{Results}\label{sec:Results}

In relatively nearby galaxies, associating UV light with recently-formed massive stars can be complicated by that also produced by old low-mass stars that have evolved away from the main sequence.  The latter are held responsible for an upturn in the UV continuum shortwards of $\sim$2500\,\AA\ until the Lyman limit at 910\,\AA\ in the bulges of disk and lenticular galaxies, as well as in the central regions of elliptical galaxies that have long ceased star formation \citep[e.g., review by][]{O'Connell1999}.
In elliptical galaxies, the difficulty in distinguishing between a young and old stellar population can be especially acute when the UV light of the galaxy is centrally concentrated and poorly resolved, as illustrated by previous efforts involving the $CLASH$ BCGs summarised below.  This difficulty underscores the pain-staking work we had to carry out to distinguish between these two stellar populations as described in Sections\,\ref{subsec:SEDs}--\ref{subsec:continuum_subtracted}, along with additional work involving also selected cluster members as described in Section\,\ref{subsec:evolved_stars}.

To estimate the contribution by evolved stars to the UV light of the $CLASH$ BCGs (restricted to just  the twenty clusters belonging to the dynamically-relaxed subsample), \citet{Fogarty2015} fitted: (i) a fifth order spline to the averaged spectral energy distributions (SEDs) spanning rest-frame optical to near-IR wavelengths of five satellite galaxies to each BCG (i.e., one-hundred satellite galaxies in all); supplemented by (ii) UV$-$IR colours (with UV measurements from GALEX and near-IR measurements in the J-band from 2MASS) of a mostly different set of BCGs studied by \citet{Hicks2010}.  In all twenty of the $CLASH$ BCGs thus examined, they find levels of UV continuum (at a significance of at least 3$\sigma$)  in excess of those estimated from evolved stars for fourteen of the BCGs (see Table 3 of \citet{Fogarty2015})



As an additional diagnostic for star formation, \citet{Fogarty2015} constructed H$\alpha$+[N\,II] images for all these BCGs by subtracting images in an off-line from an on-line filter.  To estimate the level of stellar continuum in the on-line filter, \citet{Fogarty2015} used the mean ratio in intensity between the off-line and on-line filters as measured for one or more satellite galaxies having a similar near-IR to optical color as the individual BCGs.  
In this way, \citet{Fogarty2015} identified eleven BCGs to display H$\alpha$+[N\,II] emission, including the nine found by \citet{Donahue2015} to display complex UV morphologies.  Ancillary spectra were taken for fifteen of these BCGs with the Southern Astrophysical Research Telescope ({\it SOAR}), including the eleven found to display H$\alpha$+[N\,II] emission.  Comprising slit spectra encompassing the [O\,II]$\lambda$3726\,\AA, $\lambda$3729\,\AA, [O\,III]$\lambda$4959\,\AA, $\lambda$5007\,\AA, and H$\beta$ lines, \citet{Fogarty2015} found that, for a given BCG, the star-formation rate inferred from either [O\,II] or H$\beta$ is roughly comparable (within factors of a few) with that inferred from the UV continuum (in all cases, after applying a correction for dust extinction by assuming an intrinsically flat spectrum over rest wavelengths from 1500\,\AA\ to 2800\,\AA).  The comparable star-formation rates computed whether using emission lines or UV continuum was taken to imply that the former also traces star formation, at least for the BCGs in the $CLASH$ program.  In a subsequent work examining the relationship between BCG star formation and ICM thermodynamics among the $CLASH$ clusters (again restricted only to those belonging to the dynamically-relaxed subsample), \citet{Fogarty2017} classified the nine BCGs not found to exhibit emission-line nebulae as non-star-forming (see their Fig.\,4).

\begin{figure}[hbt!]
\centering
\includegraphics[width=\columnwidth]{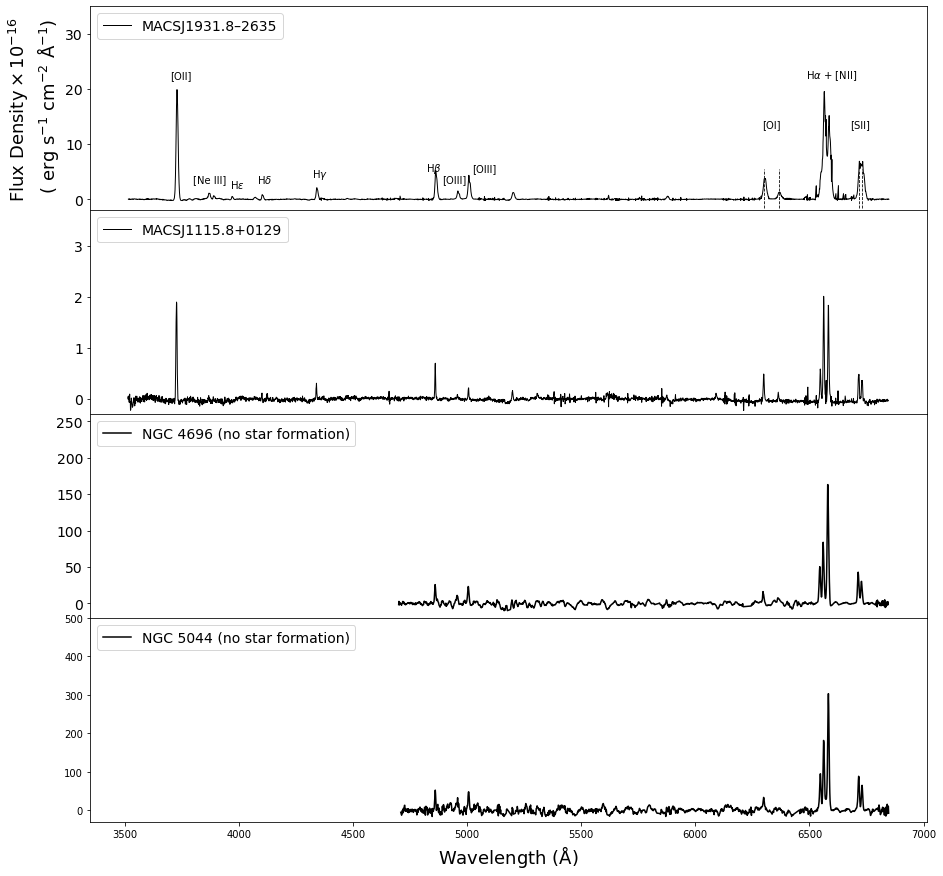}
\caption{Comparison between optical spectra of two $CLASH$ BCGs having star-formation rates that differ by an order of magnitude as inferred from their UV luminosities, with those of a BCG (NGC\,4696) and a BGG (NGC\,5044) that show no detectable star formation.  The host clusters and star-formation rates inferred from the UV luminosities of the two $CLASH$ BCGs are indicated in the legends for the respective panels in the upper two rows.  These comparisons emphasise the similarity between the optical emission-line spectra of non-star-forming BCGs and star-forming BCGs -- and therefore that the presence of optical emission-lines should not be assumed to indicate star formation, nor directly used to derive star-formation rates even in star-forming BCGs without further care.}
\label{MUSE spectra}
\clearpage
\end{figure}

In our work, we do not rely on the presence or absence of emission-line nebulae as a diagnostic for or against star formation -- for the following reason.  As mentioned earlier, the bulk spectra of optical emission-line nebulae in BCGs have long been known to be very different from H\,II regions, being characterised by low-ionisation lines \citep{Crawford1999}:  specifically, displaying bright H$\alpha$, [N\,II], and [O\,II], but very dim if at all detectable [O\,III].  In Figure\,\ref{MUSE spectra}, we compare the continuum-subtracted optical spectra of two $CLASH$ BCGs displaying complex UV morphologies -- and which have undoubtedly formed massive stars recently -- with those of NGC\,4696 (the BCG of the Centaurus cluster) and NGC\,5044 (the brightest group galaxy, BGG, of the NGC\,5044 group).  The data from which these spectra were extracted was taken with the Multi Unit Spectroscopic Explorer (MUSE) on the Very Large Telescope (VLT), and retrieved from the ESO Science Archive Facility; in all cases, the spectra were extracted over the entire region of the BGG or BCGs displaying detectable H$\alpha$ emission.  Both NGC\,4696 (at $z = 0.0097$) and NGC\,5044 (at $z = 0.0090$) display complex, filamentary, and multiphase emission-line nebulae -- including relatively cool molecular gas as detected in CO \citep{Johnstone2007, David2014} -- resembling those seen in other BGG or BCGs.  Yet, neither show any evidence for star formation (they are so close that individual star clusters ought to be detectable in images taken with the HST, just like the thousands detected in the somewhat more distant BCG of the Perseus cluster).  Instead, whether lacking any or displaying copious star formation, the optical spectra of these four galaxies are qualitatively similar: all show relatively dim [O\,III] compared with H$\alpha$ or [N\,II].  Indeed, the [O\,III]/H$\alpha$ ratio of the non-star-forming BCGs (NGC\,5044 and NGC\,4696) is significantly higher than that of the two star-forming $CLASH$ BCGs shown in the same figure.

Rather than relying on emission lines as a signature for star formation (or the lack thereof to imply no star formation over recent history), we independently assess the presence of star formation (and emission-line nebulae) among the $CLASH$ BCGs deemed by \citet{Donahue2015} to display a compact central enhancement in UV.  We start by checking whether the spectral energy distribution (SED) at the innermost regions of the BCGs is different from that farther out, which if the case would then indicate the presence of a discrete central source different from the general population of old stars (Section\,\ref{subsec:SEDs}).  We then construct colour images as an independent check for such a source, we well as to search for more spatially-extended continuum or line-emitting sources (Section\,\ref{subsec:colour_images}).  To measure quantitatively the size of the discrete central sources, as well as better delineate the morphology of any spatially-extended sources, we then create continuum-subtracted images in the relevant filters (Section\,\ref{subsec:continuum_subtracted}).  Finally, we examine ancillary data, where available, to test our understanding of the nature of the sources uncovered (Section\,\ref{subsec:ancillary}).

\subsection{Spectral Energy Distributions}\label{subsec:SEDs}

\begin{figure*}[hbt!]
\centering
\includegraphics[width=\textwidth]{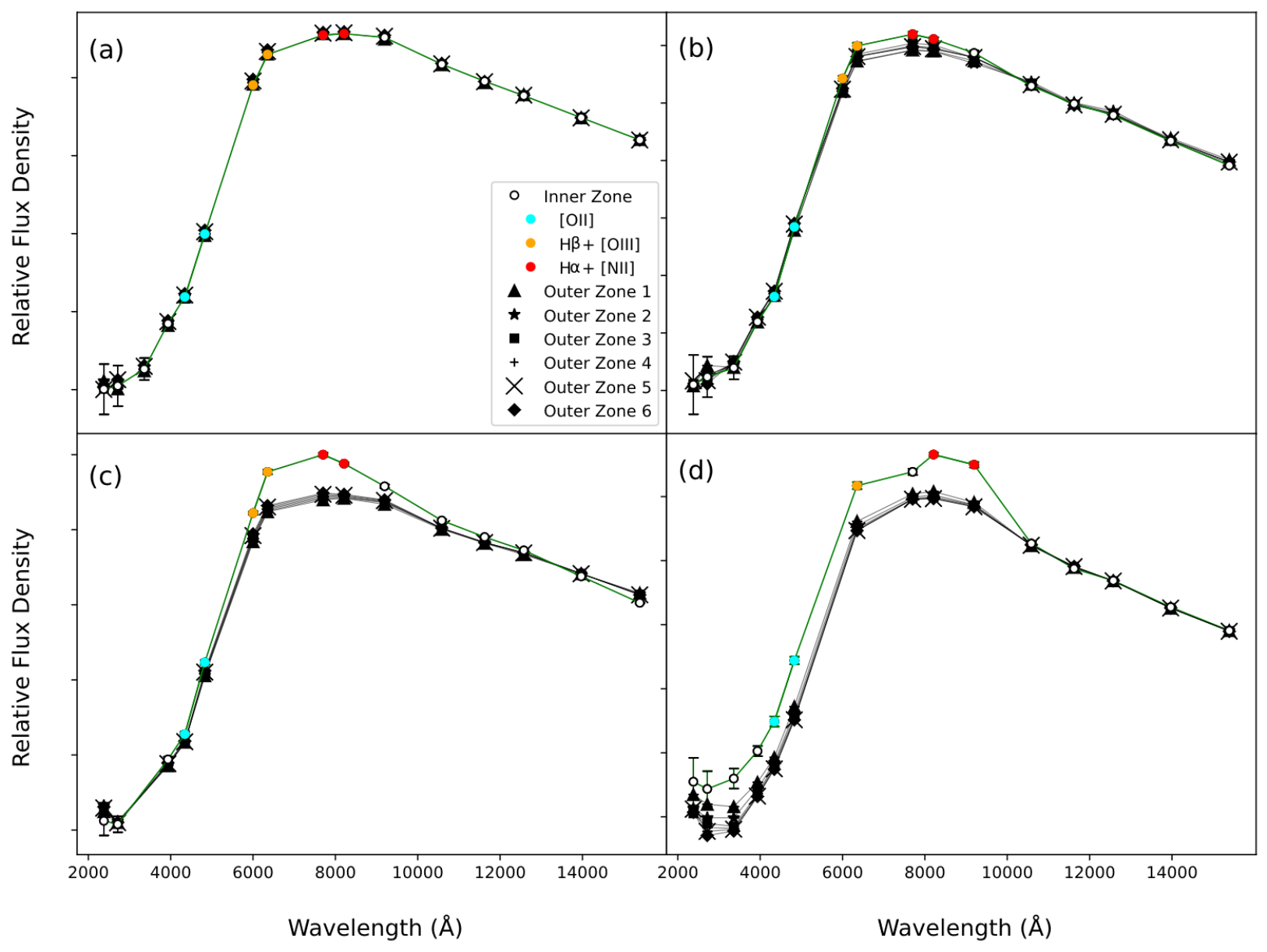}
\caption{Exemplar radial dependence in SEDs for four of the $CLASH$ BCGs listed in Table\,\ref{BCG parameters}.  In each panel, the SED of the central zone is indicated by a green curve, for which coloured data points correspond to filters encompassing emission-lines of interest as listed at the lower right corner of panel $(a)$.  The SED of surrounding annular zones are indicated by black curves, labelled zones 1--6 successively outward as indicated by the different symbols defined at the lower right corner of panel $(a)$.  See text for how the dimensions of the different zones are defined.  Error bars are plotted only for measurements from the central zone, and apart from the shortest-wavelength filters are often smaller than the symbol size; error bars for measurements from the outer zones, which are larger in area, are smaller still.
All SEDs have been normalised by the flux density within the central zone averaged over multiple filters longwards of those containing H$\alpha$+[N\,II].  These SED display either: $(a)$ no appreciable change with radius (BCG in Abell\,2261); $(b)$ an excess in filters containing H$\beta$+[O\,III] and H$\alpha$+[N\,II] at the central zone (BCG in Abell\,1423); $(c)$ an excess in filters containing [O\,II], H$\beta$+[O\,III], and H$\alpha$+[N\,II] at the central zone (BCG in Abell\,209); and $(d)$ an excess from the UV to optical at the central and closely surrounding zones (BCG in MS2137.3-2353).
}
\label{fig:SED_radius}
\clearpage
\end{figure*}

As demonstrated here, the radial dependence in the SEDs of the BCGs can provide important diagnostics on the nature of their UV continuum, and incidentally also reveal any emission lines (at rest-frame optical wavelengths) from these galaxies.   To examine any changes in their SEDs with radius, we subdivided each BCG into radial zones.  The central zone is bound by an ellipse with an ellipticity corresponding to that of the BCG, and a semi-minor axis of 0\farcs3 (i.e., minor axis equal to three times the FWHM of the PSF).  Moving outwards, we defined elliptical annular zones having equal widths of 0\farcs3 until the signal-to-noise ratio (S/N) at optical wavelengths becomes too poor to provide a meaningful measure.  An SED is extracted from each zone using SEP\footnote{https://sep.readthedocs.io} aperture photometry codes, and corrected for reddening imposed by dust in our Galaxy based on the dust maps published by \citet{Schlegel1998}.  Finally, for each BCG, the SEDs extracted from the different zones are normalised to their average intensities at the four longest wavelengths in the near-IR for clusters at $z < 0.5$, the three longest wavelength filters in the near-IR for those at $0.5 < z < 0.89$, and just the two longest wavelength filters in the near-IR for the single cluster at $z = 0.89$; in all cases, so as to avoid filters encompassing the H$\alpha$+[N\,II] lines.  Because the passbands of the near-IR filters employed in the $CLASH$ program are very broad (all much broader than those in the UV or optical), any emission lines redshifted into these filters are likely to make a near-negligible contribution to the intensity in these filters compared with the stellar continuum.  In this way, we maximise the visibility of any differences in relative intensities between the different zones in the UV and, incidentally, also optical.

\begin{figure*}[hbt!]
\centering
\includegraphics[width=\textwidth]{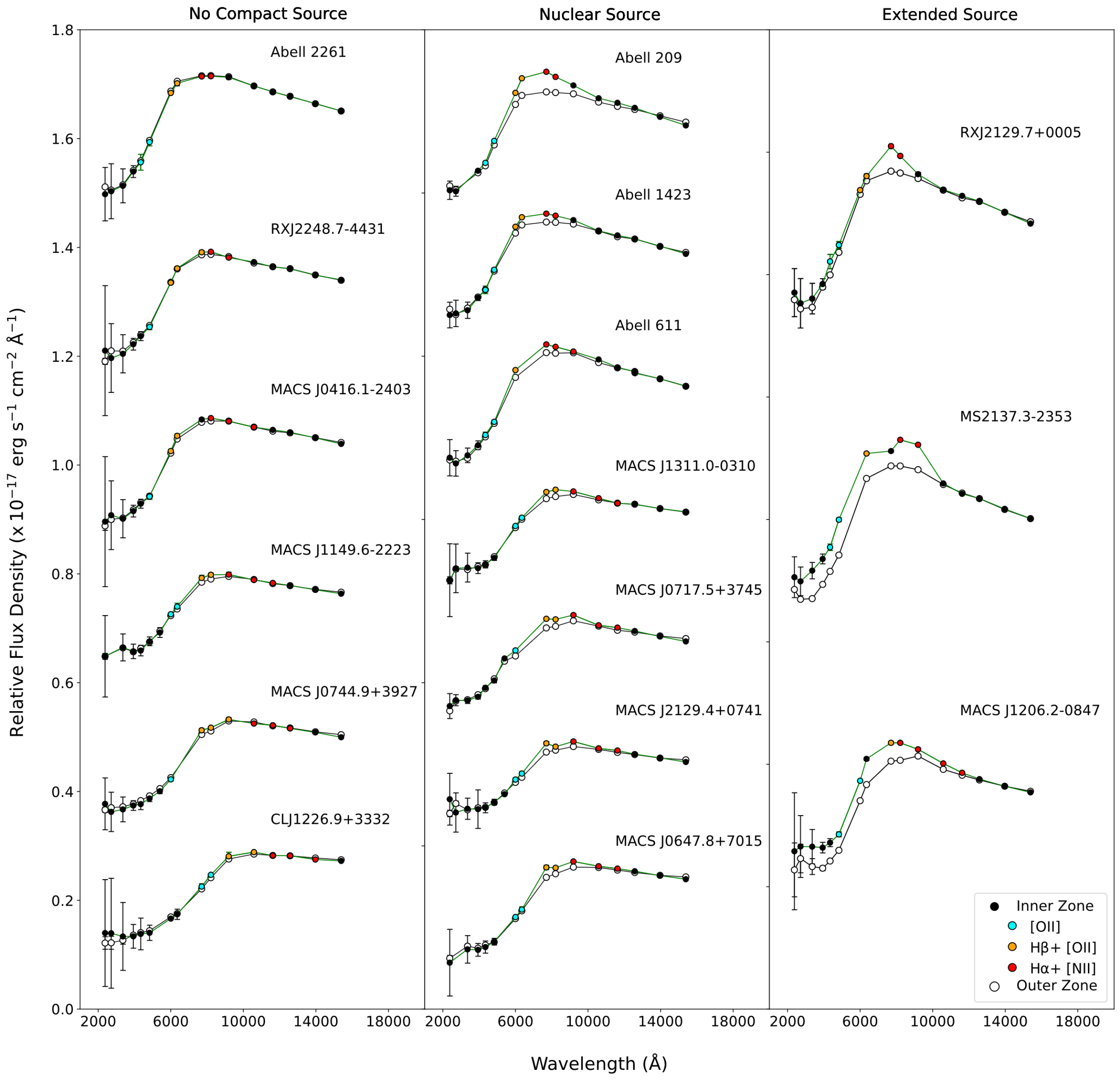}
\caption{SED of the central zone and surrounding annular zone for each of the sixteen $CLASH$ BCGs that we studied.  Measurements at the central zone are indicated by unfilled circles, and the surrounding annular zone by closed circles.  The coloured symbols have the same meaning as in Fig.\,\ref{fig:SED_radius}.  Once again, error bars are plotted only for measurements from the central zone, and except at relatively short wavelengths are smaller than the symbol size; error bars for measurements from the outer zone, which is much larger in area, are much smaller.  See text for how the dimensions of the inner and outer zones are defined.  These plots are particularly good at revealing a relatively compact central source having a markedly different SED than the surrounding body of the BCG.  The first row are for the BCGs with no detectable central source, the second for BCGs with a compact central source, and the third for BCGs with an extended central source (see text).}
\label{fig:SEDs}
\clearpage
\end{figure*}

Example results are shown in Figure\,\ref{fig:SED_radius}.  In this figure, the red circles indicate filters containing H$\alpha$+[N\,II], yellow circles those containing H$\beta$+[O\,III] (both the [O\,III]$\lambda$4958.9\,\AA\ and [O\,III]$\lambda$5006.9\,\AA\ doublet), and blue circles those containing [O\,II] (both the [OII]$\lambda$3726.1\,\AA\ and O[II]$\lambda$3728.8\,\AA\ doublet) at the redshift of the individual clusters.  These particular spectral lines are highlighted because, as we shall demonstrate, they provide diagnostics for differentiating between gas exhibiting low-ionisation lines (as is characteristic of BCG nebulae) or high-ionisation lines (as is characteristic of H\,II regions or AGNs).  Figure\,\ref{fig:SED_radius}$a$ shows an example for which the BCG exhibits no detectable change in its SED with radius\footnote{Elliptical galaxies are known to exhibit a colour gradient owing to a metallicity gradient (resulting in increasingly bluer colour outwards as the metallicity decreases), and -- even in the absence of recently-formed stars and/or AGNs -- would therefore be expected to exhibit a radially-varying SED.  The plots in Figure\,\ref{fig:SED_radius} do not provide good diagnostics for colour gradients owing to the manner by which the SEDs have been normalised, but serve to highlight any central sources having different SEDs than their surroundings.  A check using filters closest to rest-frame $U - I$ for the BCG shown in Figure\,\ref{fig:SED_radius}$a$ does reveal a colour gradient in the form of increasing bluer colour outwards.}.  Such BCGs constitute (at final count, based on the more stringent comparison as explained in the next paragraph) six of the sixteen BCGs.  The remaining ten BCGs show appreciable changes in their SEDs between different zones, specifically: (i) between the central zone and the remaining outer zones (which have similar SEDs) in filters encompassing H$\beta$+[O\,III] and H$\alpha$+[N\,II], as the example in Figure\,\ref{fig:SED_radius}$b$, suggesting a line-emitting source at the BCG centre; (ii) as in the previous case but including also the filter or filters encompassing [O\,II], as the example in Figure\,\ref{fig:SED_radius}$c$; and (iii) among several of the inner zones from UV to optical wavelengths, as the example in Figure\,\ref{fig:SED_radius}$d$, suggesting a spatially extended continuum source.  As we shall demonstrate more definitively in Section\,\ref{subsec:colour_images}, the observed differences in SEDs between different zones are produced by discrete sources in the individual BCGs.

To make a more sensitive search for a compact central source having a markedly different SED than the surrounding body of the BCG, we extracted the SED of each BCG over a central zone with a radius of 0\farcs2 (i.e., diameter of twice the FWHM of the PSF) to compare with the SED of its surrounding annular zone with an outer semi-major axis of 2\arcsec\ (the approximate size as determined from growth curves in the longer-wavelength UV filters to provide the maximal S/N in these filters).  
Like before, for each BCG, the SEDs at the central and surrounding outer zones are normalized to their average intensities in multiple near-IR filters selected to avoid those encompassing H$\alpha$+[N\,II].  The results for all the BCGs are shown in Figure\,\ref{fig:SEDs}, and reveal that: (i) six BCGs (left column) have SEDs that are indistinguishable (within measurement uncertainties) between their central and surrounding annular zones; (ii) seven BCGs (middle column) exhibit a central source that produces an excess in filters encompassing H$\alpha$+[N\,II] and H$\beta$+[O\,III], as well as sometimes also [O\,II];
and (iii) three BCGs (right column) exhibit a spatially extended central source that produces an excess in the continuum from UV to optical wavelengths (as shown in Section\,\ref{subsec:colour_images}, two of these BCGs also exhibits spatially-extended line emission).

\subsection{Colour Images}\label{subsec:colour_images}

To confirm the detection or otherwise of discrete sources emitting in the continuum and/or line as indicated by Figures\,\ref{fig:SED_radius}--\ref{fig:SEDs}, as well as to assess whether the source is spatially extended, we made the following colour images for each BCG.  The ``continuum'' colour image involves the longest-wavelength rest-frame UV filter not encompassing the [O\,II] line, together with the F140W filter in the near-IR.  The ``line'' colour images involve the narrowest filter encompassing either the [O\,II], H$\beta$+[O\,III], or H$\alpha$+[N\,II] lines at the redshift of the cluster (and hence BCG), together with another filter at a longer wavelength in the near-IR corresponding to either the F125W or F140W filters for clusters at $z < 0.4$, or the F140W or F160W filters for clusters at $z > 0.4$.  Two line colour images were constructed for each set of emission lines so as to cross-check both detections and non-detections.

\begin{figure*}[htb!]
\centering
\includegraphics[width=12.7cm]{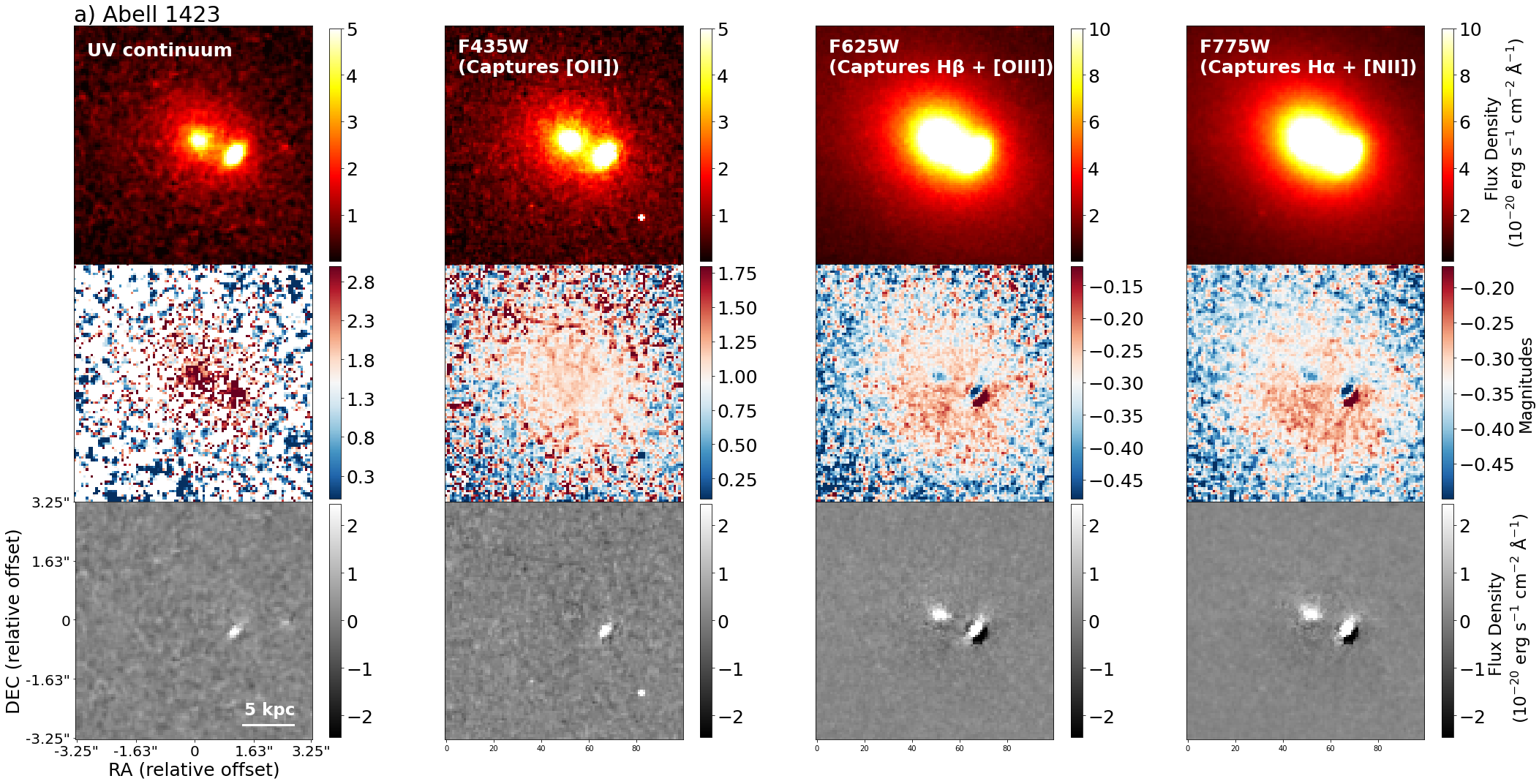} 
\includegraphics[width=12cm]{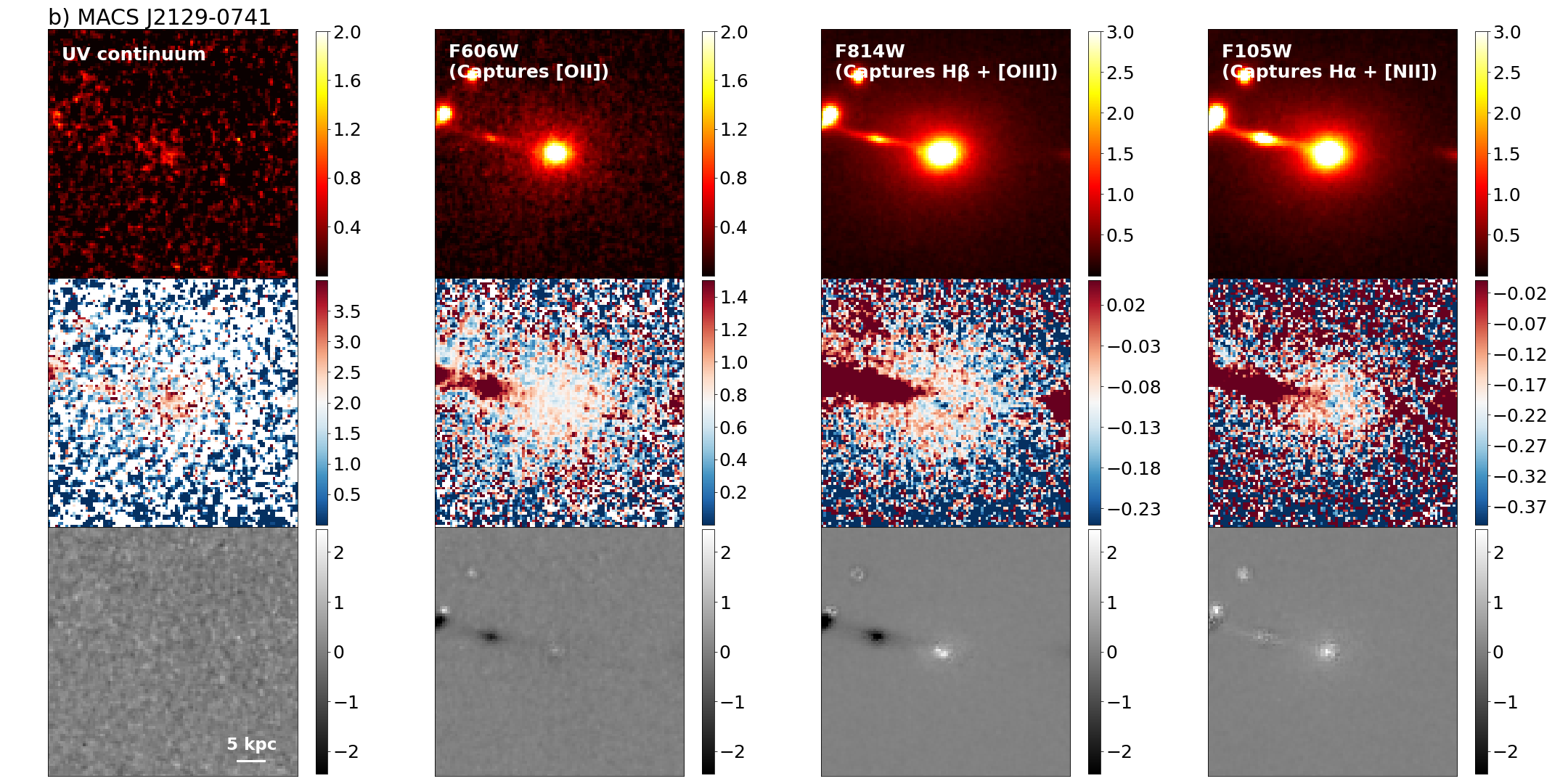} 
\includegraphics[width=12cm]{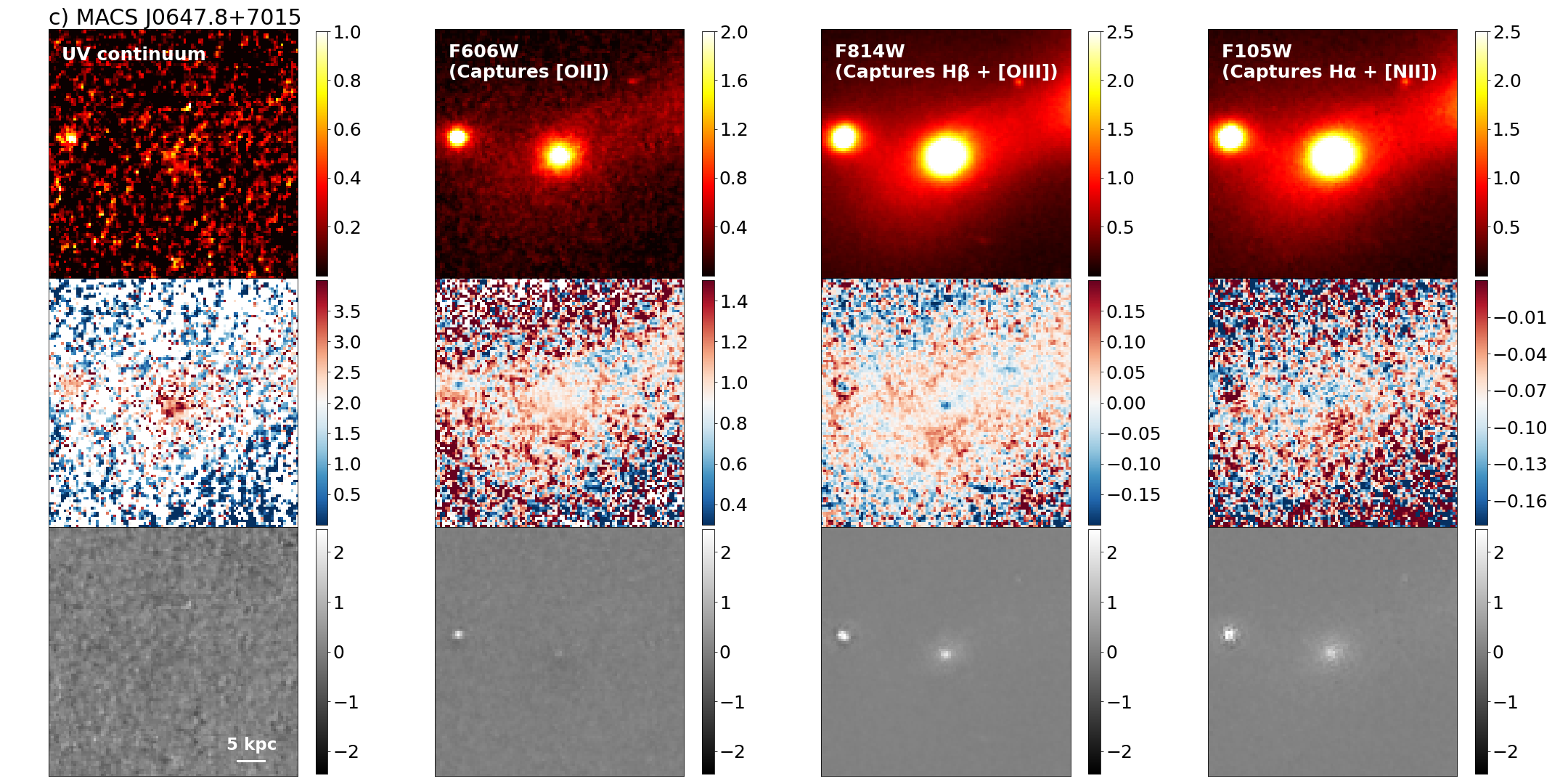}
\includegraphics[width=12cm]{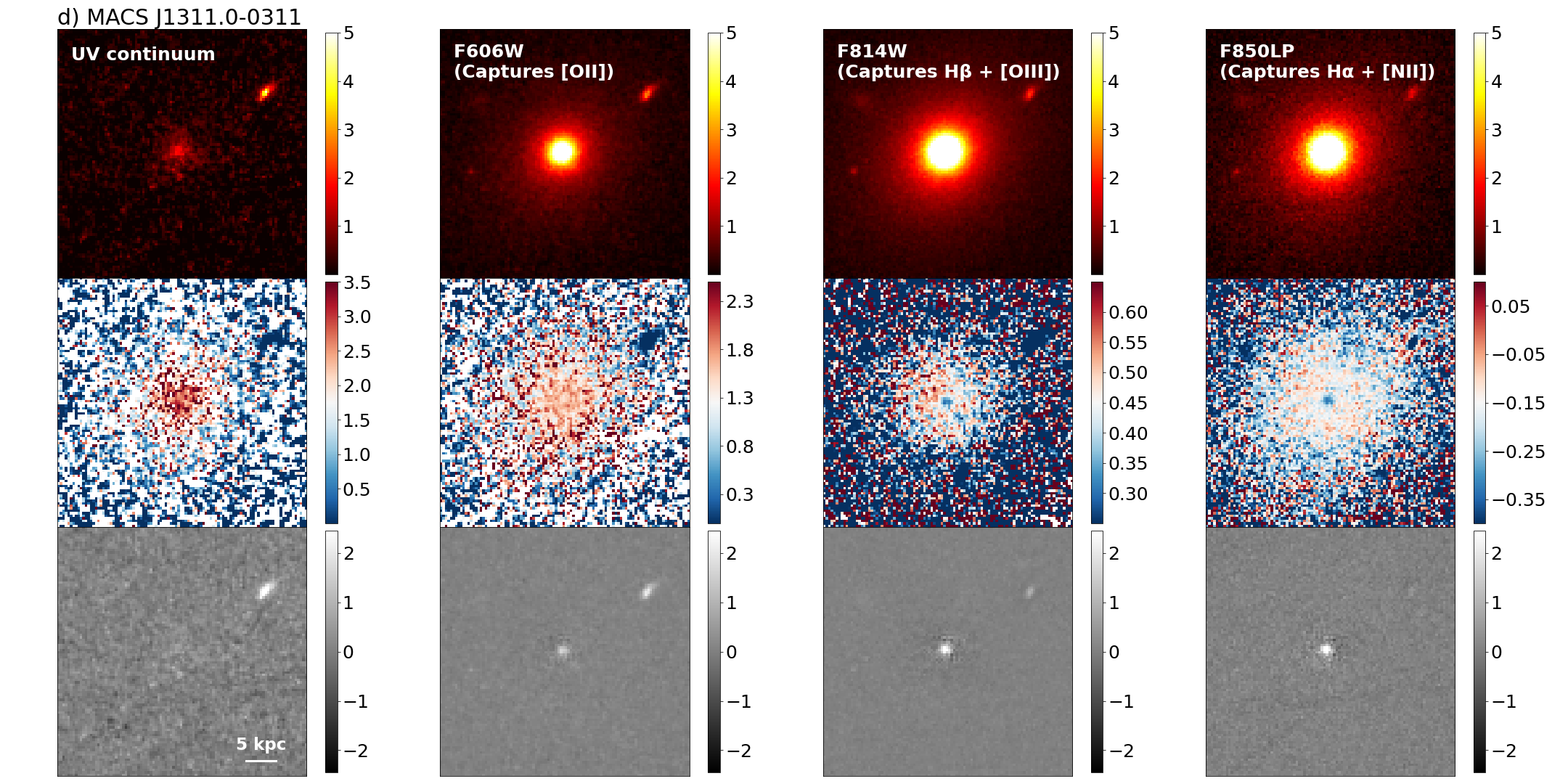} 
\end{figure*}
\begin{figure*}[htb!]
\centering
\includegraphics[width=12cm]{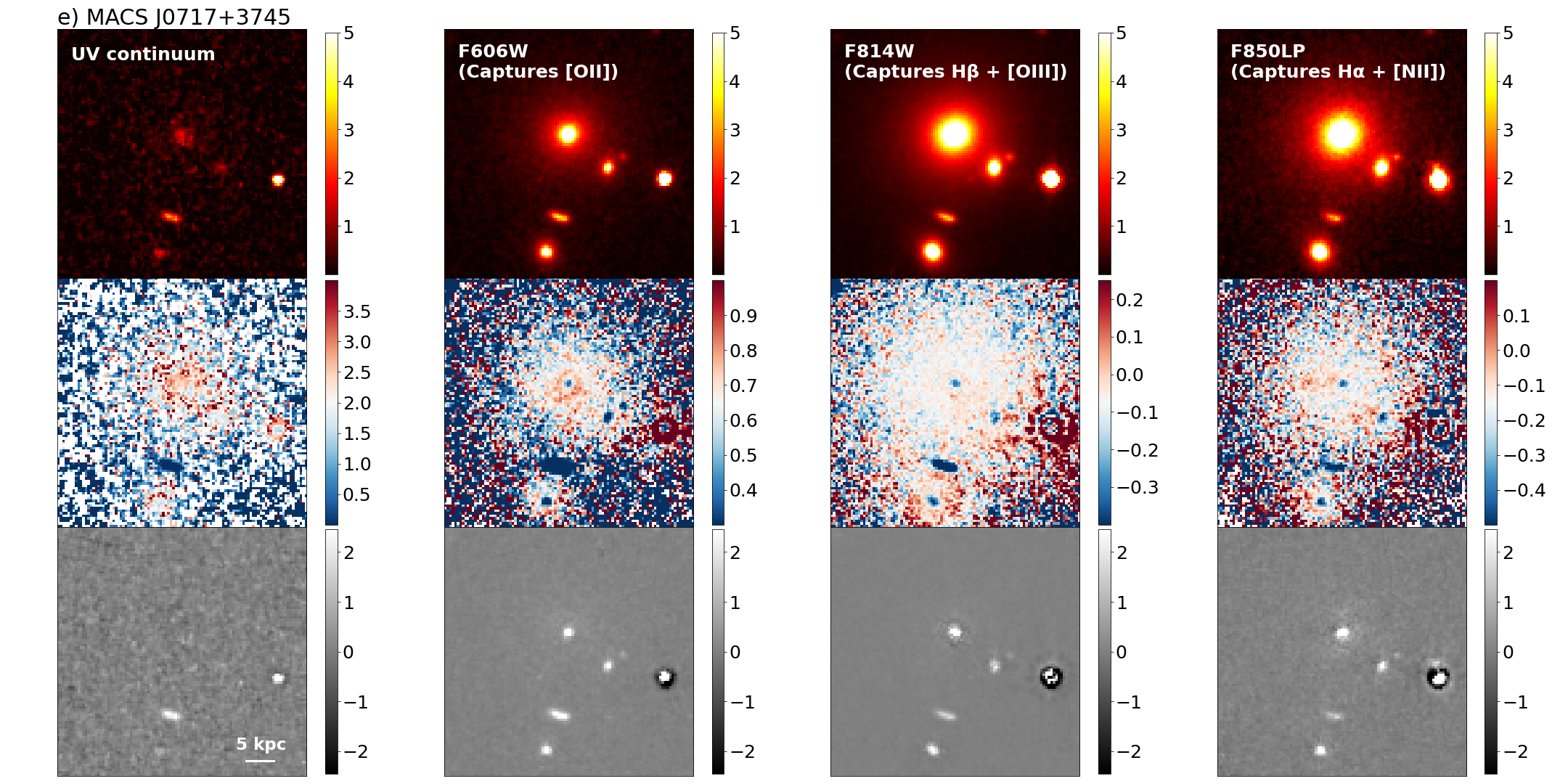} 
\includegraphics[width=12cm]{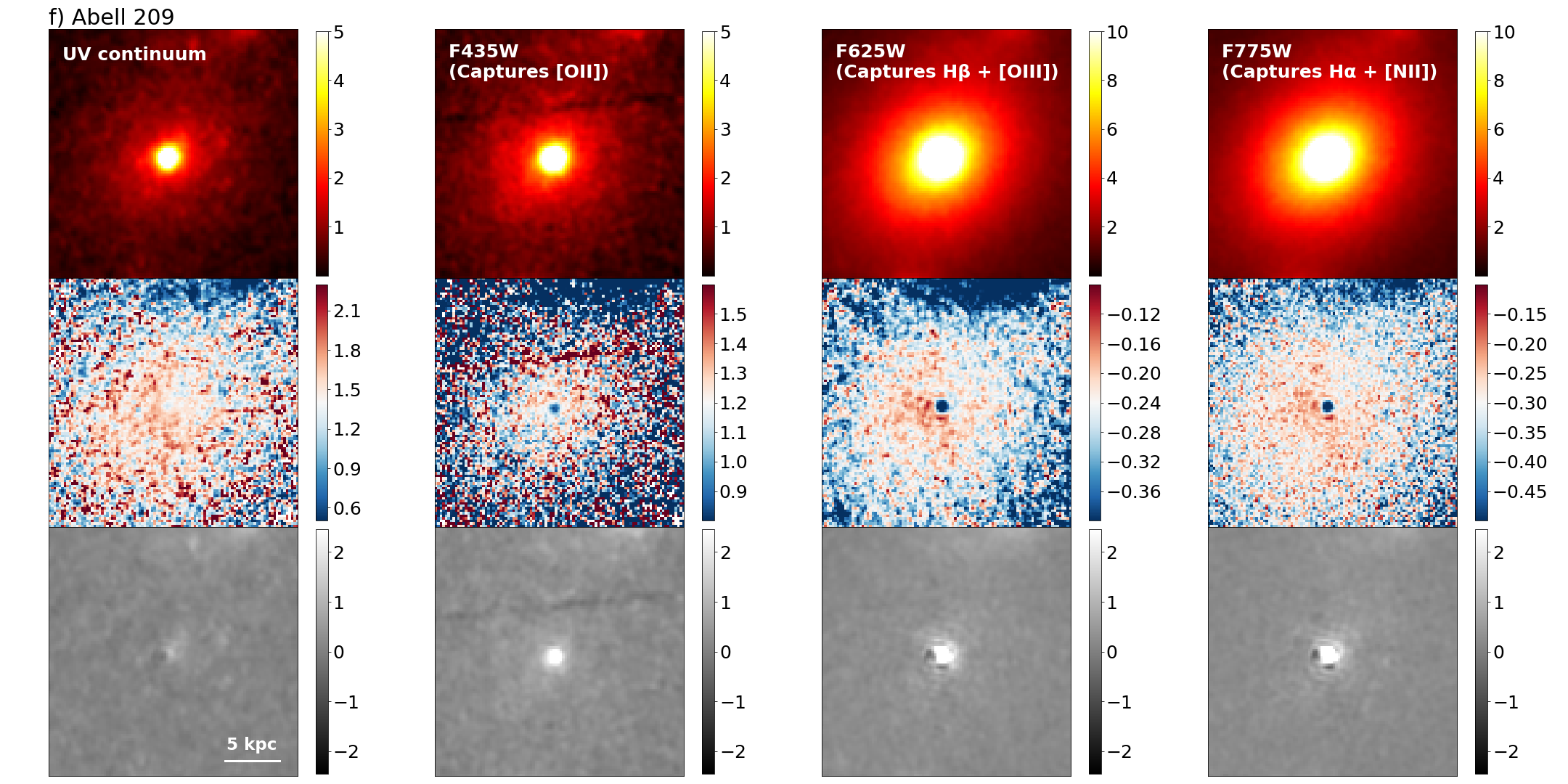} 
\includegraphics[width=12cm]{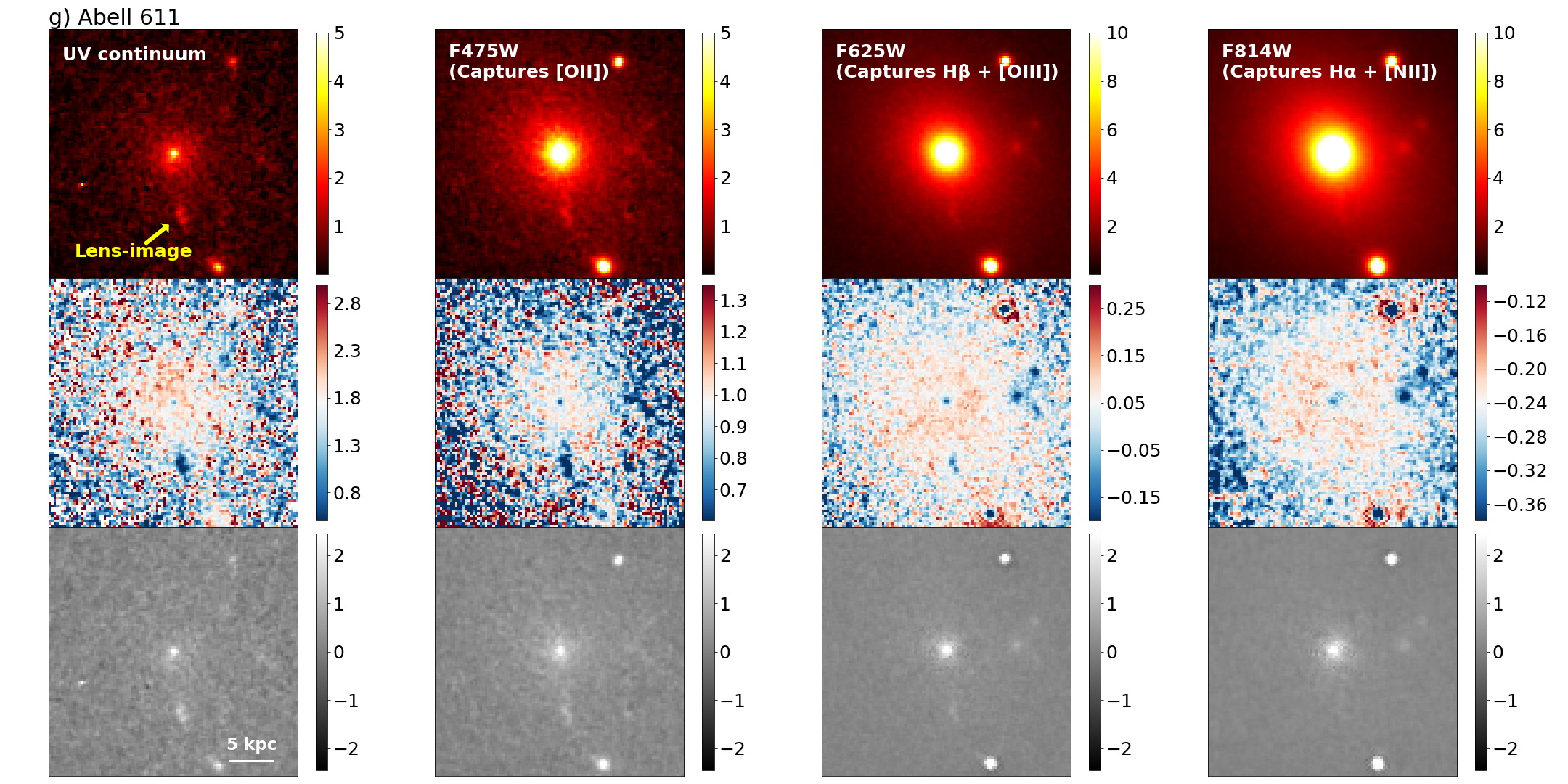}
\caption{For the cluster indicated in each block of panels from $(a)$--$(g)$, first row are images centred on the BCG in different filters, second row colour (ratio of intensities between two) images involving different filters at a shorter wavelength and the F140W filter, and third row continuum-subtracted images in different filters.  These images best highlight any discrete sources emitting in: (i) the UV continuum just shortwards of [O\,II] (first column); (ii) [O\,II] (second column); (iii) H$\beta$+[O\,III] (third column); and (iv) H$\alpha$+[NII] (fourth column).  All panels have sizes of $6\farcs5 \times 6\farcs5$ as indicated in $(a)$.  The BCGs in this figure all display a compact central source. The feature extending to the south of the BCG in Abell 611 is a lensed image of a background galaxy.}
\label{fig:colours-extended-A}
\end{figure*}

\begin{figure*}
\centering
\includegraphics[width=18cm]{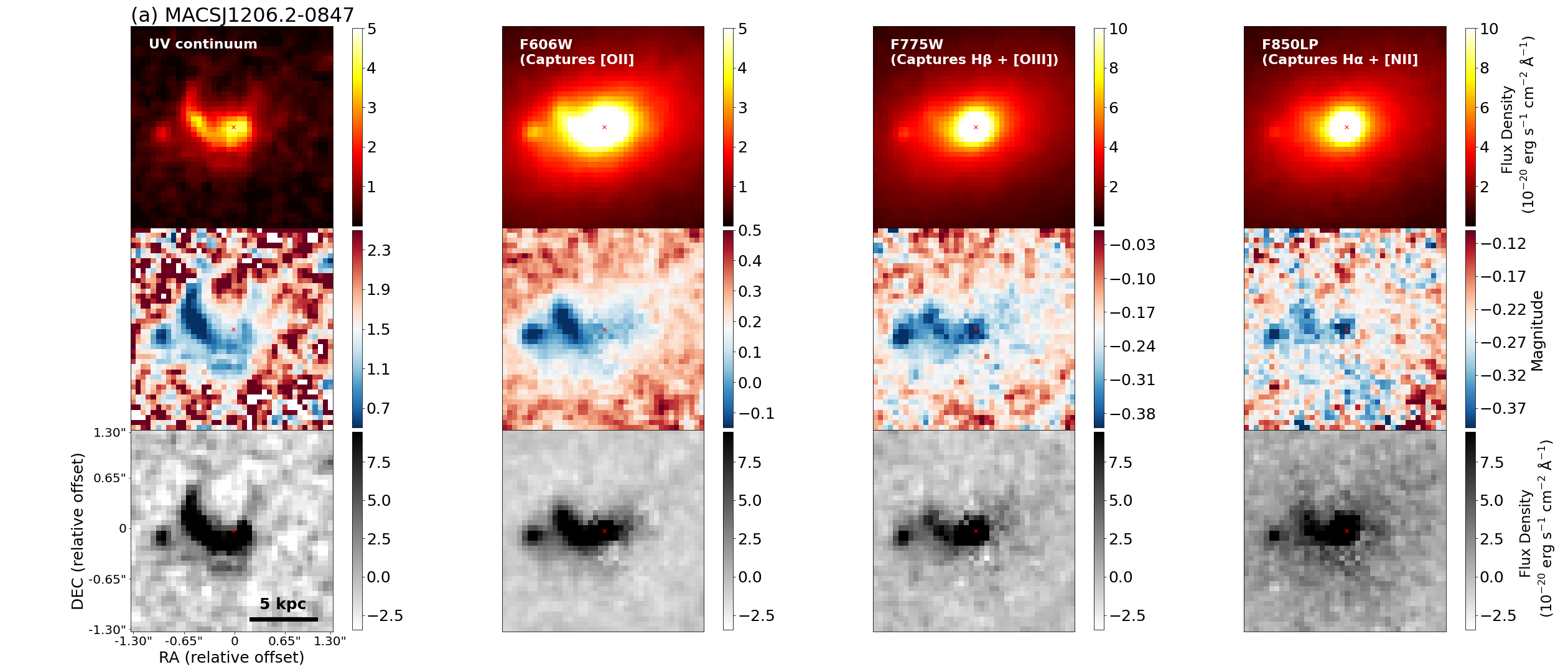}
\includegraphics[width=18cm]{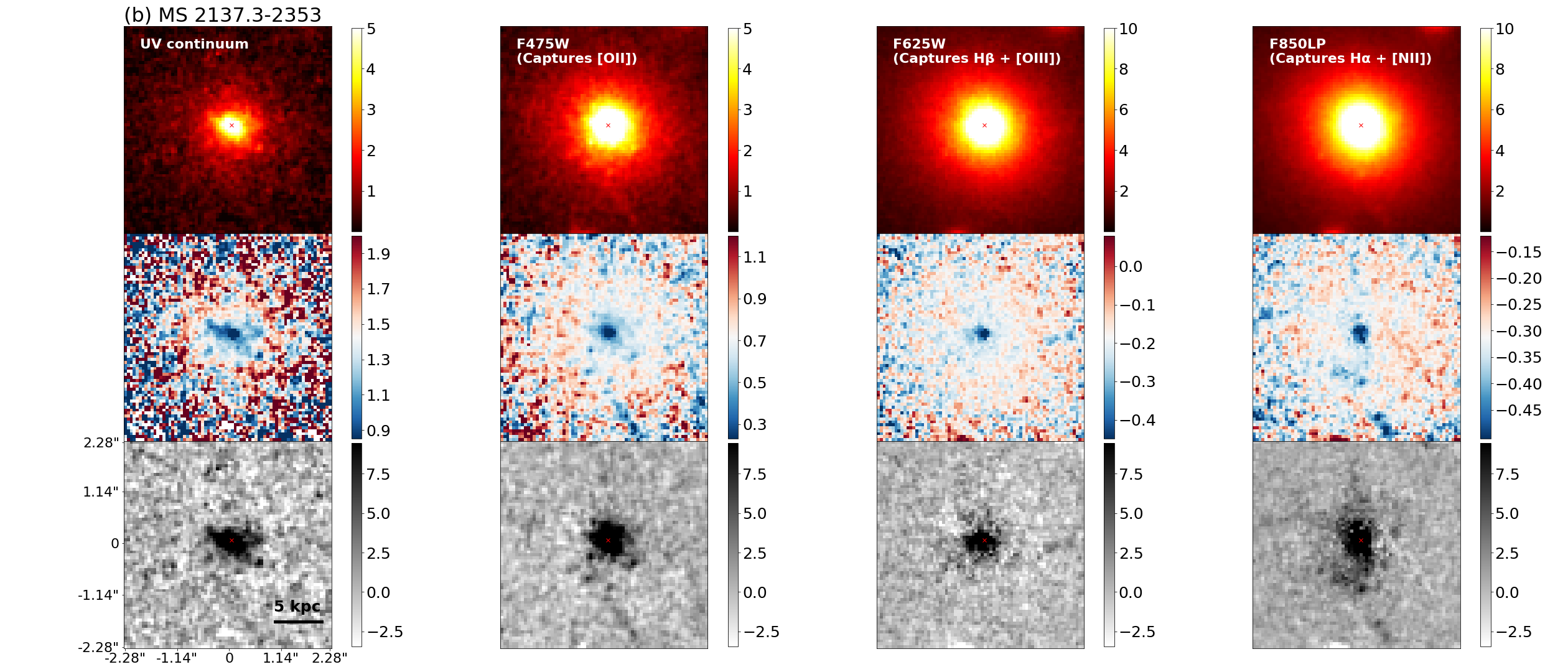}
\includegraphics[width=18cm]{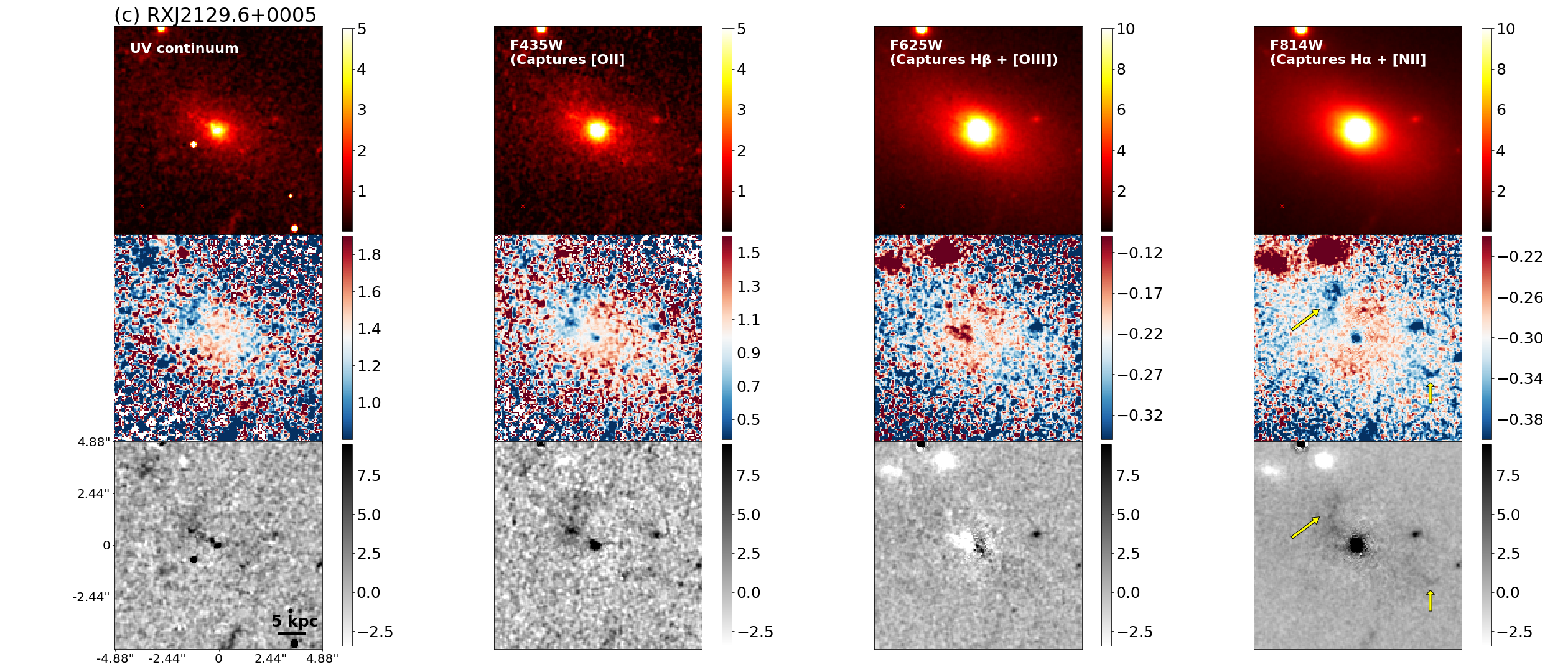}
\caption{Same as Fig.\,\ref{fig:colours-extended-A}.  Note that the block of panels associated with the different clusters have different angular sizes.  All the BCGs in this figure display spatially-extended sources. The linear absorption feature cutting across the centre of the RXC\,J2129.7+000 BCG in the continuum-subtracted image involving H$\alpha$+[N\,II] is an image artefact}
\label{fig:colours-extended-B}
\end{figure*}

The colour images made are shown in Figure\,\ref{fig:colours-extended-A} for the BCGs in the middle column of Figure\,\ref{fig:SED_radius}, and Figure\,\ref{fig:colours-extended-B} for the BCGs in the right column of Figure\,\ref{fig:SED_radius}.  Only the line colour images involving the F140W filter are shown, as the same results are obtained for the corresponding line colour images involving either the F125W or F160W filters.  These results confirm the presence of discrete sources where such sources are deduced to be present based on the SED plots of Figures\,\ref{fig:SED_radius}--\ref{fig:SEDs}.  By contrast, the colour images made for the BCGs in the left column of Figure\,\ref{fig:SED_radius} shows no discrete sources (and hence are not shown), thereby confirming their absence as indicated by the SED plots of Figures\,\ref{fig:SED_radius}--\ref{fig:SEDs}. 
Four of the BCGs (Fig.\,\ref{fig:colours-extended-A}$a$--$d$) display a relatively blue and compact central source in line colour images involving H$\alpha$+[N\,II] and H$\beta$+[O\,III], but no such apparent source in any of the other colour images.  Three other BCGs (Fig.\,\ref{fig:colours-extended-A}$e$--$g$) display a relatively blue and compact central source in all the line colour images, but not in the continuum colour image.  
Finally, three BCGs (Fig.\,\ref{fig:colours-extended-B}) display a relatively blue and spatially extended source (or sources) in all the colour images.  Most obvious among the last example is the BCG in MACS\,J1206.2-0847 (Fig.\,\ref{fig:colours-extended-B}$a$), which displays a continuum source that extends to the east of centre at rest-frame optical as well as north-east at rest-frame UV.  This BCG was not identified by \citet{Fogarty2015} to be engaged in star formation. 
\citet{Donahue2015} had earlier suggested that this feature may be a gravitationally-lensed image of a background galaxy, which as we show in Section\,\ref{subsec:ancillary} is not the case.  The extended nature of the discrete sources in the remaining two BCGs (Fig.\,\ref{fig:colours-extended-B}$b$--$c$) are better defined in continuum-subtracted images, which we shall described next.

\subsection{Continuum-Subtracted Images}\label{subsec:continuum_subtracted}

To better define the morphologies and measure the sizes of the discrete sources apparent in the colour images shown in Figure\,\ref{fig:colours-extended-A}--\ref{fig:colours-extended-B}, we made continuum-subtracted images involving the same filters as the colour images. Furthermore, while the colour images have the immediate advantage of revealing features that are either bluer or redder (or both) than the overall colour of the BCG, the continuum-subtracted images offers the advantage of lower noise fluctuations thus better revealing dimmer features.
In each case, the F140W image in the near-IR was scaled so as to best subtract the (axially symmetric) continuum light at the innermost region of the BCG produced by its old stellar population.  This procedure can leave appreciable residuals at the outer regions of the BCG owing to a colour gradient, as well as residuals at cluster members or galaxies lying in projection owing to their different colours as can be clearly seen in some of the images. The resulting continuum-subtracted images are shown in Figures\,\ref{fig:colours-extended-A}--\ref{fig:colours-extended-B} below their corresponding colour images involving the same filters. 


\begin{figure*}[hbt!]
\centering
\includegraphics[width=18cm]{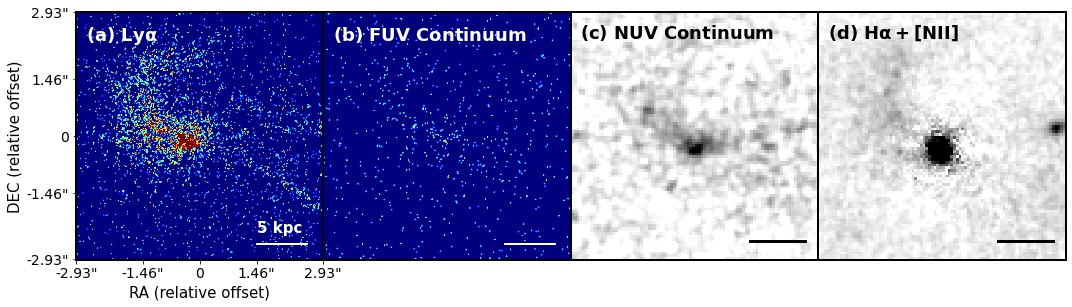}
\caption{Images of the BCG in RXC\,J2129.7+0005 as observed with the $HST$ in the: $(a)$ F140LP filter, encompassing the Ly-$\alpha$ line; $(b)$ F165LP filter, containing only far-UV (FUV) continuum; $(c)$ three of the shortest-wavelengths filters in the $CLASH$ program combined, containing only the near-UV (NUV) continuum; and $(d)$ the same continuum-subtracted image in H$\alpha$+[NII] as shown in Fig.\,\ref{fig:colours-extended-B}$c$.  The spatially-extended source in this BCG exhibits both hydrogen line emission and the UV continuum.}
\label{fig:FUV}
\end{figure*}

\begin{figure*}[htb!]
\centering
\includegraphics[width=17.8cm]{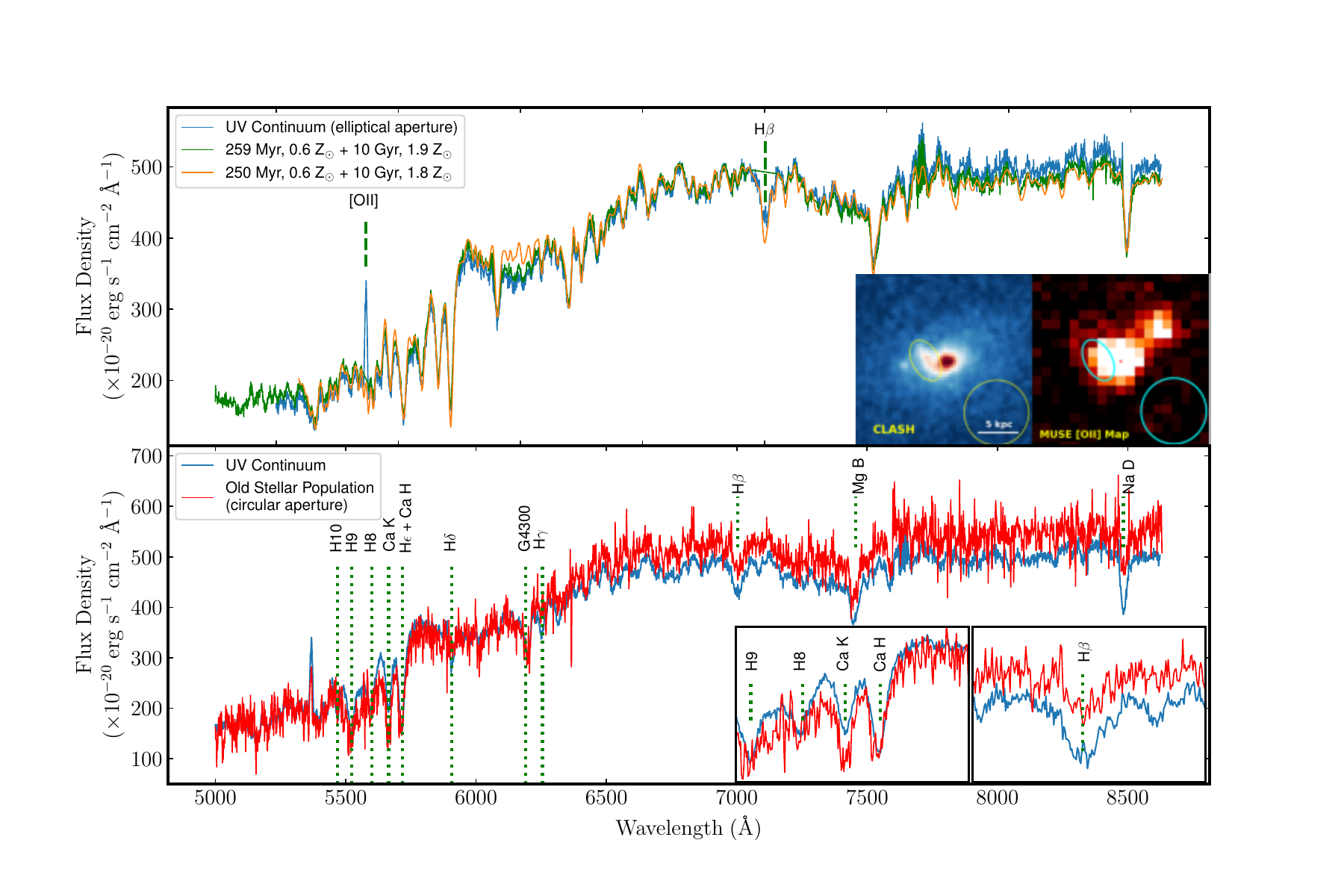}
\caption{Upper panel shows spectrum (blue) extracted over the elliptical aperture shown in the two inserts at the lower right corner of this panel, enclosing UV continuum towards the BCG of MACS\,J1206.2-0847.  The left insert shows its UV image from the $HST$ (same as in Fig.\,\ref{fig:colours-extended-B}$a$), and the right insert an [OII] image made from the MUSE-VLT. The [OII] and H$\beta$ lines at the redshift of the cluster are labelled. Two model spectra are shown, both comprising a composite young and old stellar population with ages and metallicity as indicated at the upper left corner and coloured: (i) orange for the best model fit to the SED shown in Fig.\,\ref{fig:SED-fitting} extracted over a somewhat larger region covering the same feature; and (ii) green for the best model fit to the  spectrum shown here.
Lower panel shows the same spectrum (blue) to be compared with the spectrum (red) extracted over a circular aperture, as shown also in both inserts in the upper panel, enclosing just the old stellar stellar population of the BCG.  Spectral lines of interest at the redshift of the cluster are labelled.  Majority of the remaining unlabelled lines (both upwards and downward spikes) in both panels are imperfectly subtracted sky lines. The region spanned by the UV continuum exhibits deeper Balmer absorption lines and an inversion of the Ca\,II H\&K lines  (left insert), indicating the presence of a younger stellar population, along with [O\,II]  and H$\beta$ lines (right insert), implicating a genuine associated with the BCG rather than a lensed feature at a higher redshift.}
\label{fig:MUSE}
\end{figure*}

The same features seen in the colour images are seen also in the continuum-subtracted images, affirming the reality of these features.  In addition, the continuum-subtracted images reveal features not clearly apparent if at all detectable in the colour images, specifically a central source in the filter encompassing [OII] in three BCGs (Fig.\,\ref{fig:colours-extended-A}$b$--$d$), a central source in the UV continuum in one BCG (Fig.\,\ref{fig:colours-extended-A}$f$), and relatively dim and spatially-extended features in the BCGs of MS2137.3-2353 (Fig.\,\ref{fig:colours-extended-B}$b$) and RXC\,J2129.7+0005 (Fig.\,\ref{fig:colours-extended-B}$c$).  Like before, we verified that the BCGs not exhibiting discrete features in any of the colour images also do not exhibit discrete features in the continuum-subtracted images.

The seven BCGs shown in Figure\,\ref{fig:colours-extended-A} each exhibit only a single discrete source at their centres:\,\,these sources are all detected in the continuum-subtracted images involving H$\alpha$+[N\,II] and H$\beta$+[O\,III], in all but one case also the continuum-subtracted images involving [OII], and in two cases also the UV continuum.  The three BCGs in Figure\,\ref{fig:colours-extended-B} each exhibit spatially-extended sources detectable from the UV to optical.  In the case of the BCG in MS2137.3-2353 (Fig.\,\ref{fig:colours-extended-B}$b$), a filament curling to the south is apparent in the continuum-subtracted (and also the colour) images involving the UV continuum, [O\,II], and H$\alpha$+[N\,II], but not H$\beta$+[O\,III].  Furthermore, a filament extending to the north can be seen in the continuum-subtracted (just barely the colour) image involving H$\alpha$+[N\,II] and marginally also [O\,II], but not H$\beta$+[O\,III] or UV continuum.  For the BCG in RXC\,J2129.7+0005 (Fig.\,\ref{fig:colours-extended-B}$c$), a filament curling to the north-east apparent in the colour images involving H$\alpha$+[N\,II], [O\,II], and UV continuum is more clearly apparent in the continuum-subtracted images.  In both the colour and continuum-subtracted images involving H$\beta$+[O\,III], this filament is detectable only at its inner region and there only in absorption.  As we will show in Section\,\ref{subsec:nebula}, this filament emits in H$\alpha$+[N\,II], [O\,II], and UV continuum, but not detectably in H$\beta$+[O\,III] whereby dust extinction arising within this filament causes it to appear in absorption; by contrast, this dust does not entirely suppress emission in the UV continuum or the other emission lines.
A filamentary feature is apparent also to the south-west of this BCG, detectable only in H$\alpha$+[N\,II].

\subsection{Ancillary and Other Data}\label{subsec:ancillary}

\citet{O'Dea2010} reported observations of the BCG in RXC\,J2129.7+0005 with the $HST$ in the F140LP and F165LP filters, which span shorter UV wavelengths than the filters employed in the $CLASH$ program.  At the redshift of this cluster, the Ly-$\alpha$ line is redshifted into the F140LP filter but lies below the cutoff in the F165LP filter.  Figure\,\ref{fig:FUV} shows the images in the F140LP (Fig.\,\ref{fig:FUV}$a$) and F165LP (Fig.\,\ref{fig:FUV}$b$) filters that we extracted from the Mikulski Archive for Space Telescopes (MAST), along with images in the UV constructed from the three shortest wavelength filters employed in the $CLASH$ program combined (Fig.\,\ref{fig:FUV}$c$), as well as the same continuum-subtracted image in H$\alpha$+[N\,II] (Fig.\,\ref{fig:FUV}$d$) as shown in Figure\,\ref{fig:colours-extended-B}$c$.  Whereas the image in Ly-$\alpha$ (Fig.\,\ref{fig:FUV}$a$) closely resembles that in H$\alpha$+[N\,II]  (Fig.\,\ref{fig:FUV}$c$), only the very brightest parts of the line-emitting feature around the centre are detectable in the UV contained in the F165LP filter (Fig.\,\ref{fig:FUV}$b$), although much of this feature is detectable at the longer UV wavelengths observed in the $CLASH$ program (Fig.\,\ref{fig:FUV}$d$).  These images demonstrate definitely that the filament curling to the north-east of center emits in both line and continuum.

As mentioned in Section\,\ref{subsec:colour_images}, \citet{Donahue2015} suggest that the feature curling to the north-east from the centre of the BCG in MACS\,J1206.2-0847 (see Fig.\,\ref{fig:colours-extended-B}$a$) is a gravitationally-lensed image of a background galaxy.  No lens model constructed for this cluster \citep[e.g.,][]{Caminha2017}, however, has identified the aforementioned feature as a lensed image.  MACS\,J1206.2-0847 has been observed with the Multi Unit Spectroscopic Explorer ($MUSE$) on the Very Large Telescope (VLT) as reported by \citet{Richard2021}.  We have retrieved the $MUSE$ data for this cluster from the ESO Science Archive Facility.  In Figure\,\ref{fig:MUSE}, we show an image of the [OII] linemap of the BCG (right insert at the upper panel of Fig.\,\ref{fig:MUSE}), as well as the same image in UV continuum as shown in Figure\,\ref{fig:colours-extended-B}$a$ (right insert at the upper panel of Fig.\,\ref{fig:MUSE}).  We also show the $MUSE$ spectrum extracted over an aperture enclosing the filament of interest, as well as another $MUSE$ spectrum showing the old stellar population in the BCG.  As can be seen, the spectrum at the UV filament exhibits emission lines that, if at the redshift of the cluster, corresponds to the (spectrally unresolved) [O\,II] doublet and H$\beta$ (in which case H$\alpha$+[N\,II] would be redshifted out of the spectrum covered by $MUSE$).  Furthermore, absorption lines from multiple elements can be seen at the same wavelengths from both spectra. The higher Balmer absorption lines are deeper relative to continuum at the UV filament and an obvious CaII H\&K lines inversion is present (see left insert in lower panel), indicating (as demonstrated explicitly in Section\,\ref{subsec:star_formation}) the presence of a relatively young stellar population (i.e., containing earlier-type stars) at UV filament.  These results confirm that the feature enclosed belongs to the BCG.

\subsection{Sizes of Compact Central Sources}\label{subsec:sizes}

\begin{figure*}[htb!]
\centering
\includegraphics[width=\textwidth]{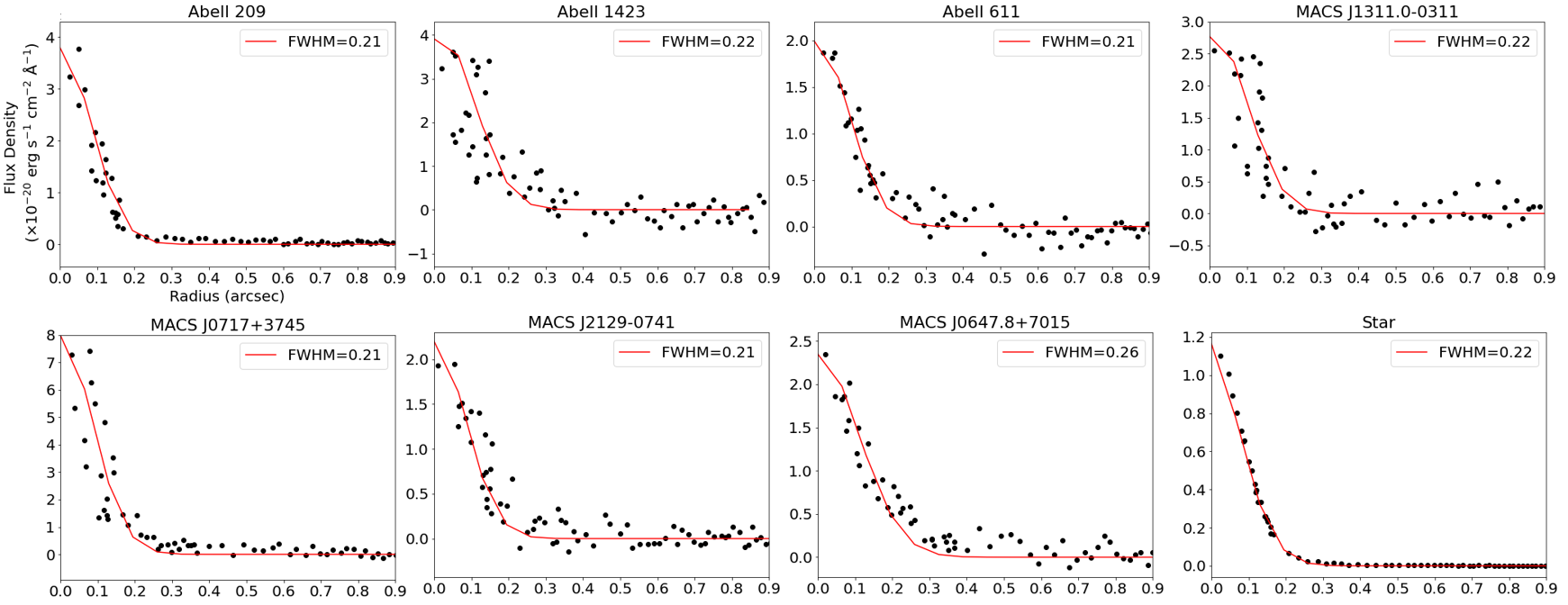}
\caption{Azimuthally-averaged radial profiles of the BCGs shown in Fig.\,\ref{fig:colours-extended-A}, each of which exhibits a compact central source, along with that of a foreground star towards the cluster Abell\,209 (last column, lower row).  These profiles were extracted from the continuum-subtracted images in H$\alpha$+[N\,II] shown in Fig.\,\ref{fig:colours-extended-A}.  A Gaussian fit to each profile (red curve) yields a FWHM near 0\farcs2, indicating that none of the sources are spatially resolved and hence referred to as nuclear sources.}
\label{fig:profile}
\end{figure*}

The seven BCGs shown in Figure\,\ref{fig:colours-extended-A} all display a compact central source suggestive of a nuclear source.  Figure\,\ref{fig:profile} shows the azimuthally-averaged
radial light profiles of these BCGs as derived from the continuum-subtracted images in H$\alpha$+[NII] shown in Figure\,\ref{fig:colours-extended-A}, along with that of a foreground star in one of their host clusters for comparison.  
A one-sided Gaussian fit to each of these radial light profiles yields a FWHM near 0\farcs2 (physical sizes of 0.7--1.7\,kpc given the range in redshifts of these BCGs), indicating that none of the central sources are spatially resolved.  These compact central sources are therefore referred to hereafter as nuclear sources.  

\section{Interpretation}  \label{sec:Interpretation}

To make the following description easier to follow, we direct the reader to Table\,\ref{BCG activity} where the results for all the BCGs in the $CLASH$ program are summarised.  Their host clusters are arranged in order of increasing $K_0$ as inferred by \citet{Cavagnolo2009} and, where not available (for MACS J2129.4-0741 and MACS J0416.1-2403), by \citet[][]{Donahue2014}.  In this table, black symbols refer to previously published studies, whereas red symbols to our study; ``---'' indicates where no specific search for the relevant feature has been made, and ``?'' indicates an uncertain result. Previous identifications of $CLASH$ BCGs exhibiting star formation combine studies made by \citet{Donahue2015} and \citet{Fogarty2017}, and those exhibiting emission lines are as reported by \citet{Fogarty2015}.  Previous identifications of BCGs exhibiting radio AGNs are those made by \citet{Yu2018} in nineteen of the $CLASH$  clusters (all of which are among the dynamically-relaxed sample), or by \citet{Xie2020} for the BCG in RXC\,J2248.7-4431 (also included in the dynamically-relaxed sample), all based on observations with the Very Large Array (VLA).  For the remaining five BCGs, we examined archival images from the NRAO VLA Sky Survey (NVSS) and, where observed, Faint Images of the Radio Sky at Twenty-cm (FIRST) also made with the VLA; we find none of these BCGs to display detectable radio AGNs, for which we determine $3\sigma$ upper limits based on the noise level in the corresponding maps. For one of these clusters, MACSJ0717.5+3745, a bright radio relic crosses its BCG \citep{Rajpurohit2021}, making the identification of any radio AGN in this galaxy problematic.  Nuclear sources are identified in the manner previously described in Section\,\ref{subsec:sizes}.


\begin{deluxetable*}{lcccccc}[ht!]
\decimals
\tabletypesize{\footnotesize}
\tablewidth{0pt}
\tablecolumns{7}
\tablecaption{$CLASH$ BCG activity}
\tablehead{
\colhead{Cluster Name} & \colhead{Redshift} & \colhead{K$_0$} & \colhead{Star} & \colhead{Emission} & \colhead{Nuclear} & \colhead{Radio} \vspace{-0.2cm} \\
\colhead{} & \colhead{} & \colhead{(keV cm$^2$)} & \colhead{Formation} & \colhead{Lines} & \colhead{Source} & \colhead{AGN} }
\startdata
MACS\,J1423.8+2404 & 0.545 & 10$\pm$5 & \checkmark & \checkmark & $-$ &  \checkmark \\
MACS\,J0329.6-0211 & 0.45 & 11$\pm$3 & \checkmark & \checkmark & $-$ &  \checkmark \\
RXC\,J1347.5-1145 & 0.451 & 12$\pm$20 & \checkmark & \checkmark & $-$ & \checkmark \\
Abell 383 & 0.187 & 13$\pm$2 & \checkmark &\checkmark & $-$ & \checkmark \\
MACS\,J1931.8-2635 & 0.352 & 14$\pm$4 &\checkmark & \checkmark & $-$ & \checkmark \\
MS2137.3-2353 & 0.313 & 15$\pm$2 & \checkmark$\!\!$\color{red}\checkmark & \checkmark$\!\!$\color{red}\checkmark & \color{red}? & \checkmark \\
MACS\,J1115.8+0129 & 0.352 & 15$\pm$3 & \checkmark & \checkmark & $-$ & \checkmark \\
MACS\,J1532.8+3021 & 0.345 & 17$\pm$2 & \checkmark & \checkmark & $-$ & \checkmark \\
MACS\,J0429.6+0253 & 0.399 & 17$\pm$4 & \checkmark & \checkmark & $-$ & \checkmark \\
RXC\,J2129.7+0005 & 0.234 & 21$\pm$4 & \checkmark$\!\!$\color{red}\checkmark & \checkmark$\!\!$\color{red}\checkmark & \color{red}? & \checkmark \\
MACS\,J1720.2+3536 & 0.391 & 24$\pm$3 & \checkmark & \checkmark & $-$ & \checkmark \\
MACS\,J0744.9+3927 & 0.686 & 42$\pm$11 & \color{red}$\times$ & \color{red}$\times$ & \color{red}$\times$ & \checkmark \\
MACS\,J1311.0-0310 & 0.494 & 47$\pm$4 & \color{red}$\times$ & \color{red}$\times$ & \color{red}\checkmark & $\times$ \\
Abell 2261 & 0.224 & 61$\pm$8 & \color{red}$\times$ & \color{red}$\times$ & \color{red}$\times$ & \checkmark \\
Abell 1423 & 0.213 & 68$\pm$13 & \color{red}$\times$ & \color{red}$\times$ & \color{red}\checkmark & $\times$ \\
MACS\,J1206.2-0847 & 0.44 & 69$\pm$10 & \color{red}\checkmark & \color{red}$\times$ & \color{red}? & \checkmark \\
Abell 209 & 0.206 & 106$\pm$27 & \color{red}$\times$ & \color{red}$\times$ & \color{red}\checkmark & $\times$ \\
Abell 611 & 0.288 & 125$\pm$18 & \color{red}$\times$ & \color{red}$\times$ & \color{red}\checkmark & \checkmark \\
CL\,J1226.9+3332 & 0.89 & 166$\pm$45 & \color{red}$\times$ & \color{red}$\times$ & \color{red}$\times$ & \checkmark \\
RXC\,J2248.7-4431$^c$ & 0.348 & 170$\pm$20 & \color{red}$\times$ & \color{red}$\times$ & \color{red}$\times$ & \checkmark \\
MACS\,J2129.4-0741 & 0.570 & 200$\pm$100 & \color{red}$\times$ & \color{red}$\times$ & \color{red}\checkmark & \color{red}$\times$ \\
MACS\,J0717.5+3745 & 0.548 & 220$\pm$96 & \color{red}$\times$ & \color{red}$\times$ & \color{red}\checkmark & \color{red} ? \\
MACS\,J0647.8+7015 & 0.584 & 225$\pm$50 & \color{red}$\times$ & \color{red}$\times$ & \color{red}\checkmark & \color{red}$\times$ \\
MACS\,J1149.6+2223 & 0.544 & 280$\pm$40 & \color{red}$\times$ & \color{red}$\times$ & \color{red}$\times$ & \color{red}$\times$  \\
MACS\,J0416.1-2403 & 0.396 & 400$\pm$100 & \color{red}$\times$ & \color{red}$\times$ & \color{red}$\times$ & \color{red}$\times$ \\
\enddata
\tablecomments{Black symbols refer to previously published studies (see text), whereas red symbols to our study.  ``---'' indicates where no specific search for the relevant feature has been been made, and ``?'' indicates an uncertain result.}
\end{deluxetable*}
\label{BCG activity}
\vspace{-0.6cm}

\begin{figure*}[htb!]
\centering
\includegraphics[width=11.2cm]{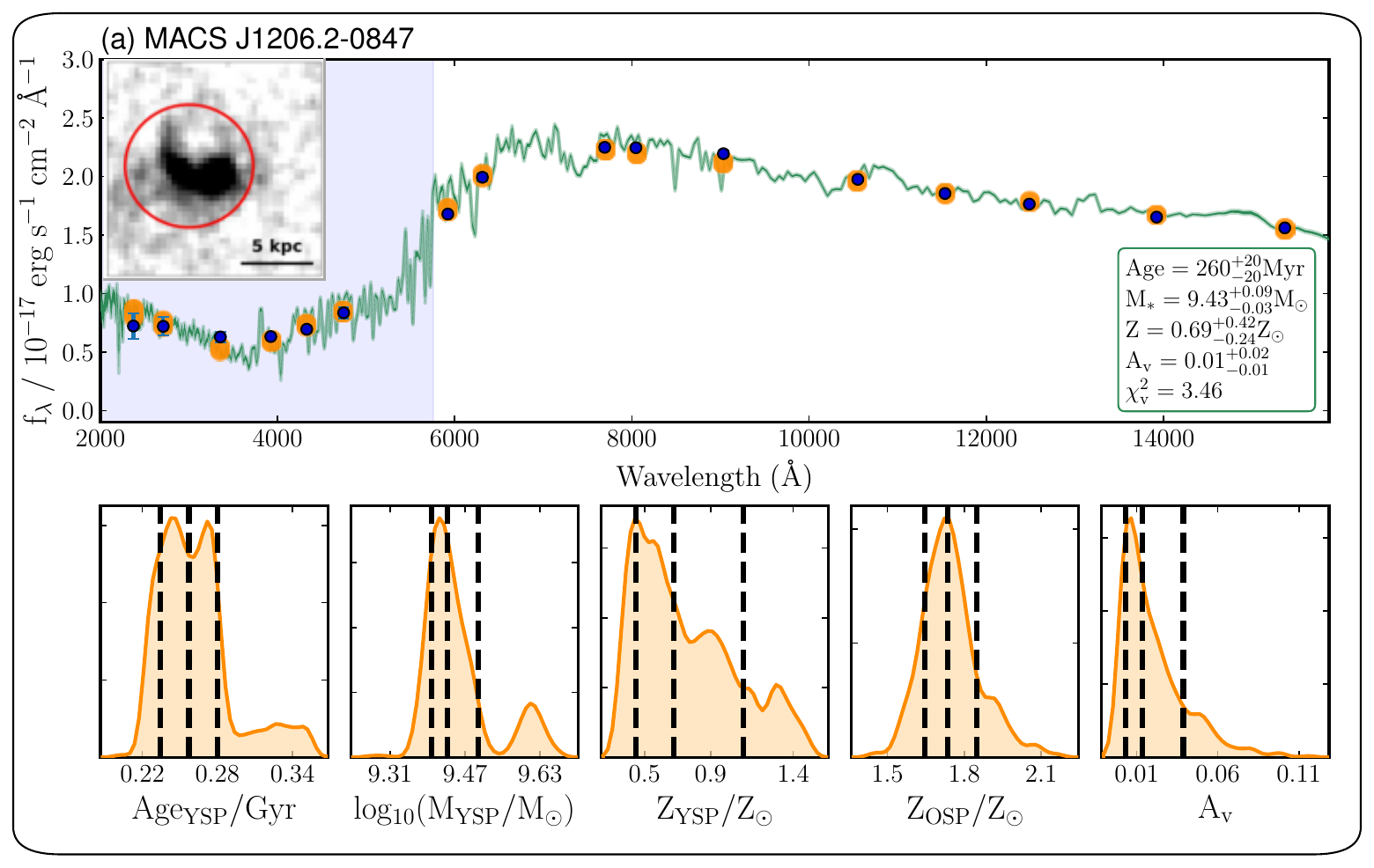}
\includegraphics[width=11.2cm]{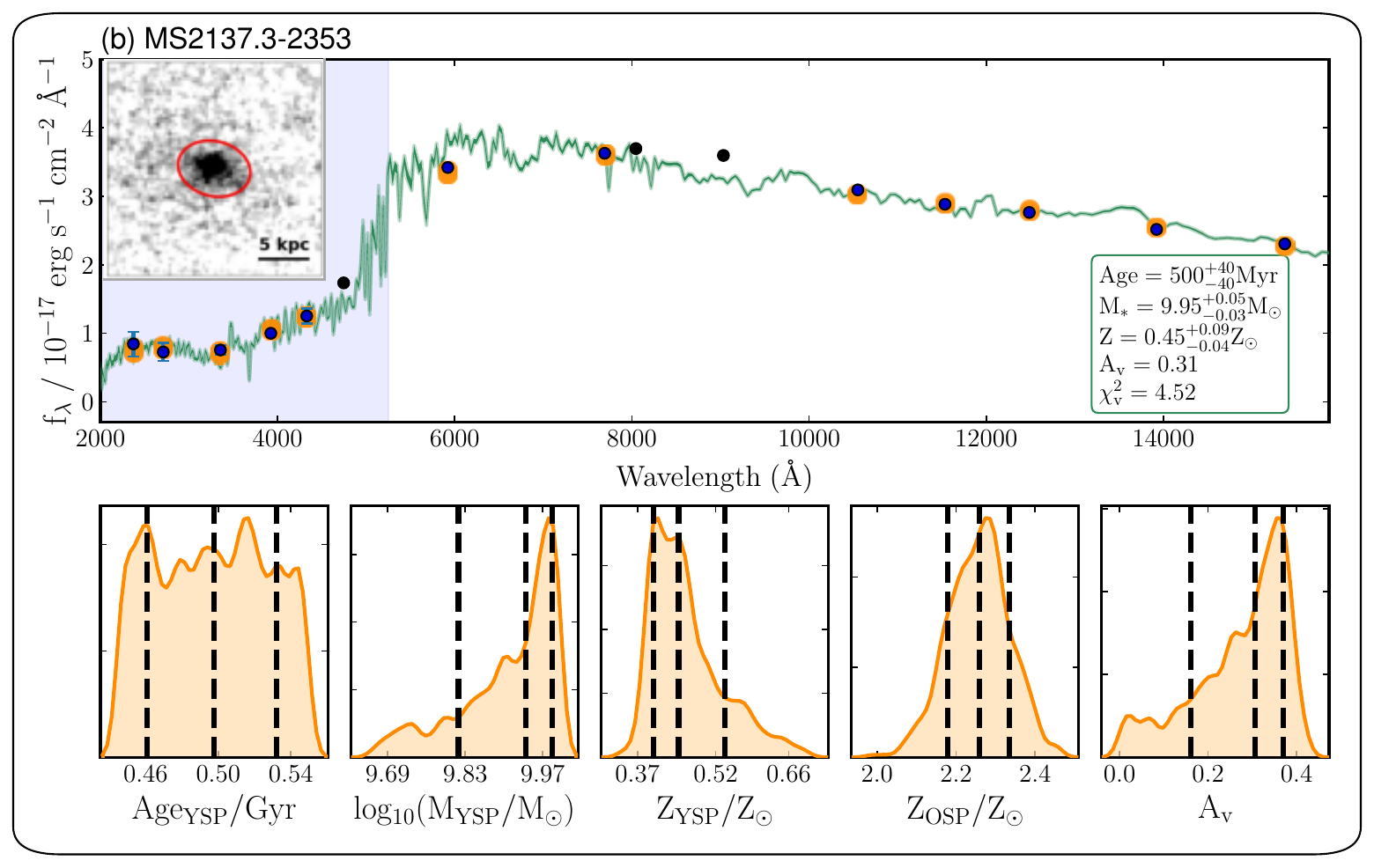}
\includegraphics[width=11.2cm]{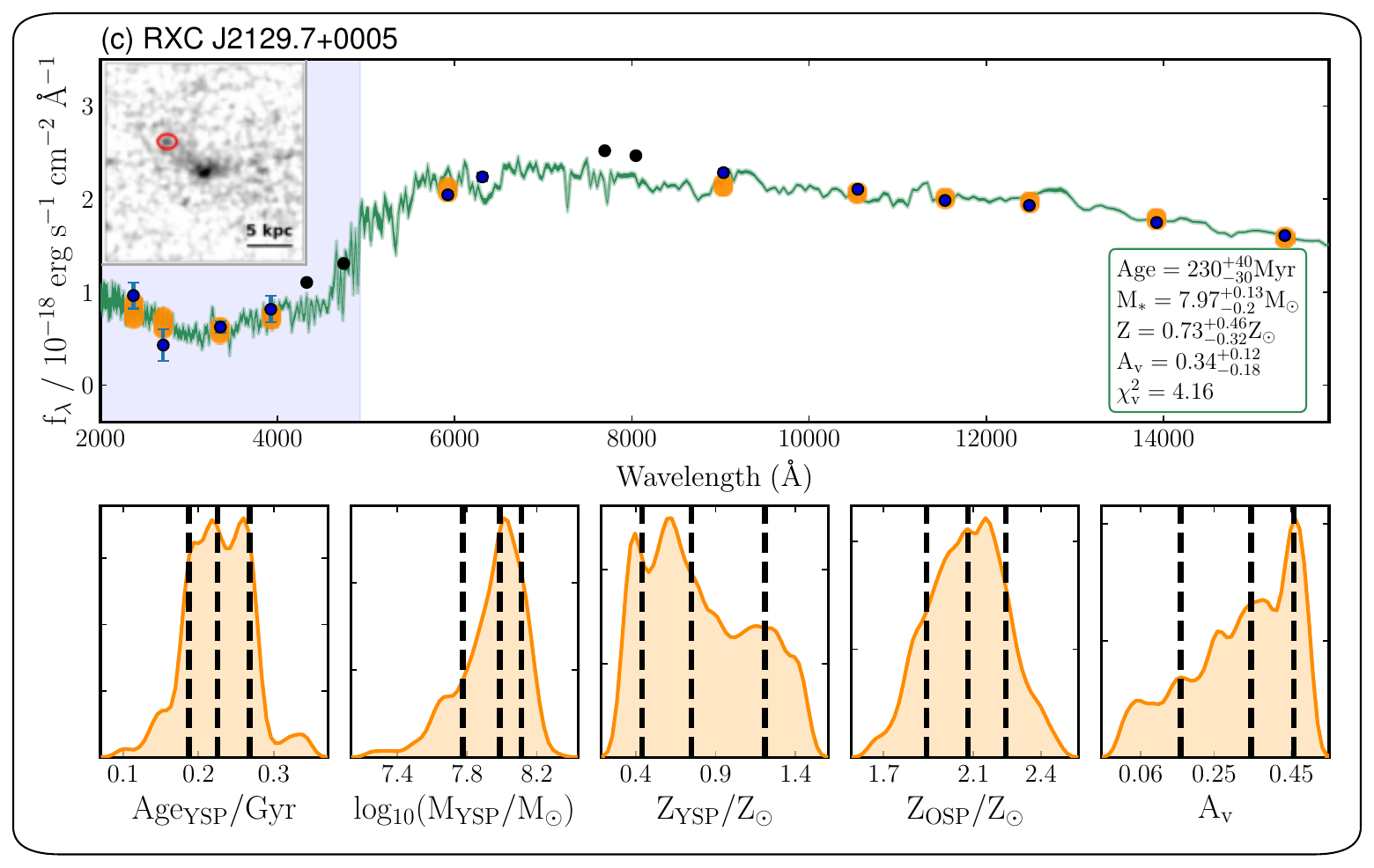}
\caption{In each block of panels, upper row shows best-fit composite model spectrum (green curve) to the spatially-integrated SED (blue data points) within the aperture shown in the insert, corresponding to the same UV image shown in Fig.\,\ref{fig:colours-extended-B} for each BCG.  The model spectra convolved by the bandpass of the individual $CLASH$ filters are shown as orange data bands, having vertical extents spanning the 16th and 84th percentile of the best-fit likelihood distribution in the fitted parameters.  Black data points left out of the model fit owing to nebular line emission (filters containing [O\,II] and H$\alpha$+[NII]) have no accompanying orange bands (see text). Purple bands indicate those filters spanning UV wavelengths at the rest-frame of the BCG. The composite model spectra each includes two stellar populations, having a best-fit likelihood distribution in age, mass, and metallicity for the younger stellar population, as well as in metallicity for the older stellar population, shown in the lower row of each block of panels; the age of the old stellar population is fixed at a look-back time of 10\,Gyr.}
\label{fig:SED-fitting}
\end{figure*}

\subsection{Star Formation}   \label{subsec:star_formation}
Among the sixteen BCGs studied in this paper, only three exhibit complex UV morphologies as shown in Figure\,\ref{fig:colours-extended-B}.
We now investigate whether the SEDs of these BCGs can be reproduced by recently formed stars along with a pre-existing ancient stellar population, and if so the overall ages and masses of the recently formed stars.  

Figure\,\ref{fig:SED-fitting} shows, for each of the three BCGs in Figure\,\ref{fig:colours-extended-B}, the aperture (a red circle or ellipse) used to extract the spatially-integrated SED (black data points) shown in the same figure.  To guard against any nuclear emission not related to relatively young stars, we masked the central region of the BCG in MACS\,J1206.2-0847 using a circular mask with a diameter equal to twice the FWHM of the PSF.  We could not apply a similar mask to the BCG in MS2137.3-2353 as much of the discrete UV source in this galaxy would then be masked out; instead, the lack of a distinct brightening at the very centre of this BCG suggests that any genuine nuclear emission does not contribute much to the extracted SED.  For the BCG in RXC\,J2129.7+0005, we considered only the outer portion of its discrete UV source, as the inner portion is clearly affected by dust extinction (apparent in Figure\,\ref{fig:colours-extended-B}$c$ as pointed out in Section\,\ref{subsec:continuum_subtracted}, and demonstrated explicitly at the end of this subsection).  

For fitting stellar population synthesis models to the SEDs shown in Figure\,\ref{fig:SED-fitting}, we co-add the contributions by: (i) an old stellar population that formed at a fixed lookback time of 10\,Gyr and allowed to have super-solar metallicities, $Z$, over the range $1 {\rm \, Z_\sun} < Z < 2.5 {\rm \, Z_\sun}$, as is characteristic of the old stellar populations in massive elliptical galaxies \citep[e.g.,][]{Loubser2009}; and (ii) a coeval young stellar population constrained to have an age somewhere in the range 0--1\,Gyr, beyond which its UV light should have faded to obscurity, and sub-solar metallicities over the range $0.3 {\rm \, Z_\sun} < Z < 1 {\rm \, Z_\sun}$, the lower limit of which is just below the approximate metallicity of the ICM in rich galaxy clusters \citep[e.g.,][and references therein]{Mantz2017}. We allow for dust extinction, constrained over a relatively wide range spanning $0 \le A_V \le 2$, with a dust law as given by \citet{Calzetti2000}.  The fitting was performed using the publicly-available code Bayesian Analysis of Galaxies for Physical Inference and Parameter EStimation \citep[Bagpipes;][]{Carnall2018}.  As the name suggests, this code uses a Bayesian approach for SED fitting that yields the probability distributions of the free parameters selected by the user for the fit.  A prior can be given to each of the free parameters (i.e., weights to different values), such as a user-defined Gaussian function; in our work, we used a uniform prior (i.e., equal weights to all values) for all the free parameters mentioned above. The Bagpipes program adopts an initial mass function as prescribed by \citet{Kroupa2002}.  For more details on this program, see \citet{Carnall2018}.  

\begin{figure*}[hbt!]
\centering
\includegraphics[width=\textwidth]{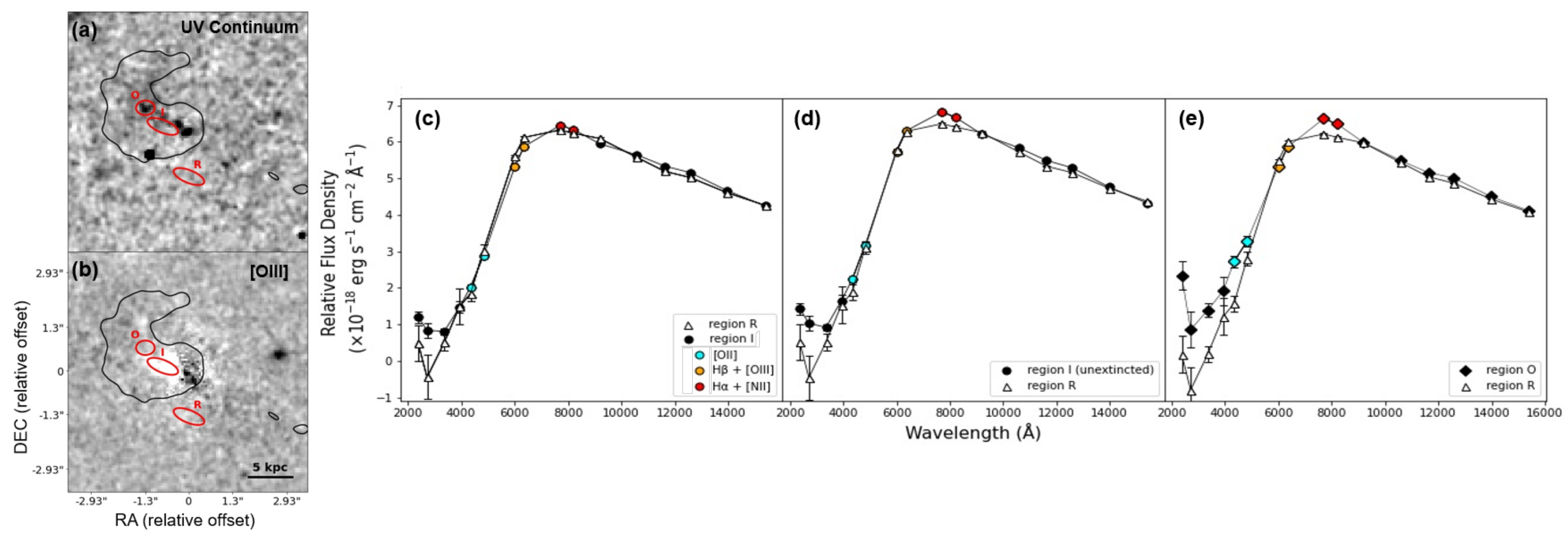}
\caption{$(a)$--$(b)$ Same continuum-subtracted images involving filters in the UV and H$\beta$+[O\,III] for the BCG in RXC\,J2129.7+0005 as shown in Fig.\,\ref{fig:colours-extended-B}$c$.  Ellipses enclose regions over which spatially-integrated SEDs were extracted, enclosing the inner region of the UV source (aperture I) that appears in absorption in the H$\beta$+[O\,III] image, the outer region of the UV source (aperture O) that is not detected in the H$\beta$+[O\,III] image, and a region away from the UV source (aperture R) enclosing just the old stellar population in the BCG.  Black contour indicates the outermost detectable extent of the Ly-$\alpha$ emission shown in Fig.\,\ref{fig:FUV}$(a)$.  $(c)$ SEDs extracted over apertures I and R normalised to their average intensities at the four longest wavelengths, showing relatively lower flux densities in filters containing H$\beta$+[O\,III] at I compared with R.  $(d)$ SEDs at same two apertures, except that extracted over I has been corrected for an extinction of $A_V = 0.07$ based on the dust extinction law of \citet{Calzetti2000}.  Now, the dust-corrected flux densities in filters containing H$\beta$+[O\,III] at I is similar to that at R, and in filters containing H$\alpha$+[N\,II] higher at I than at R.  $(e)$ SEDs extracted at apertures O and R normalised to their average intensities at the four longest wavelengths.  The SED at aperture O is similar to the extinction-corrected SED at aperture I, demonstrating that the inner region of the UV source is more strongly affected by dust extinction than its outer region.}
\label{fig:dust}
\end{figure*}


The best-fit model spectra are plotted in orange over the measured SEDs (black data points) in Figure\,\ref{fig:SED-fitting}. The corresponding model SEDs, derived by integrating the model spectra over the bandpasses of the individual $CLASH$ filters, are shown as orange points.  Most reassuringly, the best-fit model spectrum to the SED of the UV source in the BCG of MACS\,J1206.2-0847 also provides a good fit to its $MUSE$ spectrum as shown by the orange trace in the upper panel of Figure\,\ref{fig:MUSE}.  In fitting these model spectra, we had to guard against any emission-line nebulae not associated with the relatively young stellar population, as is clearly evident in, most especially, the BCG of MS2137.3-2353 (see Section\,\ref{subsec:nebula}).  In an initial fit that included all data points in the measured SEDs, we found that the best-fit model spectra and hence their corresponding model SEDs disagreed with the measurements in filters encompassing H$\alpha$+[N\,II] and [O\,II] for the BCGs in MS2137.3-2353 (Fig.\,\ref{fig:SED-fitting}$b$) and RXC\,J2129.7+0005 (Fig.\,\ref{fig:SED-fitting}$c$); these data points, corresponding to those with no associated orange points in Figure\,\ref{fig:SED-fitting}, were therefore excluded in a second round of fitting. The histograms in Figure\,\ref{fig:SED-fitting} show the probability distribution function of the free parameters, constituting the metallicities of the two stellar populations, along with the age and total initial mass of the younger stellar population (the total initial mass of the older stellar population, also a free parameter, is not shown as it is not of relevance in our study).  The values of the best-fit parameters and their uncertainties are indicated above each histogram.  In keeping with expectations, the best-fit models require highly super-solar metallicities ($\sim$2.0--2.5\,$Z_\sun$) for the older stellar populations in all three BCGs.  By contrast, the same models prefer sub-solar metallicities ($\sim$0.4\,$Z_\sun$) for the younger stellar populations.

All the best-fit models require relatively low dust extinctions ($A_V \lesssim 0.3$).  The best-fit age of the younger stellar population in the BCG of MACS\,J1206.2-0847 is $\sim$250\,Myr, and the total initial mass, $M_*$, of this population is ${\rm log} \, M_* = 9.42^{+0.04}_{-0.03} \rm \, M_\sun$.  By comparison, the total stellar mass contained in this galaxy is \mbox{$(4.93 \pm 1.39) \times 10^{11} \rm \, M_\sun$ \citep{Burke2015}}; hence, the younger stellar population comprises $\sim$0.5\% of the total stellar mass.  For the BCG in MS2137.3-2535, the best-fit age of its younger stellar population is $\sim$620 Myr and the total initial mass of this population ${\rm log} \, M_* = 9.8^{+0.02}_{-0.03} \rm \, M_\sun$.  By comparison, the total stellar mass contained in this galaxy is \mbox{$(8.38 \pm 0.42) \times 10^{11} \rm \, M_\sun$ \citep{Burke2015}}, and so its younger stellar population comprises $\sim$0.6\% of the total stellar mass.  As for the BCG in RXC\,J2129.7+0005, the best-fit age of its younger stellar population is $\sim$180 Myr and the total initial mass of this population ${\rm log} \, M_* = 7.35^{+0.09}_{-0.12} \rm \, M_\sun$; the latter is a lower limit as the inner portion of the UV source affected by dust extinction has been excluded.  The total stellar mass contained in this galaxy is \mbox{$(2.21 \pm 0.32) \times 10^{11} \rm \, M_\sun$ \citep{Burke2015}}, and so the younger stellar population comprises at least 0.02\% of the total stellar mass.

At this point, we stress that the inferred ages and therefore masses of the younger stellar population should only be regarded as characteristic ages and masses.  Different sightlines toward the younger stellar population may well have different SEDs; moreover, a given sightline may contain overlapping stellar populations having different ages.   Our primary aim is to demonstrate that the spatially-integrated SEDs toward the UV sources in these three BCGs can be produced by a combination of a young and an old stellar population -- and hence that these galaxies are or were recently engaged in star formation. In the Appendix, we show model SED fits for the BCGs in RXC\,J2129.7+0005 and MS2137.3-2535 for which the old stellar population is modelled in the same manner as mentioned above, whereas the young stellar population is assumed to have formed: (i) in a single burst less than 10\,Myr ago; or (ii) at a constant rate over the past 100\,Myr (as is approximately presumed when converting between UV continuum and star-formation rate using the \citet{Kennicutt1998} relationship).  We now include filters containing [O\,II] and H$\alpha$+[NII] (i.e., all filters) in the fits, as emission lines from H\,II regions can elevate the brightnesses in these (and other relevant) filters.  While we allowed the dust extinction to be freely fit, we also made model SED fits by setting the dust extinction at the respective values inferred by \citet{Fogarty2017} for the BCGs in RXC\,J2129.7+0005 and MS2137.3-2535 using Calzetti attenuation law.  All the best-fit model SEDs thus derived significantly under-predict the brightnesses of one or more filters in the rest-frame optical, and over-predict the brightnesses of the two rest-frame UV filters at the shortest wavelengths.  These model SED fits, despite in all likelihood being over simplistic, do collectively indicate that the young stellar population involved is predominantly too old to be still associated with H\,II regions, and therefore unlikely to be primarily responsible for elevating the brightnesses in the filters containing the aforementioned emission lines.   

\subsection{Emission Lines}  \label{subsec:nebula}

Among the nine BCGs in the $CLASH$ sample that obviously display spatially-extended and complex UV morphologies, \citet{Fogarty2015} find all those to exhibit H$\alpha$+[N\,II] emission.  Among the remaining sixteen BCGs studied in this paper, \citet{Fogarty2015} identified those in MS2137.3-2353 (Fig.\,\ref{fig:colours-extended-B}$b$) and RXC\,J2129.7+0005 (Fig.\,\ref{fig:colours-extended-B}$c$) to exhibit H$\alpha$+[N\,II] emission.  Note that the continuum subtracted H$\alpha$+[N\,II] images shown by \citet{Fogarty2015} for these two BCGs (see their Fig.\,3) may differ from those that we made (Fig.\,\ref{fig:colours-extended-B}$b$--$c$).
We attribute these differences to the different images used for the continuum subtraction: whereas we always used images in the F140W filter for the continuum subtraction, \citet{Fogarty2015} used images in filters adjacent to and either shortwards or longwards of the filter encompassing H$\alpha$+[NII] (see their Table\,2).  We selected the F140W filter so as to best subtract light from the old stellar population, which dominates the light in the near-IR, in the individual BCGs.  (We chose the F140W filter rather than the F160W filter so as to provide a direct comparison with the colour images shown in Figs.\,\ref{fig:colours-extended-A}--\ref{fig:colours-extended-B}, which also use the F140W filter.)

We caution that the manner in making continuum-subtracted line images can leave residuals corresponding to imperfectly-subtracted continuum at the locations of relatively young stars -- the very locations where residuals can be easily mistaken for genuine line emission, which in turn is attributed to H\,II regions associated with newly-formed stars.  Both \citet{Fogarty2015} and ourselves scale the continuum in the off-line filter so as to best subtract the main body of the BCG from the on-line filter.  This procedure better subtracts the light from an old stellar population that dominates the light from the BCG, but may not entirely remove (or, in the procedure employed by \cite{Fogarty2015}, potentially over-subtract) light from a younger stellar population that has a different SED than the old stellar population. These difficulties highlight the importance of checking that any features detected in the continuum-subtracted images also are detected in the corresponding colour images, as performed in Figure\,\ref{fig:colours-extended-B}.  Furthermore, a comparison of these images across adjacent filters is able to reveal whether any features detected correspond to continuum (from young stars) or line emission: in both MS2137.3-2353 (Fig.\,\ref{fig:colours-extended-B}$b$) and RXJ\,2129.6+005 (Fig.\,\ref{fig:colours-extended-B}$c$), filamentary features visible in emission in both the continuum-subtracted and colour images containing the [OII] and H$\alpha$+[NII] lines are not visible in the image containing the [OIII] line bracketed by the two aforementioned images, indicating that these features correspond to line emission.


Among the nine BCGs in the $CLASH$ program that exhibit spatially-extended and complex UV morphologies, \citet{Fogarty2015} find that emission in their continuum-subtracted H$\alpha$+[N\,II] images extend well beyond their discrete UV-continuum sources (see their Fig.\,3).  There seems little doubt that these particular BCGs exhibit genuine line emission.   In our work, the continuum-subtracted images that we made in multiple filters encompassing different spectral lines, as well as in the UV continuum, provide greater insights on the nature of the line emission in the BCGs studied.  The BCG in MS2137.3-2353 (Fig.\,\ref{fig:colours-extended-B}$b$) exhibits a filament to the south of its center that is detected in the UV continuum, [O\,II], and H$\alpha$+[N\,II], but not in H$\beta$+[O\,III].  The spectral properties of this filament is characteristic of that exhibited by emission-line nebulae in BCGs, displaying low-ionisation lines having different relative intensities (in particular, relatively dim if at all detectable in [O\,III]) compared with those of H\,II regions (see Section\,\ref{sec:Results}) -- even when spatially coincident with UV continuum.  

As mentioned above and shown in Figure\,\ref{fig:SED-fitting}, we considered only the outer portion of the discrete UV source in the BCG of RXC\,J2129.7+0005 when making a model fit to its SED.  After subtracting light from the old stellar population in the BCG, this outer portion also is detected in filters encompassing [O\,II] and H$\alpha$+[N\,II], but not H$\beta$+[O\,III] (see Fig.\,\ref{fig:colours-extended-B}$c$).  By contrast, the inner portion is detected only in the UV continuum and filters encompassing H$\alpha$+[N\,II], is not apparent in filters encompassing [O\,II], and appears in absorption in filters encompassing H$\beta$+[O\,III].  Here we demonstrate that the entire source likely emits in the UV continuum, [O\,II], and H$\alpha$+[N\,II], but not detectably in H$\beta$+[O\,III]; dust at the inner portion of this source suppresses both the line emission and UV continuum, and creates a silhouette against background stars in the BCG in filters encompassing H$\beta$+[O\,III].

In Figure\,\ref{fig:dust}$a$, we show elliptical apertures enclosing the inner (aperture I) and outer (aperture O) portions of the UV source, as well as an elliptical aperture enclosing an adjacent region (aperture R) away from the source.  The spatially-integrated SEDs extracted over apertures I and R, normalised to their average intensities at the four longest wavelengths, are shown in Figure\,\ref{fig:dust}$b$:\,\,the SED of aperture I shows a clear excess in the UV relative to that of aperture R.  In Figure\,\ref{fig:dust}$c$, the SED extracted over aperture I has been corrected for a wavelength-dependent extinction of $A_V = 0.07$ based on the dust extinction law of \citet{Calzetti2000}, and now shows a clear excess also in filters encompassing H$\alpha$+[N\,II] apart from the UV continuum.  Figure\,\ref{fig:dust}$d$ shows the corresponding normalised SEDs extracted over regions O and R.  As can be seen, the extinction-corrected SED extracted over aperture I closely resembles the SED extracted over aperture O, demonstrating that the absorption seen in filters containing H$\beta$+[OIII] at aperture I is caused by dust extinction.


Despite their generally different spectral characteristics than H\,II regions, H$\alpha$+[N\,II] emission from BCGs is often used to infer their star-formation rates \citep[e.g., as in the work by][]{Fogarty2015}.  We emphasise care in using H$\alpha$+[N\,II] as a diagnostic for star formation, even when these galaxies exhibit relatively young stellar populations as traced in UV continuum, because a portion if not the bulk of the line emission may not be associated with H\,II regions.  Indeed, as demonstrated in Section\,\ref{subsec:star_formation}, the stars traced in the UV continuum may mostly have ages of hundreds of Myr, far too old to produce H\,II regions.  

\begin{figure*}[ht!]
\centering
\includegraphics[width=18cm]{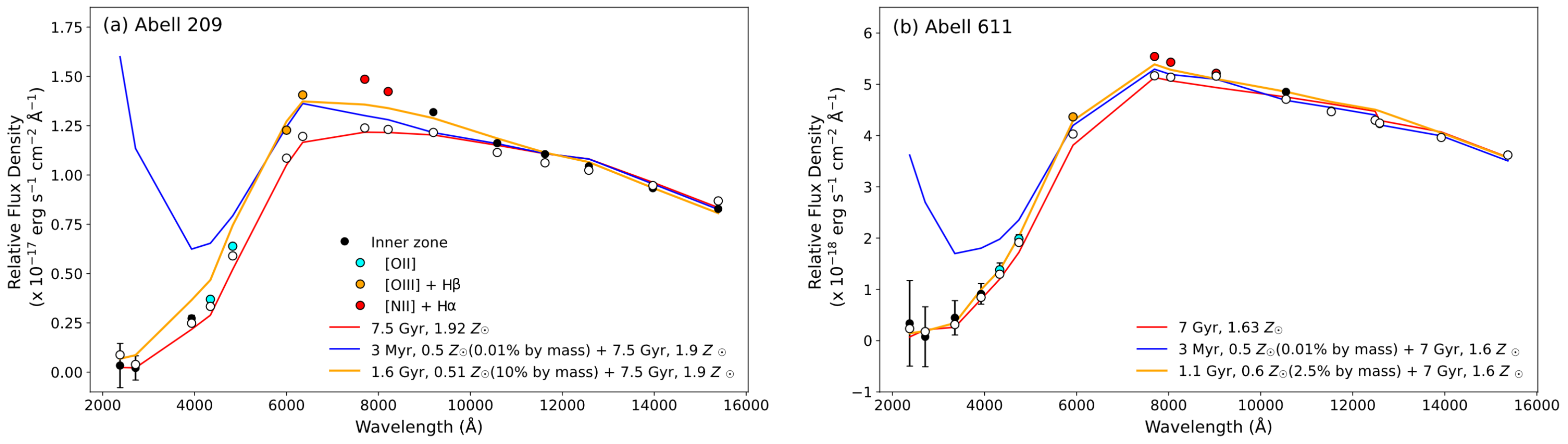}
\caption{Exemplar SEDs of BCGs that exhibit nuclear sources, which are detected in filters encompassing H$\alpha$+[N\,II], H$\beta$+[O\,III], [O\,II], and UV (filter centred at 3900\,\AA).  Filled circles are data extracted over a central circular zone containing the nuclear source, and unfilled circles those extracted over a surrounding annular zone containing the main body of the BCG; these data points are the same as those plotted for the corresponding BCGs in Fig.\,\ref{fig:SEDs}.  Model SEDs are plotted as colour curves for stellar populations having physical parameters as specified by the legends in the lower right corner of each panel.  The SED of the surrounding annular zone (unfilled circles) is well reproduced by a model SED comprising an old stellar population formed at a fixed look-back time of 10\,Gyr and having super-solar metallicities (red curve).  To this we add a younger stellar population (age of 3\,Myr) to fit the SED of the central zone (filled circles), finding that a very young stellar population still exhibiting a H\,II region produces a UV continuum far in excess of that observed (blue curve).  The addition of an intermediate age stellar population (1.6\,Gyr) also cannot reproduce the SED of the central zone in Abell\,209, suggesting its nuclear source corresponds to an AGN.  On the other hand, the addition of an intermediate age stellar population (1.1\,Gyr) can reproduce the SED of the central zone in Abell\,611, as well as the SEDs of the central zones in the other five BCGs shown in Fig.\,\ref{fig:colours-extended-A}.  The nuclear sources in all but one of these BCGs can therefore be produced by stellar populations that are significantly younger (ages in the range 0.5--2.8\,Gyr) than the surrounding stars, although their compact morphologies would seem to favour AGNs.
}
\label{fig:old+young}
\end{figure*}

\subsection{Nuclear Sources} \label{subsec:nuclear}
Seven of the BCGs exhibit discrete nuclear sources (see Fig.\,\ref{fig:colours-extended-A} and Fig.\,\ref{fig:profile}).  These source are detected only in the colour as well as continuum-subtracted images encompassing H$\alpha$+[N\,II] and H$\beta$+[O\,III] in one case (Fig.\,\ref{fig:colours-extended-A}$a$), also relatively dimly in [O\,II] in three cases (Fig.\,\ref{fig:colours-extended-A}$b$--$d$), at comparable brightness in all three sets of emission lines in one case (Fig.\,\ref{fig:colours-extended-A}$e$), along with the UV continuum in two cases (Fig.\,\ref{fig:colours-extended-A}$f$--$g$).  Their detection primarily in filters encompassing bright emission lines associated with either H\,II regions or AGNs would seem to suggest an origin associated with either or both sources.

To test whether these nuclear sources could be associated with newly-formed stars and their H\,II regions, we fit model stellar populations to the measured SEDs of these BCGs as extracted from the same two apertures as in Figure\,\ref{fig:SEDs}:\,\,a central circular zone enclosing the nuclear source, and a surrounding elliptical annular zone encompassing the main body of the BCG.  The latter constrains the parameters of the old stellar population comprising the main body of the BCG, to which we will add a young stellar population to fit the SED measured towards the nuclear source.  Like before, the age of the old stellar population is fixed to a look-back time of 10\,Gyr and its metallicity freed to span solar to super-solar values over the range $1 {\rm \, Z_\sun} < Z < 2.5 {\rm \, Z_\sun}$.
Figure\,\ref{fig:old+young} shows example results of model SEDs for the old stellar population (red curves) fitted to the SEDs extracted over annular zones surrounding the central circular zone (unfilled circles) for the BCGs in Abell\,209 and Abell\,611. 
As can be seen, the model SEDs thus fitted closely reproduce the measured SEDs of the old stellar population in both these BCGs.

To the SED extracted from the central circular zone for each of the two BCGs shown in Figure\,\ref{fig:old+young}, we fit a composite model SED comprising a young stellar population in addition to the same old stellar population determined from the best-fit model SED to the surrounding annular zone.  The young stellar population is assumed to have an age of 3\,Myr, at which time emission lines from H\,II regions peak in brightness, and a metallicity of $Z = 0.5 {\rm \, Z_\sun}$, close to that of the ICM and that inferred for the young stellar populations in the three BGCs found to contain relatively young stars (see Fig.\,\ref{fig:SED-fitting}).  This choice was motivated by the detection of the nuclear source in the BGCs of both Abell\,209 and Abell\,611 in filters encompassing H$\alpha$+[N\,II], H$\beta$+[O\,III], [O\,II], and UV (specifically, the longest wavelength filter in the UV centred at 3900\,\AA).  In the model SED fit, we tuned the mass of the young stellar population so as to approximately reproduce the measured brightness in filters encompassing H$\alpha$+[N\,II] (red filled circles).  The results are shown by the blue curves in Figure\,\ref{fig:old+young}, to be compared with the SEDs measured over the central zone as indicated by filled circles.   The composite model SEDs bear no resemblance to the measured SEDs, predicting a UV continuum far in excess of that actually observed.  These model SED fits imply that the nuclear sources in these BCGs cannot be produced by recently formed stars still surrounded by H\,II regions.

Next, we considered the possibility that the nuclear sources are associated with less youthful stellar populations no longer surrounded by H\,II regions.  Like before, we fitted a composite model SED comprising a relatively young stellar population -- but now with an age to be freely fit and a metallicity constrained to the range $0.5 \, {\rm Z_\sun} \leq Z \leq 1.5 \, {\rm Z_\sun}$ -- in addition to an old stellar population having an age and metallicity as adopted previously.  As shown by the orange curves in Figure\,\ref{fig:old+young}, such a model SED provides a poor fit to the measured SED of the nuclear source in Abell\,209, but a reasonable match to that in Abell\,611.  In the latter case, the age of the younger stellar population is 1.1\,Gyr and its total initial stellar mass $2.2 \times 10^{8}\, \rm M_{\odot}$.  The SED of the nuclear source in the other five BCGs also can be satisfactory fit by a composite SED whereby the younger stellar population has an age within the range 0.5--2.8\,Gyr and a total initial stellar mass spanning the range $10^{7} \rm \,$--$10^{10}\, \rm M_{\odot}$.  The nuclear sources in all but one of the BCGs in Figure\,\ref{fig:colours-extended-A} may therefore comprise stellar populations that are significantly younger than the surrounding stars, although their compact morphologies would seem to favour AGNs.

\subsection{UV continuum from old stars}  \label{subsec:evolved_stars}

\begin{figure*}[htb]
\centering
\includegraphics[width=16cm]{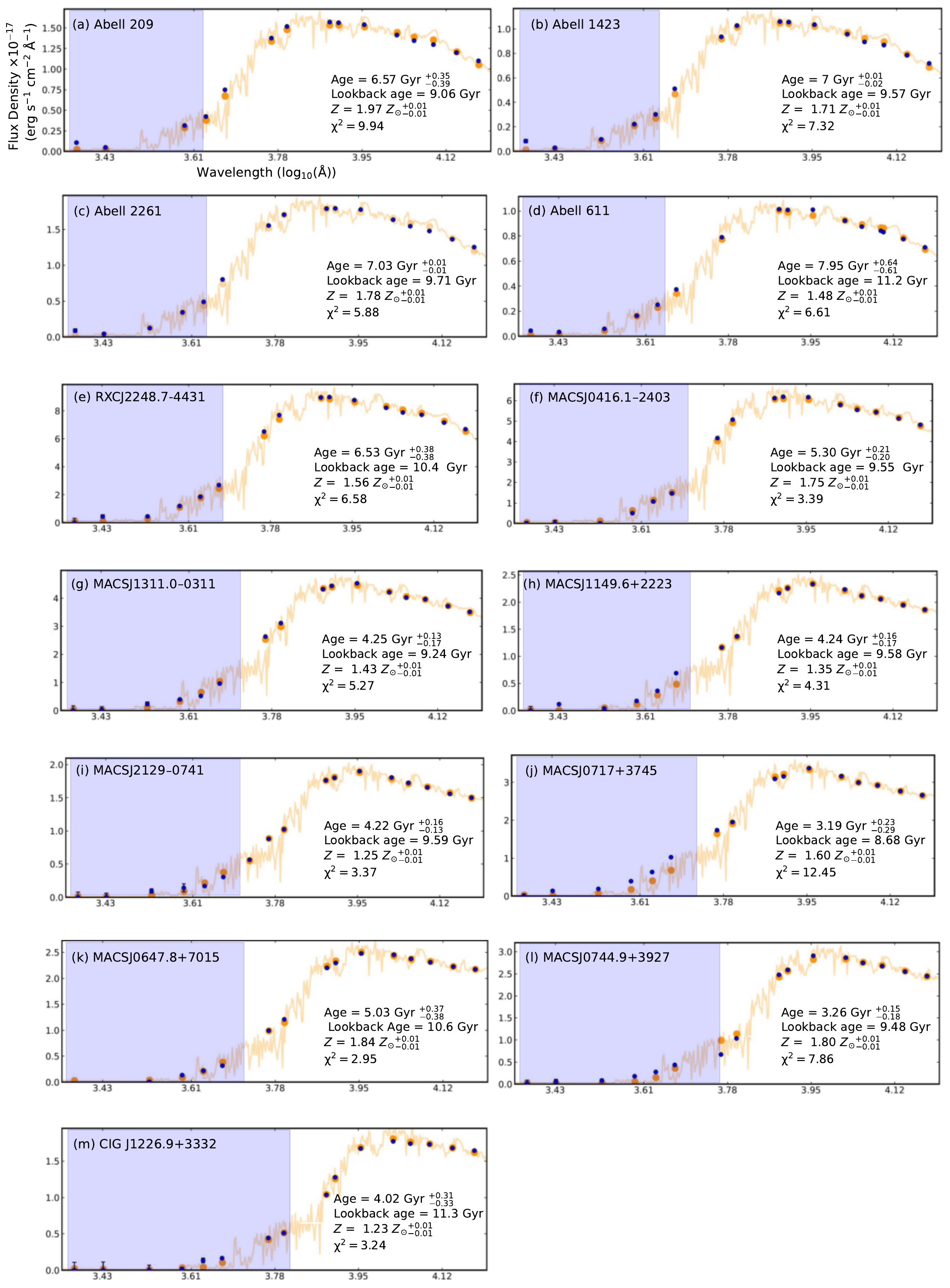}
\caption{SEDs of the thirteen BCGs that display smooth and diffuse UV apart from the nuclear source detected in seven cases as shown in Fig.\,\ref{fig:colours-extended-A}.  These SEDs were extracted over an elliptical aperture having a semi-major axis of 2\arcsec, ellipticity, and position angle as listed in Table\,\ref{BCG parameters} centred on the respective BCGs.  A central mask of radius 0\farcs2 has been applied to the seven BCGs that possess nuclear sources.  Yellow curves show best-fit model spectra corresponding to a single stellar population constrained to have an age with a look-back time between 8\,Gyr and 13\,Gyr, and a metallicity of between $1 {\rm \, Z_\sun}$ and $2.5  {\rm \, Z_\sun}$.  The orange points correspond to the model spectra convolved by the passbands of the $CLASH$ filters.  Purple bands indicate those filters spanning UV wavelengths at the rest-frame of the BCG.  The SEDs of all these BCGs can each be adequately represented by an old stellar population having a super-solar metallicity, indicating that their UV is produced entirely by old stars.}
\label{fig:oldbcgs}
\end{figure*}

\begin{figure*}[ht!]
\centering
\includegraphics[width=17cm]{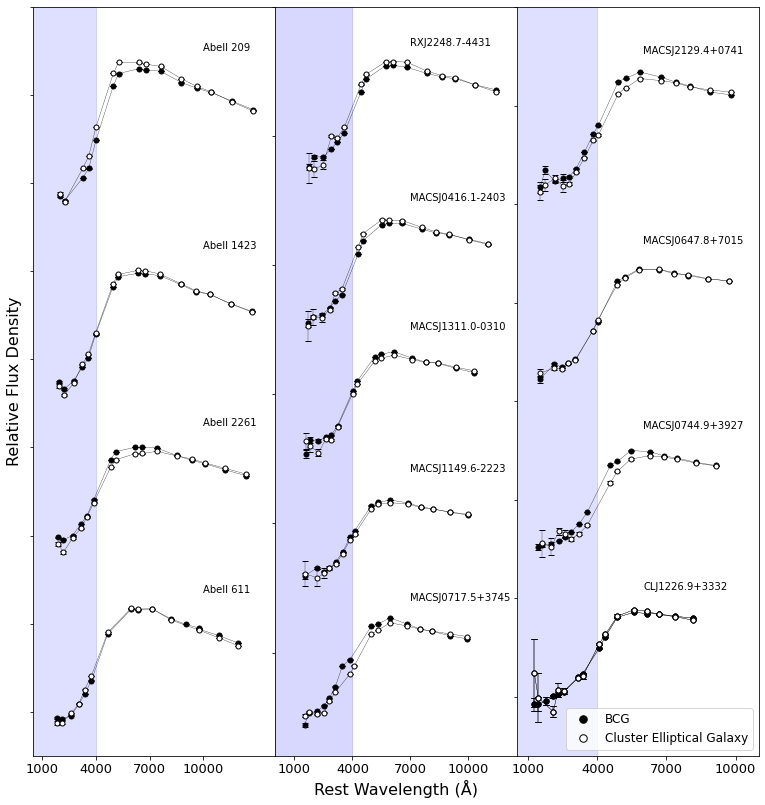}
\caption{Rest-frame SEDs of the BCGs that exhibit smooth and diffuse UV around their centres apart from any nuclear source (filled circles, same as Fig.\,\ref{fig:oldbcgs}), along with the SEDs of the second-ranked galaxies in the same clusters (unfilled circles).  A central mask of radius 0\farcs2 has been applied to the seven BCGs in Fig.\,\ref{fig:colours-extended-A}, as well as to the second-ranked cluster member galaxies, that possess nuclear sources.  The SEDs of the pair of galaxies in each cluster have been normalised to their average intensities at the four longest wavelengths.  Both the BCG and the second-ranked galaxy in each cluster exhibit similar SEDs, including in their rest-frame UV as indicated by the purple bands, indicating that their SEDs are compatible with an old stellar population having a super-solar metallicity as demonstrated in Figure\,\ref{fig:oldbcgs} for the BCGs.}
\label{fig:oldbcgs members}
\end{figure*}

Other than the three BCGs shown in Figure\,\ref{fig:colours-extended-B}, the remaining thirteen that we studied exhibit smooth and diffuse UV continuum around their centres except for those also containing a compact nuclear source that is detectable in the UV.  
We extracted SEDs for all these thirteen BCGs over a central elliptical aperture with a semi-major axis of 2\arcsec\ according to their ellipticities and position angles for their major axes as summarised in Table\,\ref{BCG parameters}.  For the seven BCGs that exhibit nuclear sources (Fig.\,\ref{fig:profile}), we placed a mask of radius 0\farcs2 over their central regions to guard against any AGN continuum emission.  For each BCG, we then fit a model SED comprising a single stellar population constrained to have an age corresponding to a look-back time between 8\,Gyr and 13\,Gyr, and a metallicity constrained to between $1 {\rm \, Z_\sun}$ and $2.5 {\rm \, Z_\sun}$. Figure\,\ref{fig:oldbcgs} shows the model spectrum for each of these BCGs in orange.  In this figure also, we plot as orange points the model spectra convolved over the bandpasses of the $CLASH$ filters, to be compared with the measured SEDs as indicated by the black points.  As can be seen, the model SEDs are able to adequately reproduce the measured SEDs at all wavelengths including the rest-frame UV (data points enclosed within the purple bands) -- indicating that the centrally-enhanced UV continuum of all these galaxies can be explained by old stars, most likely those that have evolved away from the main sequence.


If our interpretation is correct, then other cluster member elliptical galaxies also should show centrally-enhanced UV continuum.  Figure\,\ref{fig:oldbcgs members} shows the rest-frame SEDs of all thirteen BCGs (same as those in Fig.\,\ref{fig:oldbcgs}) plotted as filled circles along with the corresponding SEDs of the second-brightest galaxy in each of these clusters plotted as unfilled circles.  Like for the BCG, the SEDs of the second-ranked cluster members were extracted from a central elliptical aperture with a semi-major axis of 2\arcsec\ and an ellipticity as well as position angle for the major axis as determined from a S\'ersic fit to these galaxies; for those exhibiting a detectable nuclear source, we masked their central region over a radius of 0\farcs2.  As can be seen, all the second-ranked cluster members exhibit UV continuum at a relative intensity (i.e., compared to their intensities in the optical and near-IR) similar to that of the BCGs hosted by the same clusters.  This comparison indicates a common source for the centrally-enhanced UV continuum in all these galaxies, namely an old stellar population having a super-solar metallicity as indicated by the SED fits shown in Figure\,\ref{fig:oldbcgs}.


\section{Discussion}  \label{sec:Discussion}
Among the sixteen BCGs studied here that \citet{Donahue2015} found to exhibit apparently simple UV morphologies characterised by a relatively compact central enhancement, we show that three exhibit spatially-extended and complex UV at their inner regions (see Fig.\,\ref{fig:colours-extended-B}) produced by recently formed stars (confirmed by model fits to their SEDs in Fig.\,\ref{fig:SED-fitting}).
\citet{Fogarty2015} argue for recent star formation in two of these BCGs (Fig.\,\ref{fig:colours-extended-B}$b$--$c$), but not the third (Fig.\,\ref{fig:colours-extended-B}$a$), based in part on the detection of emission lines from the former two as we also confirm.  Combined with the other nine BCGs in the $CLASH$ program that \citet{Donahue2015} found to display obviously complex UV morphologies tracing newly-formed stars, twelve BCGs in total exhibit recent if not ongoing star formation (see summary in Table\,\ref{BCG activity}).  We shall henceforth refer to these particular BCGs as sf-BCGs.  

By contrast, the smooth and diffuse UV of the remaining thirteen BCGs are entirely consistent with light from old stellar populations (see Fig.\,\ref{fig:oldbcgs} and also Fig.\,\ref{fig:oldbcgs members}).  These particular BCGs are therefore designated as non-sf-BCGs.  Seven of these BCGs exhibit spatially-unresolved nuclear sources, which in six cases have SEDs compatible with relatively young stellar populations having ages in the range $\sim$0.4--2.5\,Gyr.  These BCGs may have ceased star formation in the recent past, although we cannot rule out the possibility that all their nuclear sources correspond to AGNs.

\subsection{Connection between BCG activity and $K_0$}

Despite studies demonstrating the lack of an entropy floor as inferred from the concept of an ``excess core entropy,'' $K_0$ (see section\,\ref{subsec:connecting activity entropy}), we affirm a clear preference for sf-BCGs to be hosted by clusters having relatively low values of $K_0$.  
Specifically, all eleven of the $CLASH$ clusters having $K_0 \leq 24 \rm \, keV \, cm^2$ host sf-BCGs.  Among the remaining fourteen having $K_0 \geq 42 \rm \, keV \, cm^2$, only one hosts a star-forming BCG.  This cluster has $K_0 = 69 \pm 10 \rm \, keV \, cm^2$, well above the threshold of $K_0 \approx 30 \pm 10 \rm \, keV \, cm^2$ at which clusters hosting sf-BCGs have previously been reported.  

We confirm the detection of emission lines from the BCGs of MS2137.3-2353 (Fig.\,\ref{fig:colours-extended-B}$b$) and RXC\,J2129.7+0005 (Fig.\,\ref{fig:colours-extended-B}$c$), as was found also by \citet{Fogarty2015}. In both these BCGs, the portions of their nebulae away from bright continuum associated with recently formed stars, which in continuum-subtracted images can be confused with line emission, have spectral features different from H\,II regions but similar to that of optical emission-line nebulae in BCGs.  Combined with the spatially-extended line-emitting gas reported by \citet{Fogarty2015} for the nine BCGs that display obviously complex UV morphologies, all eleven clusters having $K_0 \leq 24 \rm \, keV \, cm^2$ host BCGs possess emission-line nebulae -- referring to line-emitting gas hereafter irrespective of its nature.  By contrast, we find that none of the remaining fourteen clusters having $K_0 \geq 42 \rm \, keV \, cm^2$ host BCGs possessing detectable emission-line nebulae.  The presence of any emission-line nebula in the lone sf-BCG among this group, however, is difficult to disentangle from the optical continuum associated with its recently-formed stars (Fig.\,\ref{fig:colours-extended-B}$a$).  

The aforementioned results support the previously established relationship found between star formation as well as emission-line nebulae in BCGs and $K_0$ of their host clusters. Whereas previous studies found that a large fraction of clusters having $K_0 < 30 \rm \, keV \, cm^2$ host BCGs exhibiting star formation and possessing optical emission-line nebulae, we find this to be true for all eleven of the $CLASH$ clusters having $K_0 < 30 \rm \, keV \, cm^2$.  The previous studies do involve larger samples, but use ground-based optical or space-based UV telescopes (see Section\,\ref{subsec:connecting activity entropy}) that are inferior compared with the {\it HST} as employed in the $CLASH$ program at revealing recently-formed stars.  Our results also demonstrate, however, that sf-BCGs are not restricted exclusively to clusters having $K_0 < 30 \rm \, keV \, cm^2$: among the remaining fourteen $CLASH$ clusters having $K_0 > 30 \rm \, keV \, cm^2$, one hosts a sf-BCG.  The small sample size involved indicates that sf-BCGs may not be as rare as previously thought among clusters having $K_0 > 30 \rm \, keV \, cm^2$.

\begin{figure}[htb!]
\centering
\includegraphics[width=8cm]{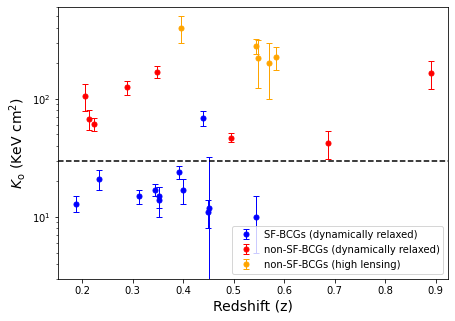}
\caption{Inferred $K_0$ (see text) versus redshift for the $CLASH$ clusters.  Blue and red symbols indicate, respectively, those hosting BCGs with and without star formation among the twenty clusters in the dynamically-relaxed subsample.  Yellow symbols indicate the five clusters in the high-lensing subsample, none of which host BCGs with detectable star formation.  A black dashed line at $K_0 = 30 \rm \, keV \, cm^2$ neatly separates clusters hosting BCGs with and without star formation in all but one case.}
\label{fig:redshift}
\end{figure}

To examine any redshift dependence in $K_0$ that might give rise to a selection effect in detecting star formation and emission-line nebulae among the $CLASH$ BCGs, Figure\,\ref{fig:redshift} shows $K_0$ versus the cluster redshift.  The blue and red symbols indicate clusters hosting sf-BCGs and non-sf-BCGs, respectively, among the dynamically-relaxed subsample; the yellow symbols indicate clusters selected for their exceptional lensing magnification, all of which host non-sf-BCGs.  The twelve clusters hosting sf-BCGs span $0.19 \lesssim z \lesssim 0.55$.  Over a similar redshift range, there are six clusters among the dynamically-relaxed subsample hosting non-sf-BCGs, most of which span $0.20 \lesssim z \lesssim 0.35$; the relatively low redshifts of these BCGs suggest no appreciable dependence in the ability to detect star formation among the $CLASH$ BCGs at redshifts of up to at least $z \sim 0.55$.  Among the dynamically-relaxed subsample, only two clusters lie at $z > 0.55$, both of which host non-sf-BCGs.  A dotted line at $K_0 = 30 \rm \, keV \, cm^2$ in Figure\,\ref{fig:redshift} neatly separates clusters hosting sf-BCGs and non-sf-BCGs in all but one case. 

\begin{figure}[htb!]
\centering
\includegraphics[width=8.5cm]{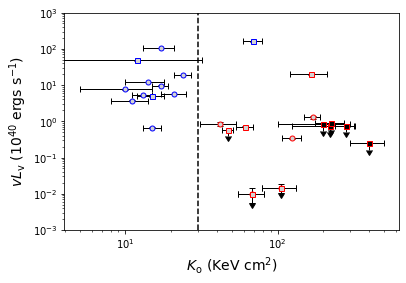}
\caption{AGN radio power of $CLASH$ BCGs versus $K_0$ of their host clusters.  All measurements and upper limits were made with the VLA, such that unfilled circles are measurements made from targeted observations at 1.5\,GHz, and unfilled squares measurements based on the NVSS or FIRST sky surveys at 1.4\,GHz, as reported by \citet{Yu2018} and \citet{Xie2020}.  Filled squares are measurements also based on the NVSS or FIRST that we made.  Colors have the same meaning as in Fig.\,\ref{fig:redshift}.  The black dashed line indicates $K_0 = 30 \rm\, keV\, cm^2$.  On the whole, BCGs hosted by clusters with $K_0 < 30 \rm\, keV\, cm^2$ have more radio-luminous AGNs, although BCGs on either sides of this divide have similar upper ranges in their AGN radio luminosities.}
\label{fig:radio}
\end{figure}

In Figure\,\ref{fig:radio}, we plot the AGN radio luminosity (or 3$\sigma$ upper limit thereof) of the $CLASH$ BCGs as measured with the VLA versus $K_0$ of their host clusters.  The measurements by \citet{Yu2018} and \citet{Xie2020} at 1.5\,GHz are indicated by unfilled circles or unfilled squares, where circles indicate targeted observations and squares indicate measurements based on the NVSS or FIRST sky surveys at 1.4\,GHz.  Measurements that we made also based on the NVSS or FIRST sky surveys at 1.4\,GHz are indicated by filled squares.  On the whole, the eleven BCGs inhabiting clusters with $K_0 \le 24 \rm \, keV \, cm^2$ exhibit more radio-luminous AGNs than those inhabiting clusters with $K_0 \ge 42 \rm \, keV \, cm^2$.  Nonetheless, clusters on either sides of this division in $K_0$ host BCGs spanning the same upper range in AGN radio luminosities.  Indeed, the most radio-luminous AGN is that in the BCG of MACS\,J1206.2-0847, the lone sf-BCG among the clusters with $K_0 \ge 42 \rm \, keV \, cm^2$.  

Overall, our results favour the preponderance of more radio-luminous AGNs among BCGs inhabiting clusters having $K_0 < 30 \rm \, keV \, cm^2$ as found by \citet{Cavagnolo2008}.  On the other hand, the incidence of radio AGNs also among BCGs inhabiting clusters having $K_0 \ge 42 \rm \, keV \, cm^2$ (at least six of fourteen cases, all of which are among the ten clusters belonging to the dynamically-relaxed subsample) indicate that radio AGNs are relatively common in BCGs irrespective of $K_0$.  Thus, the physical processes that fuel AGNs in BCGs need not necessarily be the same as those that produce emission-line nebulae and fuel star formation in these galaxies (see Section\,\ref{subsec:entropy floor} for a different pathway for fuelling AGNs). 

\subsection{What Aspect of ICM Entropy does $K_0$ actually Reflect?}\label{subsec:K_0}

The strong dependence of star formation, optical emission-line nebula, as well as preferentially more radio-luminous AGNs in BCGs on $K_0$ as derived by \citet{Cavagnolo2009} for their host clusters suggests that $K_0$ reflects a genuine physical aspect of the ICM entropy -- subsequently interpreted as an entropy floor, but for which later studies based largely on different clusters than those studied by \citet{Cavagnolo2009} find no corroborative evidence (see Section\,\ref{subsec:entropy floor}).  Recall (see Section\,\ref{subsec:core entropy}) that \citet{Cavagnolo2009} introduced $K_0$ to quantify the excess in core entropy above the best-fitting power law found at larger radii: $K(r) = K_0 + K_{100} (r/100 {\rm \, kpc})^\alpha$.  A larger $K_0$ therefore suggests either a shallower slope in the radial entropy profile (i.e., smaller $\alpha$) or a larger entropy at a radius of 100\,kpc (i.e., larger $K_{100}$), or both.

\begin{figure*}[htb!]
\centering
\includegraphics[width=18cm]{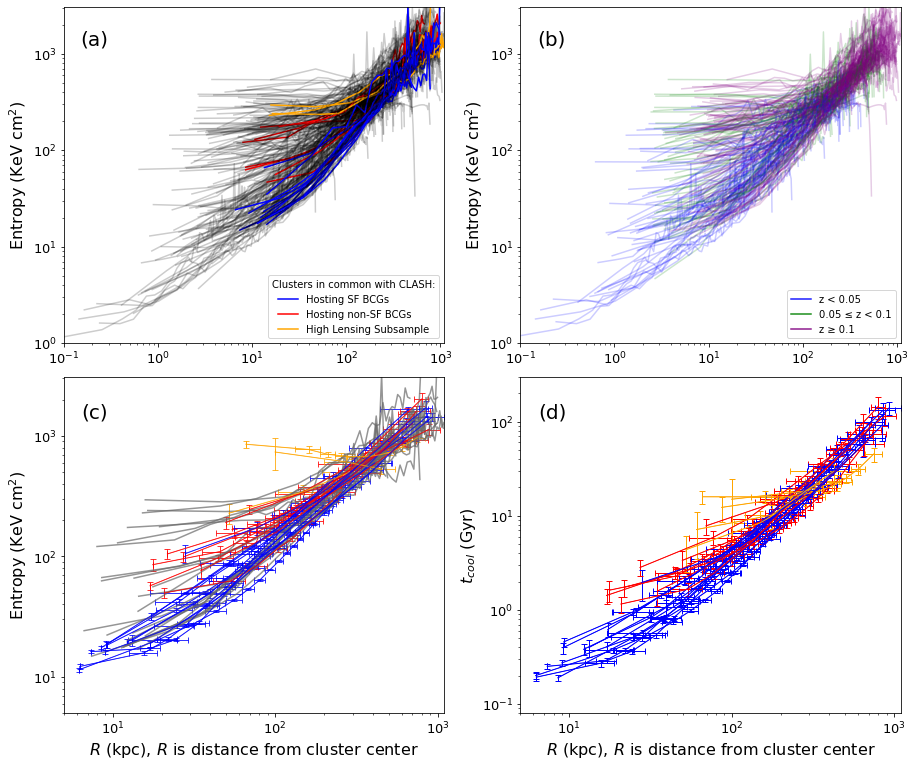}
\caption{(a) Radial entropy profiles derived by \citet{Cavagnolo2009} for 239 clusters having, at the time of their study, publicly-available data from $Chandra$.  For those in common with the $CLASH$ program, blue colors indicate clusters hosting star-forming BCGs (all among the dynamically-relaxed subsample), red colors indicate those hosting non star-forming BCGs among the dynamically-relaxed subsample, and yellow colors are the high lensing-magnification clusters.  (b) Same radial entropy profiles as in (a) but now all color coded based on three redshift bins as indicated at the lower right corner.  As can be seen, the radial entropy profiles of the clusters in the lowest redshift range tend to display less inward flattening than those of the clusters at higher redshifts, attributed (in many cases) to the manner by which these profiles were derived as explained in the text.  (c) Only for those in \citet{Cavagnolo2009} in common with the $CLASH$ program, whereby radial entropy profiles derived by \citet{Donahue2014} for the same clusters are color coded in the same manner as panel $(a)$ and those derived by \citet{Cavagnolo2009} in grey.  (d) Radial cooling timescale profiles calculated based on the radial density and temperature profiles derived by \citet{Donahue2014}, again color coded in the same manner as in panel $(a)$.}
\label{fig:entropy profiles}
\end{figure*}

In Figure\,\ref{fig:entropy profiles}$a$, we plot the radial entropy profiles derived by \citet{Cavagnolo2009} for all 239 clusters with then publicly available data (as of August 2008) in the $Chandra$ Data Archive.  The twenty-three $CLASH$ clusters in common with those studied by \citet{Cavagnolo2009} are indicated in colour: blue for those hosting sf-BCGs and red for those hosting non-sf-BCGs among clusters belonging to the dynamically-relaxed subsample, and yellow for the high-lensing subsample all hosting non-sf-BCGs.  The remaining clusters, far outnumbering the $CLASH$ clusters, are coloured grey.  As can be seen, beyond $\sim$100\,kpc, there is no overall differences in the radial entropy profiles of the different clusters.  Inwards of this radius, the slope of the radial entropy profiles becomes shallower; now, however, there is a clear distinction between $CLASH$ clusters hosting sf-BCGs and non-sf-BCGs.  Specifically, even among the dynamically-relaxed subsample alone, the clusters hosting sf-BCGs (blue curves) generally have lower entropies at 100\,kpc (lower $K_{100}$) as well as inwardly steeper slopes (larger $\alpha$) in their radial entropy profiles than those hosting non-sf-BCGs (red curves).  Providing an even greater contrast, the clusters belonging to the high-lensing subsample (yellow curves), none of which host sf-BCGs, generally have still higher $K_{100}$ and shallower inward slopes in their radial entropy profiles.

As explained in Section\,\ref{subsec:entropy floor}, studies that simultaneously solve for the radial temperature and density profiles of the ICM find no evidence for an entropy floor in the cores of galaxy clusters.  Instead, the radial entropy profiles decrease monotonically inward down to the smallest measurable radii.  If $K_0$ is entirely an artefact of the manner by which ICM temperatures and densities were derived in the era of the work reported by \citet{Cavagnolo2009}, why then is there a dependence between BCG activity and $K_0$?  In Figure\,\ref{fig:entropy profiles}$c$, we plot in color the radial entropy profiles derived by \citet{Donahue2014} for the twenty-three $CLASH$ clusters in common with those studied by \citet{Cavagnolo2009}.  Recall that, unlike \citet{Cavagnolo2009}, \citet{Donahue2014} solved simultaneously for the temperature and density of the ICM -- from which the entropy of the ICM was computed -- in concentric annular bins.  The entropy profiles shown in this figure were computed under the assumption of hydrostatic equilibrium; when the density and temperature of the ICM were derived with no assumptions made about the physical state of the ICM, their values were found to differ little for the dynamically-relaxed subsample, but can show appreciable differences for the high-lensing subsample.  As can be seen, just like for the radial entropy profiles derived by \citet{Cavagnolo2009} for a much larger sample (Fig.\,\ref{fig:entropy profiles}$a$), there is little overall difference between the radial entropy profiles of the $CLASH$ clusters beyond $r \sim 100$\,kpc irrespective of whether they host sf-BCGs or non-sf-BCGs, or whether they belong to the dynamically-relaxed or high-lensing subsamples.  Also like before, inwards of this radius, all the radial entropy profiles exhibit shallower slopes, but with clear differences between clusters hosting sf-BCGs and non-sf-BCGs or between those belonging to the dynamically-relaxed or high-lensing subsamples.  In general, among the dynamically-relaxed subsample, the entropies of the clusters hosting sf-BCGs (blue curves) decrease more steeply inward compared with those hosting non-sf-BCGs (red curves). The clearest exception to this trend is MACS\,J1206.2-0847, the cluster hosting a sf-BCG that has the shallowest slope inwards of $\sim$100\,kpc that more closely resembles, in this respect, the clusters hosting non-sf-BCGs.  In clear contrast to the dynamically-relaxed subsample, the entropies of the clusters in the high-lensing subsample flatten and may even rise inward (yellow curves).  

The corresponding radial entropy profiles derived by \citet{Cavagnolo2009} for the twenty-three clusters in common with the $CLASH$ program are shown by the grey curves in Figure\,\ref{fig:entropy profiles}$c$.  Among the dynamically-relaxed subsample, the radial entropy profiles derived by \citet{Donahue2014} and \citet{Cavagnolo2009} are most similar for those hosting sf-BCGs.  Among the same subsample but now for clusters hosting non-sf-BCGs, the radial entropy profiles derived by \citet{Donahue2014} are generally steeper than those derived by \citet{Cavagnolo2009}, and furthermore do not flatten inwards unlike for a number of these clusters as determined by \citet{Cavagnolo2009}.  By contrast, the radial entropy profiles of the high-lensing subsample derived by \citet{Donahue2014} and \citet{Cavagnolo2009} can differ considerably, presumably (at least in part) because \citet{Donahue2014} assume hydrostatic equilibrium whereas \citet{Cavagnolo2009} made no such assumption.  

The comparisons shown in Figure\,\ref{fig:entropy profiles}$c$ reveal that a higher $K_0$ primarily reflects a shallower radial entropy profile within $r \sim 100 \rm \, kpc$ than a lower $K_0$, but not, except perhaps at large $K_0$, an actual flattening in the radial entropy profile at the cluster core.   
Indeed, for clusters having sufficient X-ray counts that \citet{Cavagnolo2009} could reliably determine the temperature of their ICM inwards of a few $10 \rm \, kpc$, notably the grey curves extending inward to $r \approx 1 \rm \, kpc$ as shown in Figure\,\ref{fig:entropy profiles}$a$, the entropy can be seen to decreases inwards to the smallest measurable radii.  The dramatic flattening in the radial entropy profiles seen by \citet{Cavagnolo2009} for the majority of the clusters studied, as is apparent in Figure\,\ref{fig:entropy profiles}$a$, arises from the large outer radius of the central aperture used to infer the ICM temperature but much finer annular bins within this aperture used to infer the ICM density.  Indeed, the need for a sufficient number of X-ray counts in the central aperture to determine the ICM temperature, thus determining the radius of the central aperture, is reflected in the radial entropy profiles derived by \citet{Cavagnolo2009} as a function of cluster redshift: as can be seen in Figure\,\ref{fig:entropy profiles}$b$, only the radial entropy profiles of the lowest redshift clusters show no inward flattening, as these clusters are sufficiently close that a small central aperture can be used to determined the ICM temperature over their innermost regions.

Our explanation for what $K_0$ actually represents in the ICM entropy aligns with previous work finding no evidence for an entropy floor in the ICM \citep{Panagoulia2014,Hogan2017a,Hogan2017b,Sanders2018}.  Among generally much more nearby clusters than those studied here, \citet{Panagoulia2014}, \citet{Hogan2017a}, and \citet{Hogan2017b} find that the radial entropy profiles of these clusters become shallower inward of $\sim$100\,kpc albeit decreasing monotonically inward to the smallest measurable radii.
Furthermore, our finding that clusters hosting sf-BCGs generally have steeper radial entropy profiles inward of $\sim$100\,kpc than those hosting non-sf-BCGs echo work by \citet{Hogan2017b}, who found the same trend among clusters hosting BCGs possessing optical emission-line nebulae and those not.
\citet{Hogan2017b} conclude that the difference between clusters hosting BCGs with and without optical emission-line nebulae is related to the different cooling timescales of their ICM -- owing to their different entropies -- in the cluster core at radii $\lesssim 10 \rm \, kpc$, leading therefore to relatively strong cooling of the ICM in the former but not the latter.  We affirm this general trend in Figure\,\ref{fig:entropy profiles}$d$, where we show the cooling timescale of the ICM as a function of radius for the $CLASH$ clusters as computed from the radial entropy profiles shown in Figure\,\ref{fig:entropy profiles}$c$ (see Section\,\ref{subsec:core entropy} for computing cooling timescales from the ICM entropy).  

Our work therefore leaves us with the following perplexing question: what is the specific condition necessary for some portion of the ICM to cool catastrophically so as to deposit emission-line nebulae and fuel star formation in BCGs, as well as to, presumably, provide additional fuel to elevate the radio power of their AGNs?  Even the non-sf-BCGs among the dynamically-relaxed subsample have cooling times shorter than $\sim$3\,Gyr within the central few tens of kpc, much shorter than a Hubble time.  At least some these BCGs, in particular those inhabiting clusters with cooling times at their cores similar to clusters hosting sf-BCGs, might be anticipated to display detectable emission-line nebulae and star formation.  To put it another way, why is there such a distinct, albeit perhaps not exclusive, separation at $K_0 \simeq 30 \rm \, keV \, cm^2$ between BCGs displaying star formation as well as emission-line nebulae and those not displaying either signatures (Figure\,\ref{fig:redshift})?  Theoretical predictions that the free-fall time needs to be shorter than the cooling time to promote thermal instabilities \cite[see][]{Voit2017} finds no apparent support in the observational study by \citet{Hogan2017b}, although more work is needed in this respect.
In the next paper of this series, we shall examine in greater detail the question that this work leaves us with -- an examination that reveals why different clusters exhibit different ICM radial entropy profiles to begin with.

\section{Summary and Conclusions}\label{sec:Summary}
Our goal in this paper has been to set the stage for reassessing how star formation, emission-line nebulae, and active galactic nuclei (AGNs) in Brightest Cluster Galaxies (BCGs) are related to the thermodynamics of the intracluster medium (ICM).  Early work had established a dependence between such BCG ``activity'' and an apparent entropy floor in the ICM at the cluster core (Section\,\ref{sec:Intro}).  The notion that the ICM possesses an entropy floor was inferred from work by \citet{Cavagnolo2009}, who introduced the parameter $K_0$ to represent the typical excess of core entropy above the best-fitting power law at larger radii ($\gtrsim$100\,kpc) -- while warning that $K_0$ is not meant to represent a minimum core entropy.  The existence of an entropy floor in the ICM was then bolstered by several studies reporting that many clusters having $K_0 \lesssim 30 \rm \, keV \, cm^2$ host BCGs displaying star formation and nebulae, whereas no clusters having $K_0 > 30 \rm \, keV \, cm^2$ host such BCGs.  Furthermore, BCGs in clusters having $K_0 \lesssim 30 \rm \, keV \, cm^2$ possess preferentially more radio-luminous AGNs than those in clusters having $K_0 > 30 \rm \, keV \, cm^2$, although not all BCGs on either sides of this divide possess detectable radio AGNs.

A reassessment of how BCG activity depends on ICM entropy is needed as later studies have found no evidence for entropy floors in clusters (Section\,\ref{subsec:entropy floor}).  Our reassessment is based on the twenty-five clusters observed in the Cluster Lensing And Supernova survey with Hubble ($CLASH$) program (see brief summary of selection criteria in Section\,\ref{subsec:goals}).  This program provides deep images at high angular resolutions from UV to near-IR wavelengths, permitting an especially sensitive search for star formation and candidate optical AGNs, as well as a search for optical emission-line nebulae, in BCGs.  In addition, both the physical properties of the ICM and the mass profiles of these clusters have been studied in detail, providing the necessary information for a close examination of the dependence between BCG activity and ICM thermodynamics in these clusters.  Previous work (summarised in Section\,\ref{sec:Results}) have shown that nine of the BCGs display highly complex UV morphologies with projected linear extents of up to $\sim$100\,kpc, a clear signature of recent and perhaps ongoing star formation.  The remaining sixteen appear to exhibit simple UV morphologies characterised by a relatively compact central enhancement.  Among these sixteen BCGs, we show definitively (Sections\,\ref{sec:DataProc}--\ref{sec:Interpretation}) that three exhibit star formation at their inner regions, at least two of which also possess nebulae.  For two of these BCGs, the mass in newly-formed stars detectable in the UV is about 0.5\% of their total stellar mass (larger if the UV suffers dust extinction).  By contrast, the diffuse UV of the remaining thirteen, none of which possess nebulae, can be explained entirely by old stars.  Among these thirteen BCGs, seven exhibit spatially-unresolved nuclear sources, which in all but one case have SEDs that can be reproduced by a relatively young stellar population ($\sim$0.5--2.8\,Gyr) although this does not rule out AGNs.  A concise summary of BCG activity in all the $CLASH$ clusters can be found in Table\,\ref{BCG activity}.

Our results support the previously established dependence between star formation in BCGs and $K_0$ in their host clusters (Section\,\ref{sec:Discussion}).  All eleven clusters having $K_0 \le 24 \rm \, keV \, cm^2$ host star-forming BCGs.  On the other hand, only one of fourteen clusters having $K_0 \ge 42 \rm \, keV \, cm^2$ hosts a star-forming BCG, the first time a BCG displaying star formation has been found among clusters having $K_0 > 30 \rm \, keV \, cm^2$.  Rather than reflecting an entropy floor in the ICM, we show that $K_0$ reflects the degree to which, within $\sim$100\,kpc of the cluster center, the slope of the radial entropy profile becomes shallower rather than actually flattening (see the key plots in Fig.\,\ref{fig:entropy profiles}) -- in agreement with studies revealing no evidence for ICM entropy floors.  Clusters having lower entropies, corresponding to shorter cooling times, at a given radius over the cluster core preferentially host BCGs displaying star formation, nebulae, and more radio-luminous AGNs.  Many of the BCGs not displaying star formation or nebulae do possess radio AGNs, however, indicating multiple pathways for fuelling AGNs in BCGs.

Among the twenty-five clusters in the $CLASH$ program, twenty are classified by the $CLASH$ team as being dynamically relaxed based on their ICM properties as inferred from X-rays.  The remaining five, selected for their high lensing magnifications, are clearly disturbed by recent or ongoing mergers.  All eleven clusters hosting star-forming BCGs belong to the dynamically-relaxed subsample, accounting for just over half of the number in this subsample (Section\,\ref{sec:Discussion}).  Even among this subsample alone, BCG activity clearly depends on how steeply the ICM radial entropy profile, and hence the cooling time of the ICM,  decreases inward at the cluster core.  This dependence, however, does not explain why not all clusters having cooling timescales significantly shorter than a Hubble time host BCGs displaying emission-line nebulae or star formation, nor why different clusters possess different radial entropy profiles.  In the next paper of this series, we shall make use of the results reported here for the BCGs in the $CLASH$ program to make a more thorough examination of the link between BCG activity and ICM thermodynamics.  This more complete assessment yields not just greater insights to the dependence of BCG activity on ICM thermodynamics, but also why different clusters -- even those among the dynamically-relaxed subsample -- exhibit different ICM thermodynamical profiles.

\begin{acknowledgments}
We thank the referee for the careful reading of the manuscript and the constructive criticisms of the presentation and analyses that have resulted in a significant improvement to this paper. We are indebted to all those who planned, executed, and processed the data from the $CLASH$ program. J. L. and A. L. acknowledges support from the Research Grants Council of Hong Kong through the General Research Fund 17300620. M.D. is grateful for partial support of this work from NASA award NASA-80NSSC22K0476 and SAO/Chandra award SAO-AR1-22011X. This research employed observations made with the NASA/ESA Hubble Space Telescope and made use of archival data from the Hubble Legacy Archive, which is a collaboration between the Space Telescope Science Institute (STScI/NASA), the Space Telescope European Coordinating Facility (ST-ECF/ESAC/ESA) and the Canadian Astronomy Data Centre (CADC/NRC/CSA). This research has also made use of services of ESO Science Archive Facility. Based on the observation collected at ESO under the program ID 095.A-0181(A) and 097.A-0269(A). 

\end{acknowledgments}



%

\vspace{5mm}
\facilities{HST (ACS and WFC3), VLT (MUSE)}


\software{astropy \citep{astropy:2013, astropy:2018},  
           SEP \citep{Bertin1996, Barbary2016},
           BAPGPIPES \citep{Carnall2018},
           IMFIT \citep{Erwin2015},
           MPDAF \citep{Bacon2016}
          }

\bibliography{sample631}{}
\bibliographystyle{aasjournal}

\appendix
\vspace{-8mm}
In Section\,\ref{subsec:star_formation}, we showed that the combination of a young and old stellar population -- the former assumed to have formed in a single short-duration burst, and the latter assumed to have formed at a look-back time of 10\,Gyr -- is able to provide good fits to the measured SEDs of UV-emitting features in the BCGs of RXC\,J2129.7+0005, MS2137.3-2353, and MACS\,J1206.2-0847 (see Fig.\,\ref{fig:SED-fitting}), all of which display a compact central UV enhancement.  In this way, we derived ages of between $\sim$200\,Myr and $\sim$500\,Myr for their young stellar populations.  Far too old to be still associated with H\,II regions, the best-fit model SEDs for the BCGs in RXC\,J2129.7+0005 and MS2137.3-2353 nonetheless significantly under-predict the brightnesses of these two galaxies in filters containing [OII] and H$\alpha$+[NII], albeit not H$\beta$+[OIII].  Such excesses are characteristic spectral signatures of emission-line nebulae in BCGs (see Figure\,\ref{MUSE spectra}), therefore indicating line-emitting gas unrelated to star formation in the BCGs of RXC\,J2129.7+0005 and MS2137.3-2353.  The morphologies, at least in part, of the emission-line nebulae in both these BCGs are shown in Figure\,\ref{fig:colours-extended-B} (see panels that capture the [OII] and H$\alpha$+[NII] lines).  In the BCG of MS2137.3-2353, its emission-line nebula clearly extends well beyond the young stellar population detectable in UV continuum.  When fitting for the BCGs in RXC\,J2129.7+0005 and MS2137.3-2353, we therefore omitted filters containing [OII] and H$\alpha$+[NII] (the inclusion of these filters result in overall poorer fits). 

Instead of an emission-line nebula unrelated to star formation, we now investigate whether a stellar population sufficiently young to still produce H\,II regions can provide just as good a fit to the measured SEDs.  We therefore add, to an old stellar population modelled in the same manner as described above, a young stellar population assumed to have formed: (i) in a single burst less than 10\,Myr ago; or (ii) at a constant rate over the past 100\,Myr (as is approximately presumed when converting between UV continuum and star-formation rate using the \citet{Kennicutt1998} relationship).  We made model SED fits where the dust extinction is left to be freely fit, and corresponding model SED fits where the dust extinction is set to the value inferred by \citet{Fogarty2017}; in the latter case, setting a Gaussian prior in dust extinction corresponding to the uncertainty \citet{Fogarty2017} inferred for the dust extinction.  Unlike before, we now include filters containing [O\,II] and H$\alpha$+[N\,II] (i.e., all filters) in the model SEDs fits, as H\,II regions can elevate the brightness in these (and other) filters owing to line emission. 

\begin{figure}[p]
\centering
\includegraphics[width=16cm]{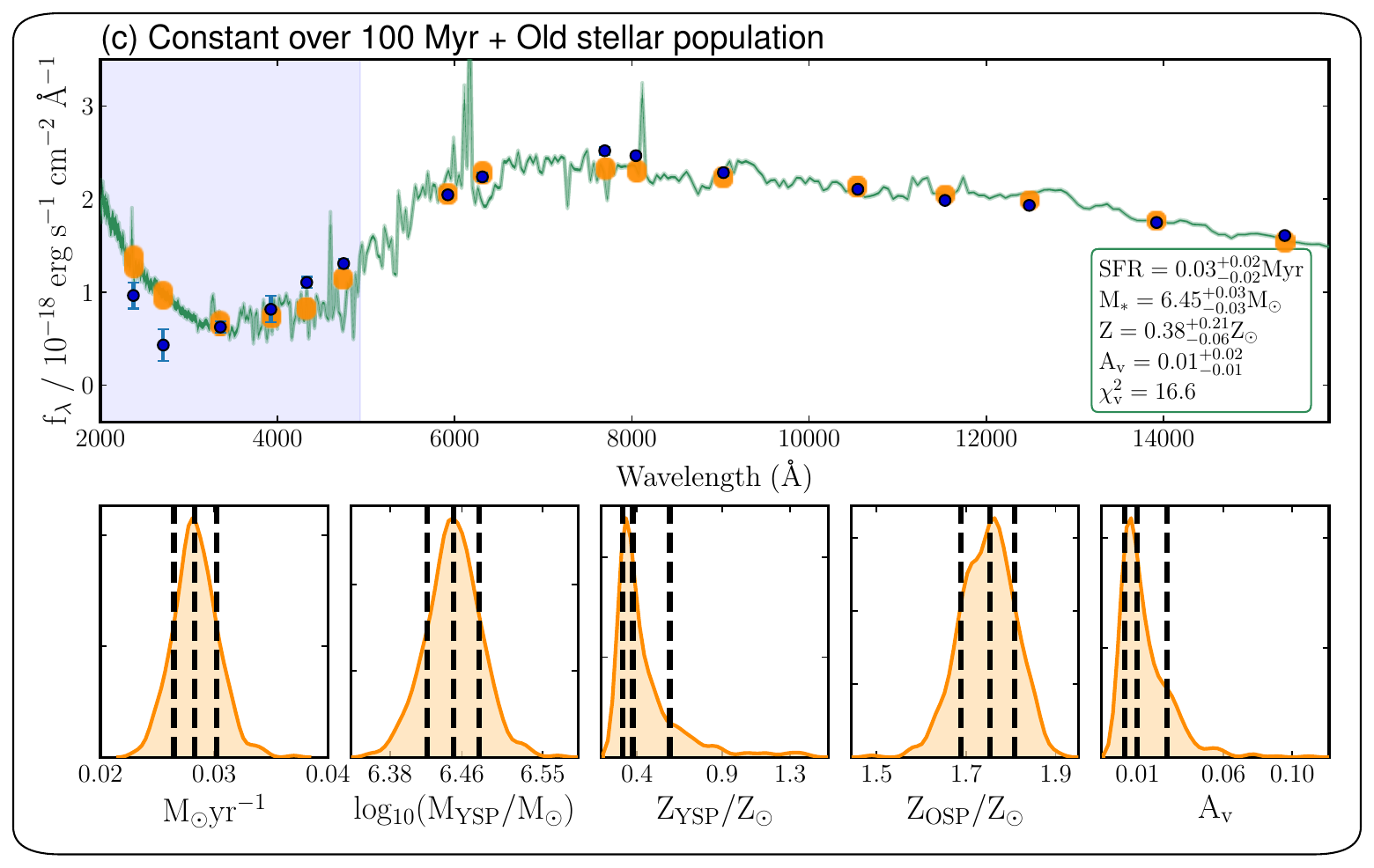}
\includegraphics[width=16cm]{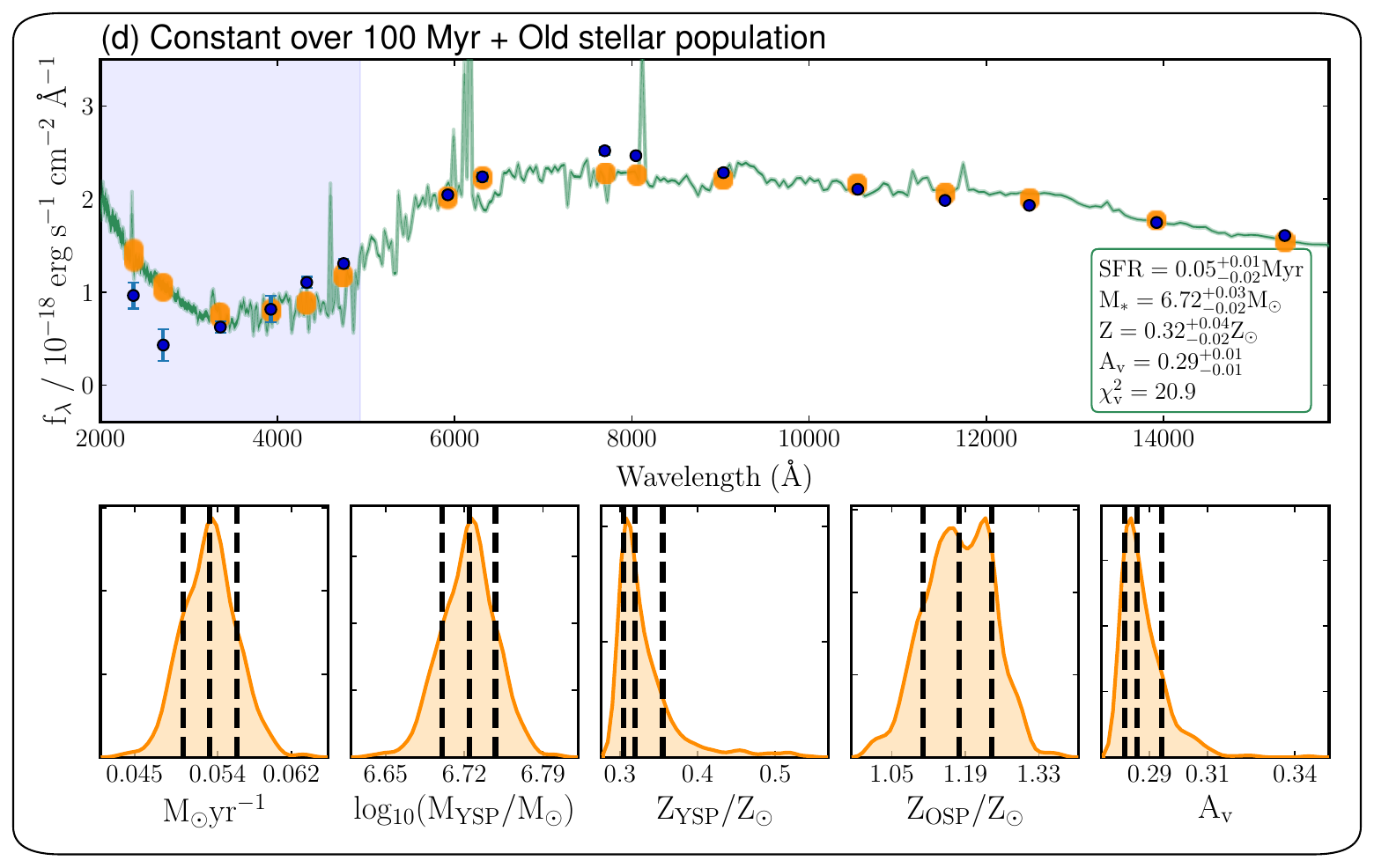}
\caption{In each block of panels, upper row shows best-fit composite model spectrum (green curve) to the spatially-integrated SED (blue data points) for the BCG in RXC\,J2129.7+0005 as extracted over the aperture shown in the insert to Figure\,\ref{fig:SED-fitting}$a$.  Orange points or bands correspond to the model spectra convolved by the bandpasses of the $CLASH$ filters, and have vertical extents spanning the 16th and 84th percentile of the best-fit likelihood distribution in the fitted parameters.  
Purple bands indicate those filters spanning UV wavelengths at the rest-frame of the BCG.  Each composite model comprises two stellar populations, having a best-fit likelihood distribution in age, mass, and metallicity for the younger stellar population, as well as in metallicity for the older stellar population, shown in the lower row of each block of panels; the age of the old stellar population is fixed at a look-back time of 10\,Gyr.  In first two blocks, the young stellar population is assumed to have formed in a single burst less than 10\,Myr ago; in the third and fourth blocks, the young stellar population is assumed to have formed a constant rate over the past 100\,Myr.  In the first and third block, dust extinction has been allowed to be freely fit.  In the second and fourth blocks, the dust extinction has been set to the value inferred by \citet{Fogarty2017} for the BCG in RXC\,J2129.7+0005 using the Calzetti attenuation law, with a Gaussian prior corresponding to the uncertainty inferred by \citet{Fogarty2017} for this dust extinction. 
}
\label{fig:rxj2129_appendix}
\end{figure}

\begin{figure}[p]
\centering
\includegraphics[width=17cm]{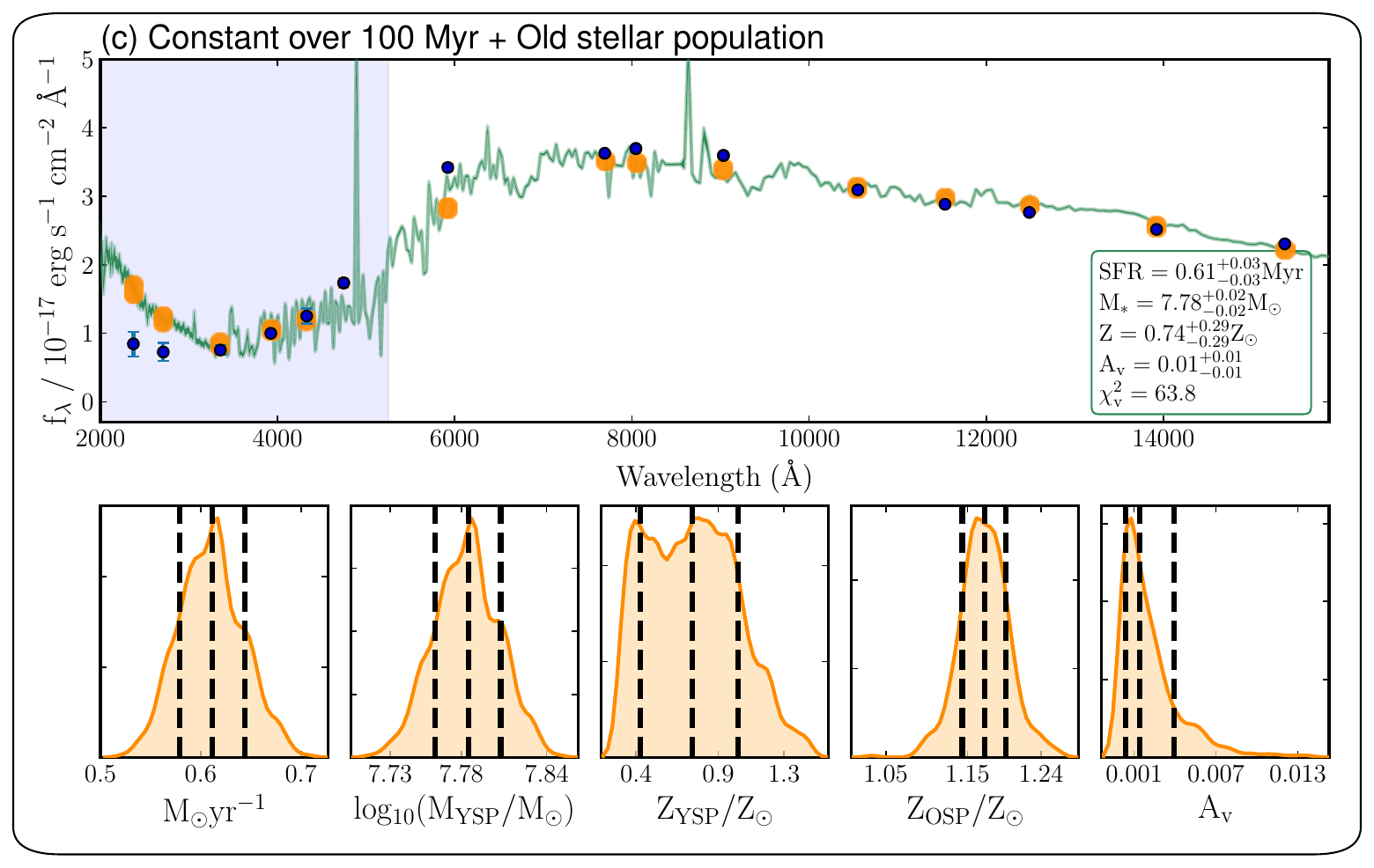}
\includegraphics[width=17cm]{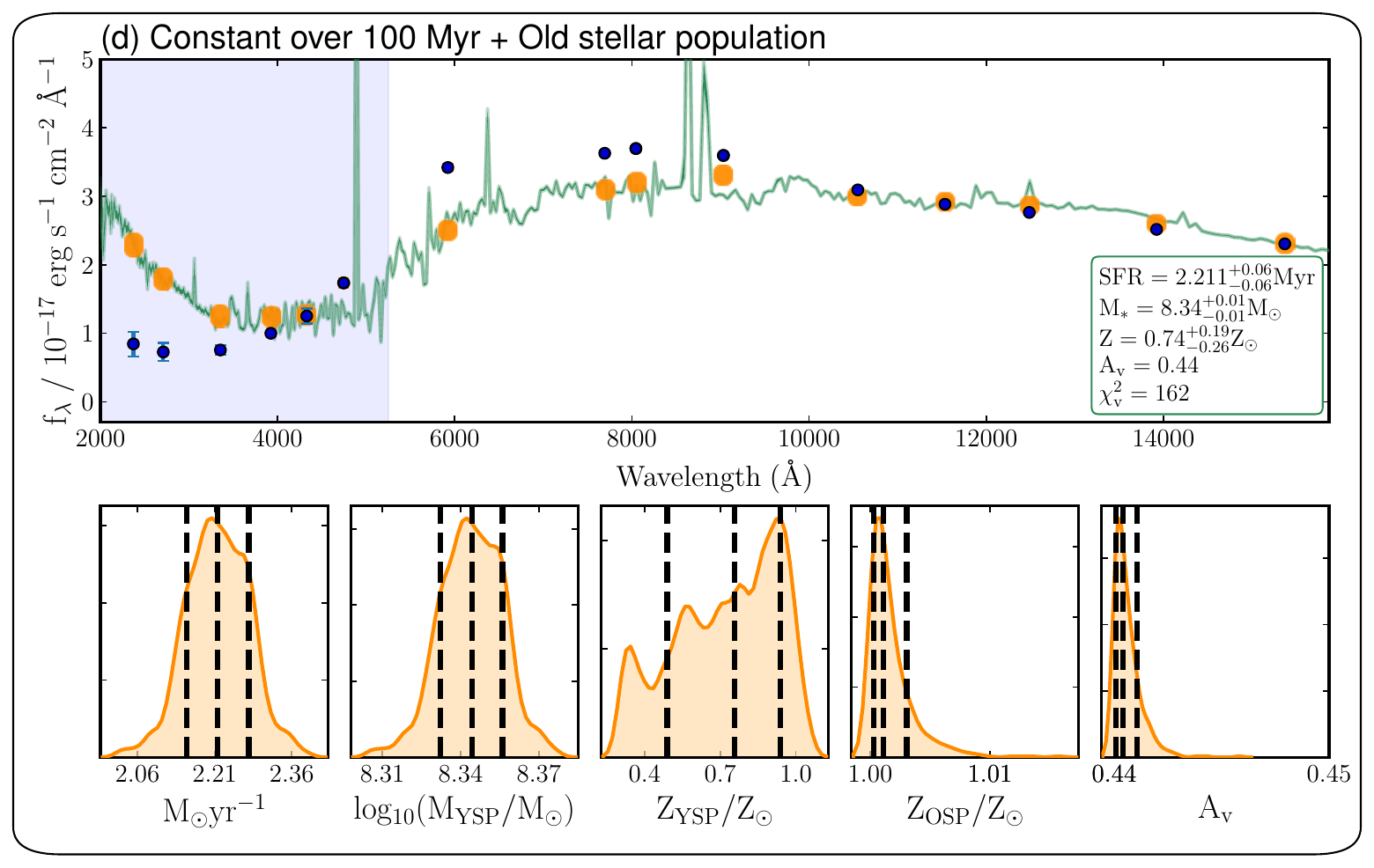}
\caption{Same as for Figure\,\ref{fig:rxj2129_appendix}, but now for the BCG in MS2137.3-2353.  Its measured SED was extracted over the aperture shown in the insert to Figure\,\ref{fig:SED-fitting}$b$.}
\label{fig:ms2137_appendix}
\end{figure}

The results are shown in Figure\,\ref{fig:rxj2129_appendix} for the BCG in RXC\,J2129.7+0005 and Figure\,\ref{fig:ms2137_appendix} for the BCG in MS2137.3-2353.
As can be seen, irrespective of the star-formation history or amount of dust extinction, all the fits are far inferior to those shown in Figure\,\ref{fig:SED-fitting} for either BCG.  In all cases, whereas the model SEDs present a good fit to the rest-frame near-IR filters (dominated by the old stellar population), they under-predict the brightnesses in one or more rest-frame optical filters and over-predict the brightnesses of the two rest-frame UV filters at the shortest wavelengths.  
Although all the model SED fits we try are likely over-simplistic (see discussion in Section\,\ref{subsec:star_formation}), they do collectively indicate that that the young stellar populations in the BCGs of RXC\,J2129.7+0005 and MS2137.3-2353 are predominantly too old to be still associated with H\,II regions, and are therefore unlikely to be primarily responsible for elevating the brightnesses of these BCGs in filters containing [O\,II] and H$\alpha$+[NII].

\begin{figure}[p]
\centering
\includegraphics[width=16cm]{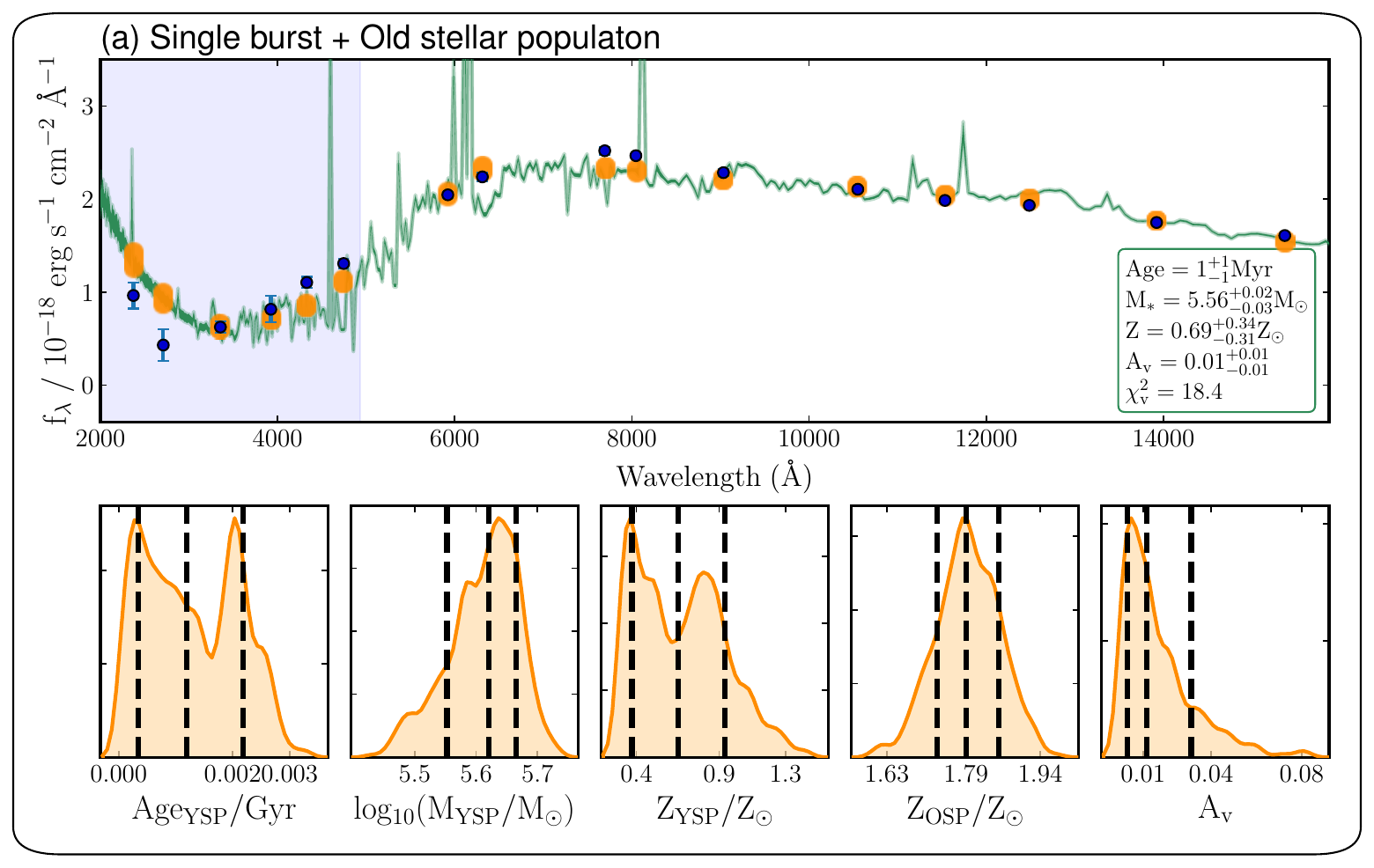}
\includegraphics[width=16cm]{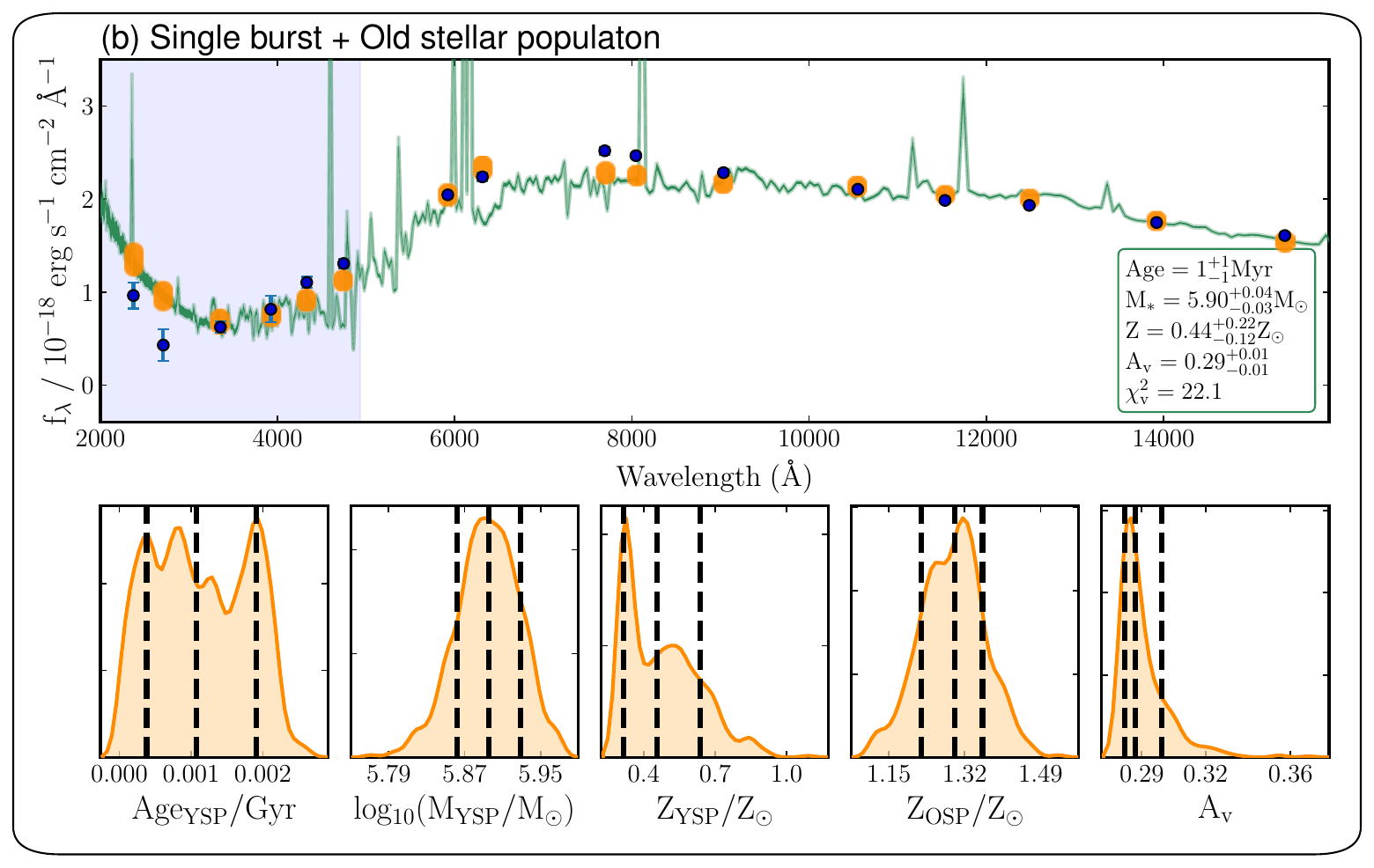}
\end{figure}

\begin{figure}[p]
\centering
\includegraphics[width=17cm]{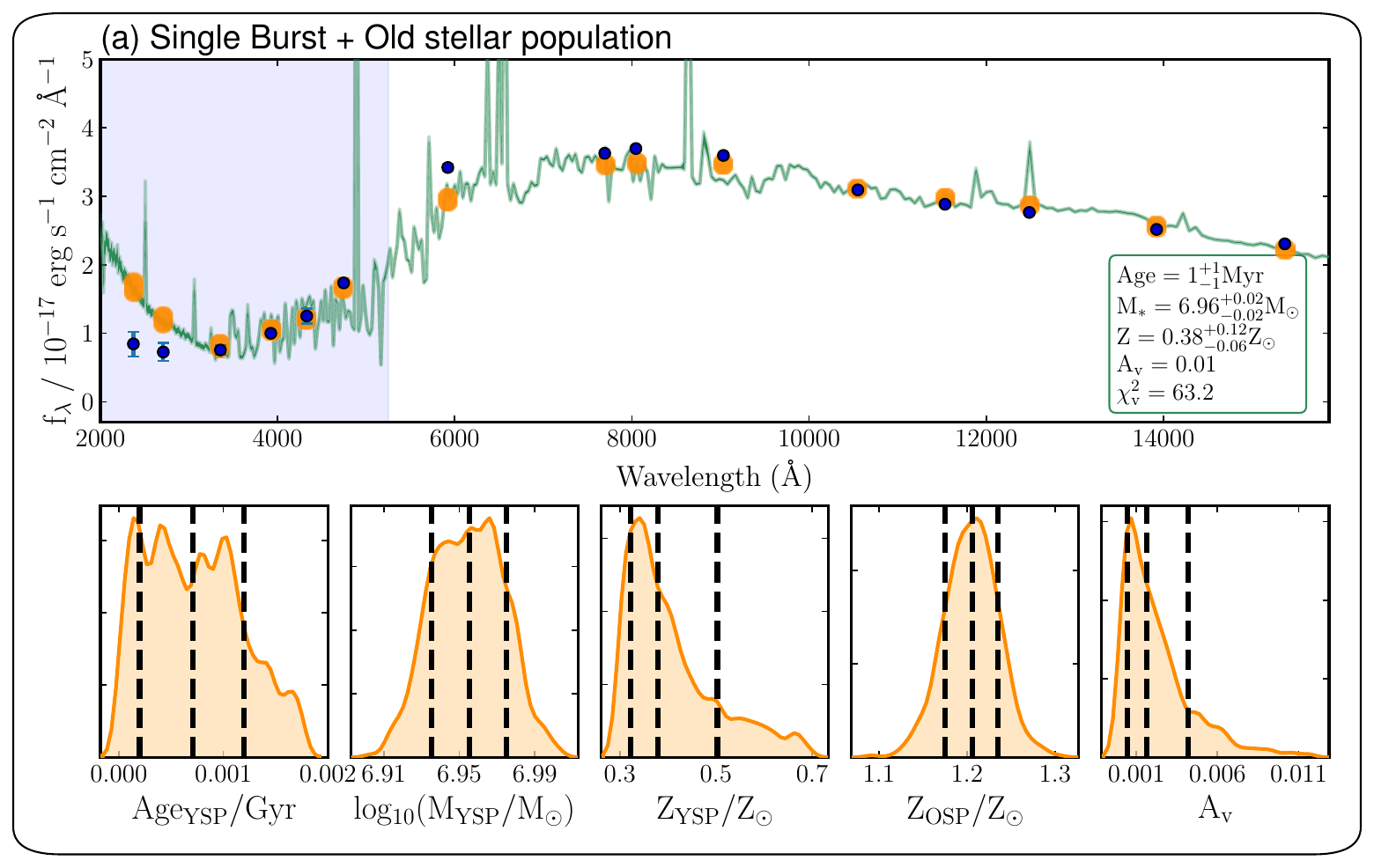}
\includegraphics[width=17cm]{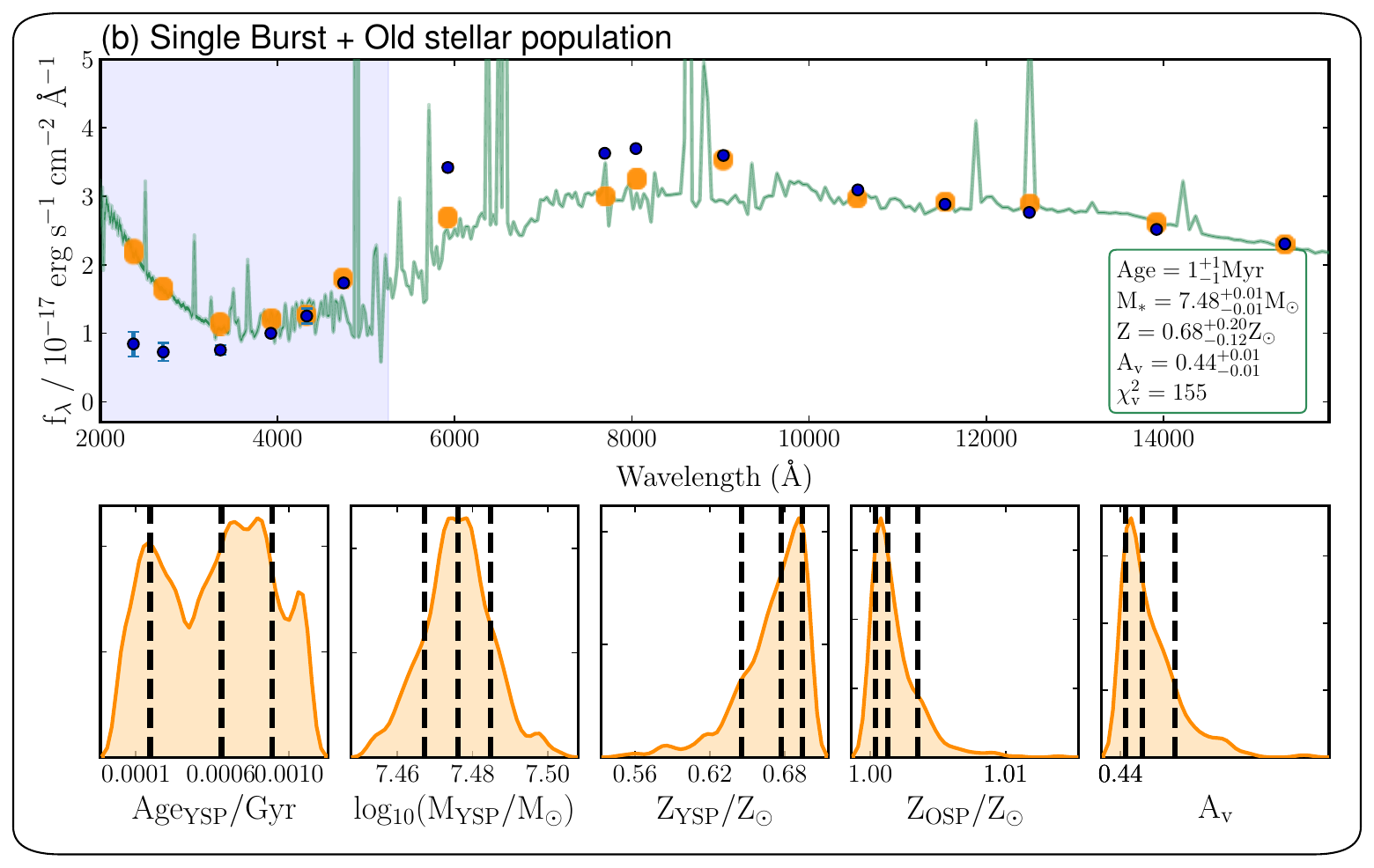}
\end{figure}

\end{document}